\documentclass{aa}
\usepackage{txfonts}
\usepackage{graphicx}
\usepackage{natbib}
\usepackage{multirow} 
\usepackage{lscape}
\usepackage{longtable}
\usepackage{amssymb,amsmath}
\usepackage{color}

\bibpunct{(}{)}{;}{a}{}{,} 

\newcommand{\micron}{\mbox{$\mu$m}}

\begin{document}
	\title{Consistent dust and gas models for protoplanetary disks IV.\\ A panchromatic view of protoplanetary disks}
	\author{O.~Dionatos\inst{1}
          \and
	  P.~Woitke\inst{2,3}
	  \and
	  M.~G{\"u}del\inst{1}
	  \and
	  P.~Degroote\inst{4}
	  \and
	  A.~Liebhart\inst{1}
	  \and
	   F.~Anthonioz\inst{5}
	   \and
	   S.~Antonellini\inst{6, 7}
	   \and
	  C.~Baldovin-Saavedra\inst{1}
	  \and 
	  A.~Carmona\inst{8}
	  \and
	  C.~Dominik\inst{9}
	  \and
	  J.~Greaves\inst{10}
	  \and
	  J.~D.~Ilee\inst{11}
	  \and
	  I.~Kamp\inst{6}
	  \and
	  F.~M\'enard\inst{5}
	  \and
	  M.~Min\inst{9, 12}
	  \and
	  C.~Pinte\inst{5, 13, 14}
	  \and
	  C.~Rab\inst{1, 6}
	  \and
	  L.~Rigon\inst{2}
	 \and
	 W.~F.~Thi\inst{15}
	 \and
	 L.~B.~F.~M.~Waters\inst{9, 12}
          }
\institute{
		University of Vienna, Department of Astrophysics,  T{\"u}rkenschanzstrasse 17,
A-1180, Vienna, Austria\\
\email{odysseas.dionatos@univie.ac.at}
	\and 
		SUPA School of Physics \& Astronomy, University of St Andrews, North Haugh, KY16 9SS, St Andrews, UK
	\and
		Centre for Exoplanet Science, University of St Andrews, St Andrews, UK
         \and
               Instituut voor Sterrenkunde, K.U. Leuven, Celestijnenlaan 200D, 3001, Leuven, Belgium
         \and
		Univ. Grenoble Alpes, CNRS, IPAG, F-38000 Grenoble, France
     	 \and
	Kapteyn Astronomical Institute, University of Groningen, Postbus 800, 9700 AV Groningen, The Netherlands
	\and
	Astrophysics Research Centre, School of Mathematics and Physics, Queen's University Belfast, University Road, Belfast BT7 1NN, UK
	\and
	IRAP, Universit\'e de Toulouse, CNRS, UPS, Toulouse, France
	\and
	Astronomical institute Anton Pannekoek, University of Amsterdam, Science Park 904, 1098 XH, Amsterdam, The Netherlands
	\and
	School of Physics and Astronomy, Cardiff University, 4 The Parade, Cardiff CF24 3AA, UK
	\and
	Institute of Astronomy, University of Cambridge, Madingley Road, Cambridge CB3 0HA, UK
	\and
	SRON Netherlands Institute for Space Research, Sorbonnelaan 2, 3584 CA Utrecht, The Netherlands
	\and
	UMI-FCA, CNRS/INSU France (UMI 3386), and Departamento de Astronomica, Universidad de Chile, Santiago, Chile
	\and
	Monash Centre for Astrophysics (MoCA) and School of Physics and Astronomy, Monash University, Clayton Vic 3800, Australia
	\and
	Max-Planck-Institut f{\"u}r extraterrestrische Physik, Giessenbachstrasse 1, 85748 Garching, Germany
             }	
\abstract
{Consistent modeling of protoplanetary disks requires the simultaneous solution of both continuum and line radiative transfer, heating/cooling balance between dust and gas and, of course, chemistry. Such models depend on panchromatic observations that can provide a complete description of the physical and chemical properties and energy balance of protoplanetary systems. Along these lines we present a homogeneous, panchromatic collection of data on a sample of 85 T Tauri and Herbig Ae objects for which data cover a range from X-rays to centimeter wavelengths. Datasets consist of photometric measurements, spectra, along with results from the data analysis such as line fluxes from atomic and molecular transitions. Additional properties resulting from modeling of the sources such as disc mass and shape parameters, dust size and PAH properties are also provided for completeness.}
{The purpose of this data collection is to provide a solid base that can enable consistent modeling of the properties of protoplanetary disks. To this end, we performed an unbiased collection of publicly available data that were combined to homogeneous datasets adopting consistent criteria. Targets were selected based on both their properties but also on the availability of data. }
{Data from more than 50 different telescopes and facilities were retrieved and combined in homogeneous datasets directly from public data archives or after being extracted from more than 100 published articles. X-ray data for a subset of 56 sources represent an exception as they were reduced from scratch and are presented here for the first time.}
{Compiled datasets along with a subset of continuum and emission-line models  are stored in a dedicated database and distributed through a publicly accessible online system. All datasets contain metadata descriptors that allow to backtrack them to their original resources. The graphical user interface of the online system allows the user to visually inspect individual objects but also compare between datasets and models. It also offers to the user the possibility to download any of the stored data and metadata for further processing.}
{}

\maketitle


\section {Introduction}
\label{sec:1}

Knowledge is advanced with the systematic analysis and interpretation of data. This statement is especially valid in fields such as contemporary astrophysics, amongst others, where observational data play a fundamental role in describing objects and phenomena on different cosmic scales. Data alone is however not sufficient; it is the accurate description of data, the evaluation of the data quality (collectively coined as metadata), and the integration of data  into large datasets that can provide a solid basis for understanding the mechanisms involved in diverse physical phenomena. Such datasets can then be analyzed consistently and systematically through meta-analysis to confirm existing and reveal new trends and global patterns. 

The study of star and planet formation, in particular, is a field that requires extensive wavelength coverage for an appropriate characterization of sources. Such coverage can only be obtained by combining data from different facilities and instruments, which, however come with very different qualities (e.g. angular and spectral resolution, sensitivity and spatial/spectral coverage). The importance of the study of protoplanetary disks is today even more pronounced when seen from the perspective of planet formation and habitability. Protoplanetary discs are indeed the places where the complex process of planet formation takes place, described by presently two competing theories. The core accretion theory \citep{Laughlin:04a, Ida:05a}, initially developed to explain our Solar System architecture, posits collisional growth of sub-micron sized dust grains up to km-sized planetesimals on timescales of 10$^5$ to 10$^7$ years, and further growth to Earth-sized planets by gravitational interactions. Once protoplanetary cores of ten Earth-masses have formed, the surrounding gas is gravitationally captured to form gas giant planets. Alternatively, gravitational instabilities in discs may directly form planets on much shorter timescales (few thousand years), but require fairly high densities and short cooling timescales at large distances from the star \citep{Boss:09a, Rice:09a}. The field  is going through major developments following recent advances in instrumentation \citep[e.g. ALMA, VLT/SPHERE][respectively]{Ansdell:16a, Garufi:17a} but also due to more complex and sophisticated numerical codes. This input challenges our understanding of disk evolution, so it becomes increasingly important to evaluate it and interpret the data in terms of physical disc properties such as disc mass and  geometry, dust size properties and chemical concentrations.  

Observations of protoplanetary discs are challenging to interpret since physical densities in the discs span more than ten orders of magnitude, ranging from about 10$^{15}$ particles/cm$^{3}$ in the midplane close to the star to typical molecular cloud densities of 10$^{4}$ particles per cm$^{3}$ in the distant upper disc regions. At the same time, temperatures range from several 1000K in the inner disc to only 10 - 20K at distances of several 100\,au. The central star provides high energy UV and X-ray photons which are scattered  into the disc where they drive various non-equilibrium processes. The exact structure of the discs is not known, but it strongly affects the excitation of atoms and molecules and therefore their spectral appearance in form of emission lines. The morphology of the inner disk regions, for example, is expected to have a direct impact on the appearance of the outer disk. An inclined inner disc geometry or a puffed up morphology will cast shadows in the outer disc regions, while gaps may allow the direct illumination of the inner rim of the outer disc. Such complex disk topologies can be understood only through multi-wavelength studies.  Emission at short wavelengths (X-ray, UV, optical) links to the high-energy processes like mass accretion, stellar activity, and jet acceleration close to the star. Intermediate wavelengths (near to mid-IR) trace the nature and distribution of dust and gas in the inner disc, while observations at longer wavelengths provide information about the total mass and chemistry of the gas and dust in the most extended parts of the disc. A better understanding of these multi-wavelength observations requires consistent models that are capable of treating all important physical and chemical processes in detail, simultaneously, in the entire disc.

In this paper we present a coherent, panchromatic observational datasets for  85 protoplanetary disks and their host stars, and derive the physical parameters and properties for a subset of 24 discs. The present collection was created as one of the two main pillars (the other being consistent thermochemical modeling) of the "\textit{DiscAnalysis}" (DIANA)\footnote{an EU FP7-SPACE 2011 funded project, http://www.diana-project.com/} project, aiming to perform a homogeneous and consistent modeling of their gas and dust properties with the use of sophisticated codes such as ProDiMo  \citep{Woitke:09a, Kamp:10a, Thi:11a, Woitke:16a, Kamp:17a}, MCFOST \citep{Pinte:06a, Pinte:09a} and MCMax \citep{Min:09a}. In the context of the DiscAnalysis project, data assemblies for each individual source along with modeling results for both continuum and line emission are now publicly distributed through the "\textit{DiscAnalysis Object Database}" (DIOD)\footnote{http://www.univie.ac.at/diana/index.php}. The basic functionalities of the end-user interface of {DIOD} is presented in Appendix~\ref{app:db}.  

\section{The Data}\label{sec:2}


The majority of the sample sources consists of Class II and III, T Tauri and Herbig Ae systems. Selected targets cover an age spread between $\sim$ 1 and 10 million years and spectral types ranging from B9 to M3. Sources were selected based on availability and overlap of good quality data across the electromagnetic spectrum. We avoided known multiple objects where disc properties are known to be modified by the gravitational interaction of the companion and that at different wavelengths and angular resolutions may appear as single objects. We also avoided highly variable objects and in most cases edge-on disc geometry, as in such configurations the stellar properties are not well constrained and often remain unknown. In terms of sample demographics, the sample consists of 13 Herbig Ae, 7 transition disks, 58 T Tauri systems along with 7 embedded (Class I) sources or systems in an edge-on configuration (Table~\ref{tab:1}).  

Most of the data presented here were retrieved from public archives but were also collected from more than 100 published articles. In a few cases, unpublished datasets were collected through private communications. An exception to the above is the X-ray data that were reduced for the purposes of this project and are presented in this paper for the first time. Datasets consist of photometric data points along with spectra, where available. Together, they provide a complete description of the spectral energy distribution (SED). Such data were assembled from more than 150 individual filters and spectral chunks observed with $\sim$50 different telescopes/facilities. Information on the gas content of disks is provided in the form of measured fluxes per transition for different atoms and molecules, and when available, as complete spectral line profiles. 

\begin{figure*}[!tbhp]
\centerline{\includegraphics[width=18cm]{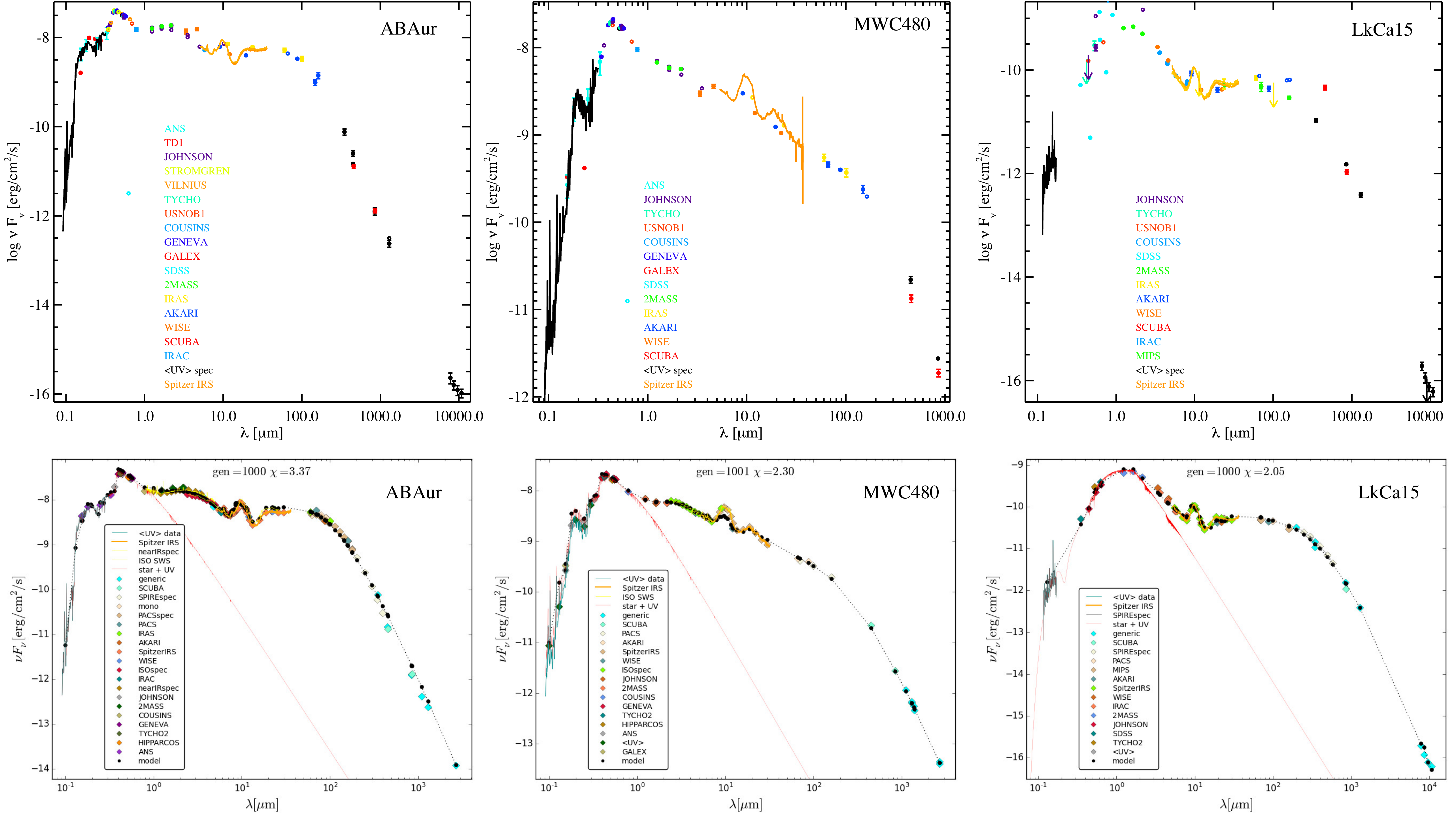}}
\caption{Example of ``raw'' collected data represented as Spectral Energy Distribution (SED) diagrams for three sources (top row). Data for AB Aur (left panel) delineate well the stellar and disk emission and show little scatter. The same is true for MWC 480 (mid panel), the Akari data points however show some deviation when compared to the Spitzer/IRS spectra. For a weaker source like LkCa 15 (right panel), the scatter is significant due to certain, not well pointed observations, and therefore the SED is not well defined. SED plots for all sources are given as online data in Fig.~\ref{fig:2}. Lower row presents the actual modeled data for the three sources, after being hand-selected for consistency.}
\label{fig:1}
\end{figure*}

A basic data quality check was performed using the following scheme: for data assembled from large surveys we propagated the original data quality flags; however, in cases that more datasets exist at the same or adjacent wavelengths, flags were modified to reflect inconsistencies and systematic (e.g. calibration) errors. In all cases links to the relevant papers are maintained so that the end-user can efficiently trace back the original data resources. An example showing different qualities of assembled data are given in the SED plots in Fig.~\ref{fig:1}, while the complete collection of SEDs for all sources is provided as online material in Fig.~\ref{fig:2}.

\onlfig{

\begin{figure*}
\centering
\includegraphics[width=0.98\textwidth]{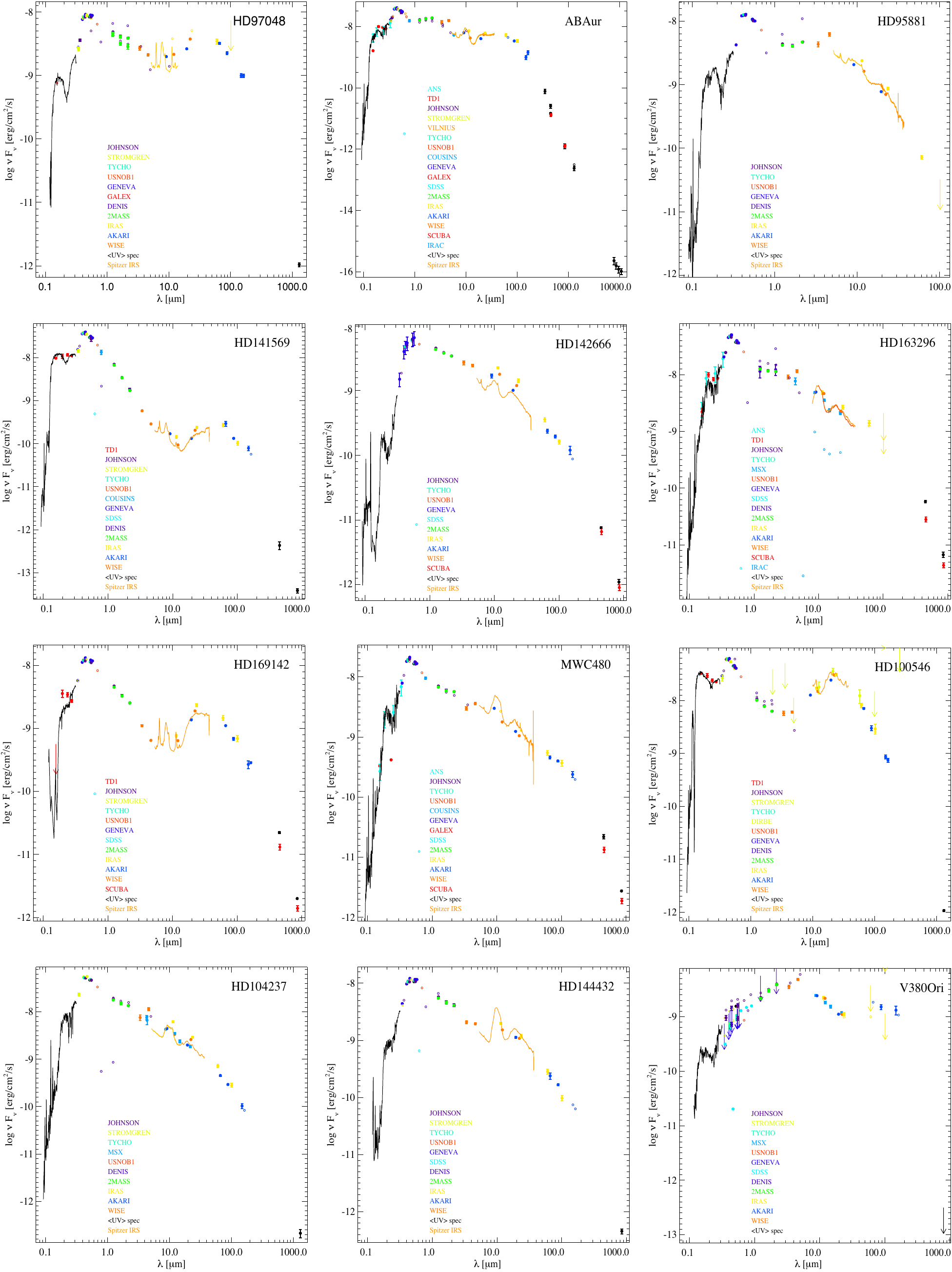}
\caption{Spectral Energy Distribution diagrams for all sources reported in Table~\ref{tab:1}} 
\label{fig:2}
\end{figure*}

\setcounter{figure}{1}
\begin{figure*}
\centering
\includegraphics[width=0.98\textwidth]{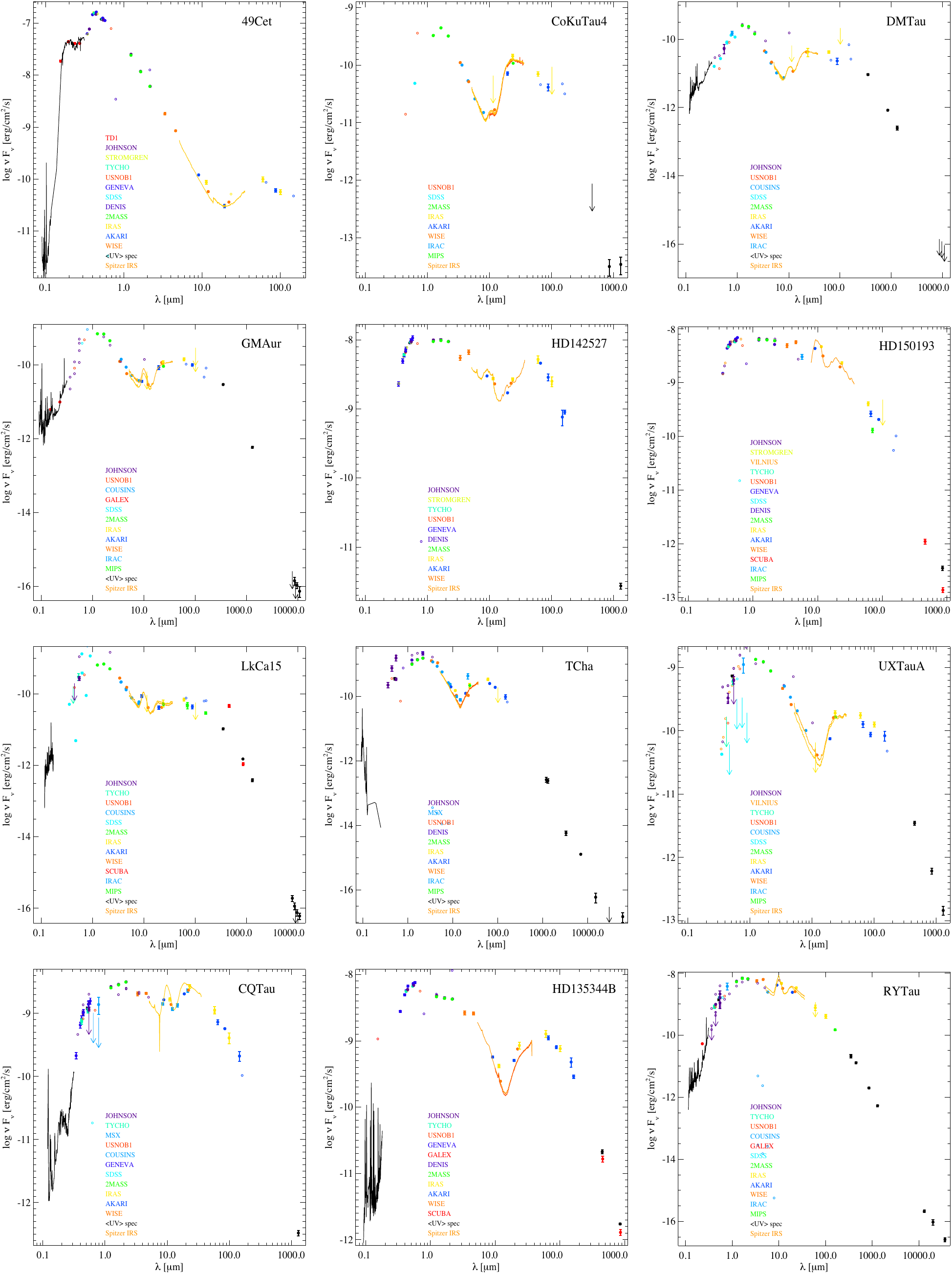}
\caption{(continued from page~\pageref{fig:2})} 
\end{figure*}

\setcounter{figure}{1}
\begin{figure*}
\centering
\includegraphics[width=0.98\textwidth]{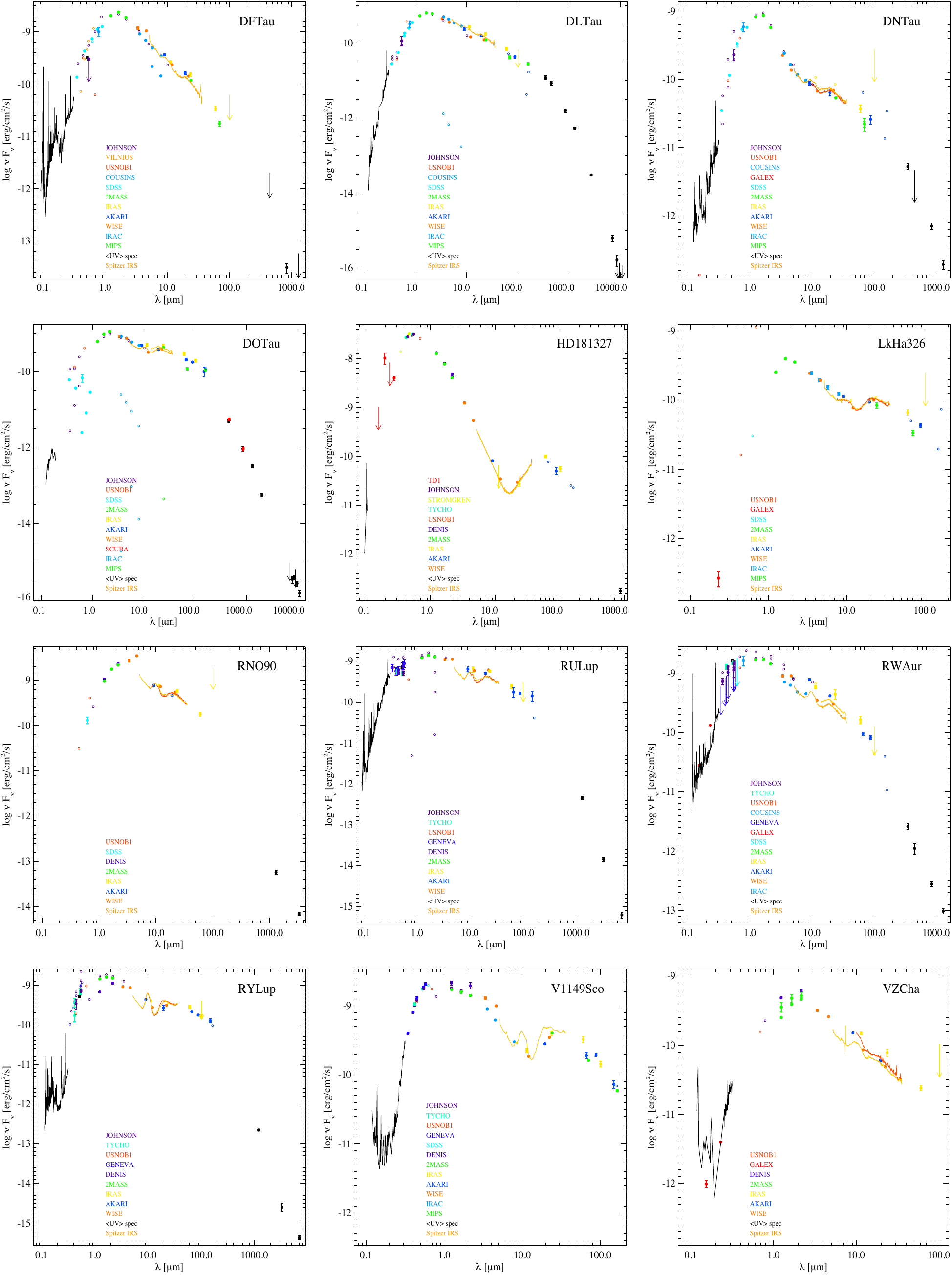}
\caption{(continued from page~\pageref{fig:2})}
\end{figure*}

\setcounter{figure}{1}
\begin{figure*}
\centering
\includegraphics[width=0.98\textwidth]{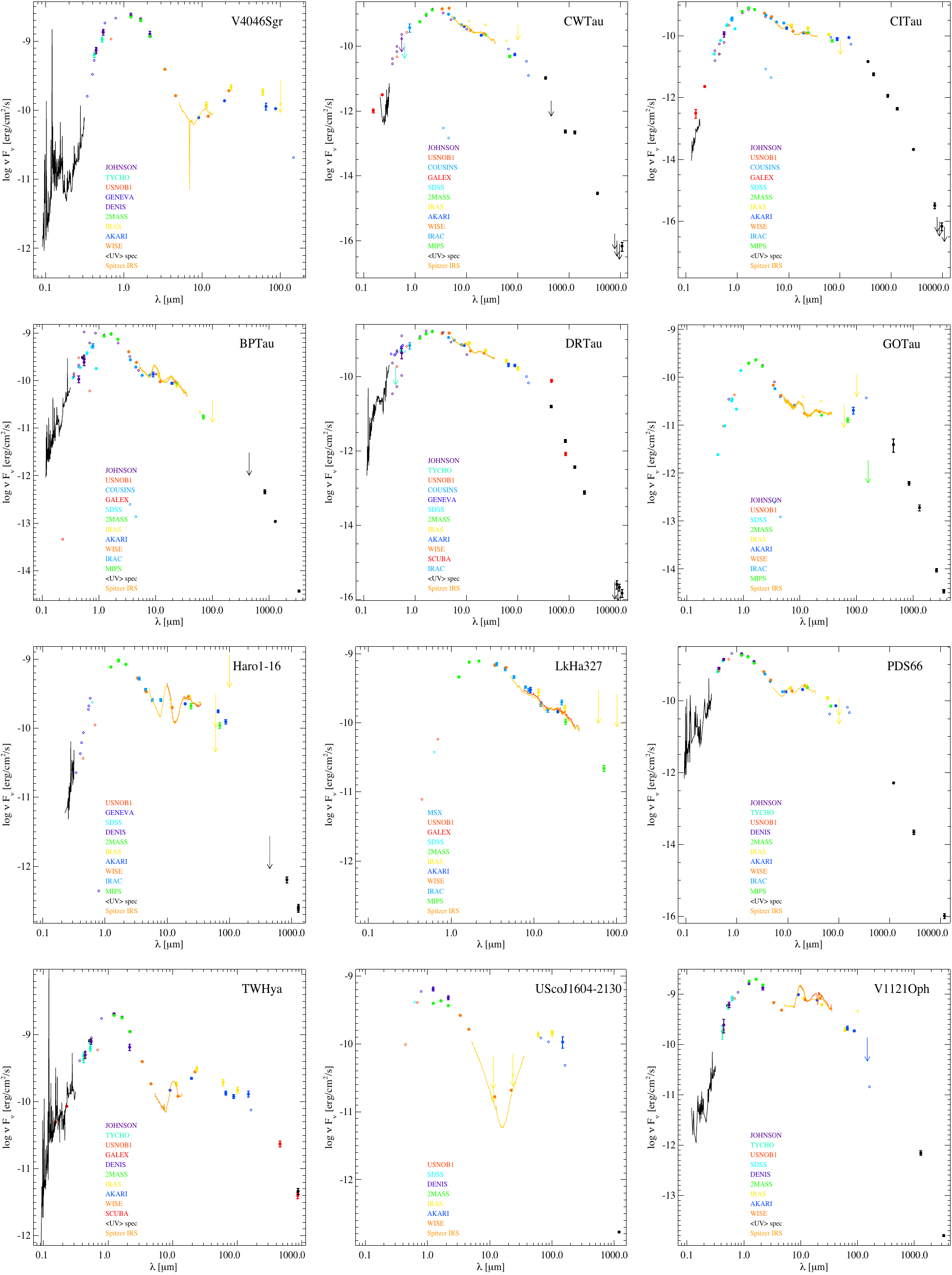}
\caption{(continued from page~\pageref{fig:2})}
\end{figure*}

\setcounter{figure}{1}
\begin{figure*}
\centering
\includegraphics[width=0.98\textwidth]{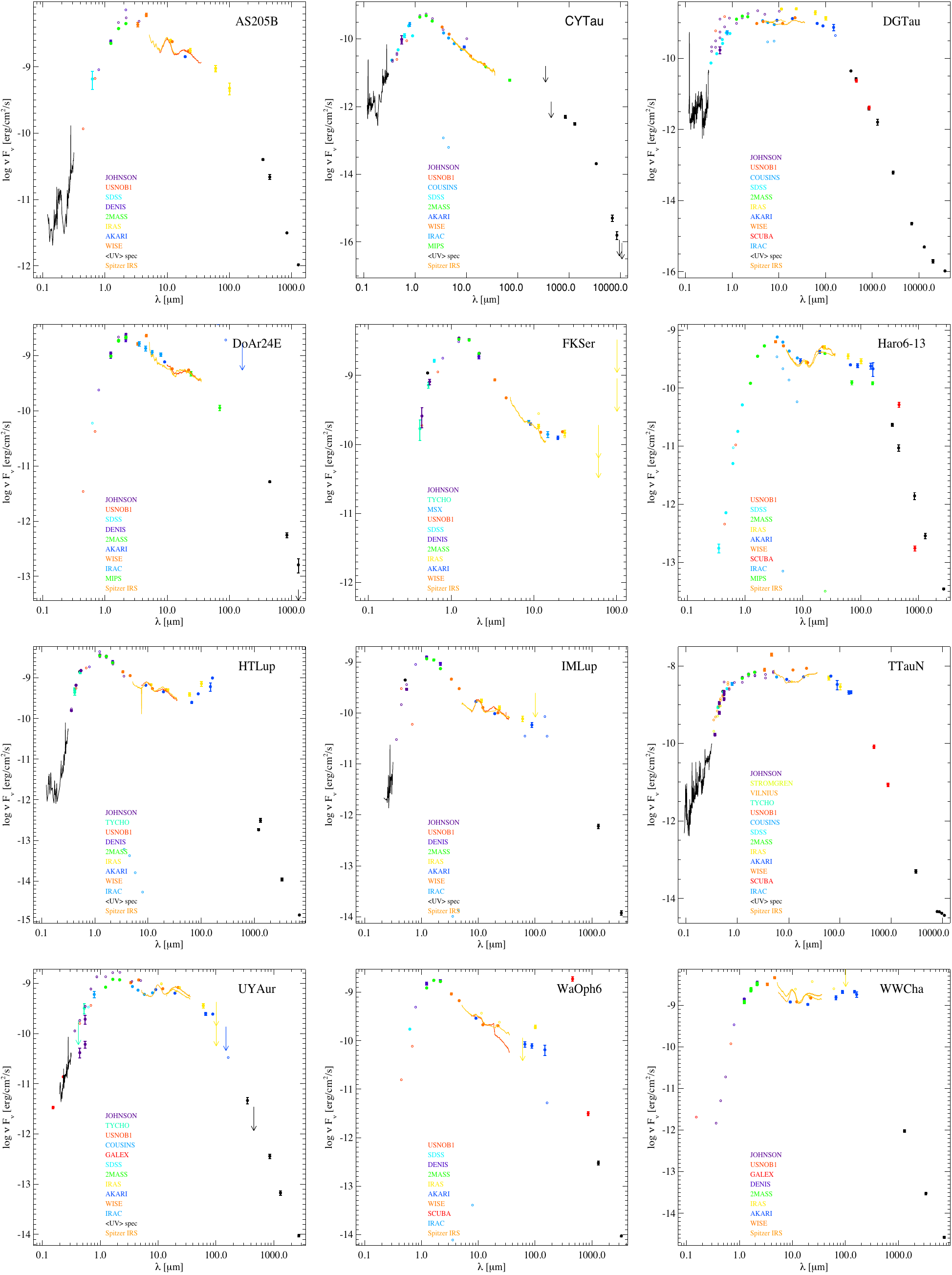}
\caption{(continued from page~\pageref{fig:2})}
\end{figure*}

\setcounter{figure}{1}
\begin{figure*}
\centering
\includegraphics[width=0.98\textwidth]{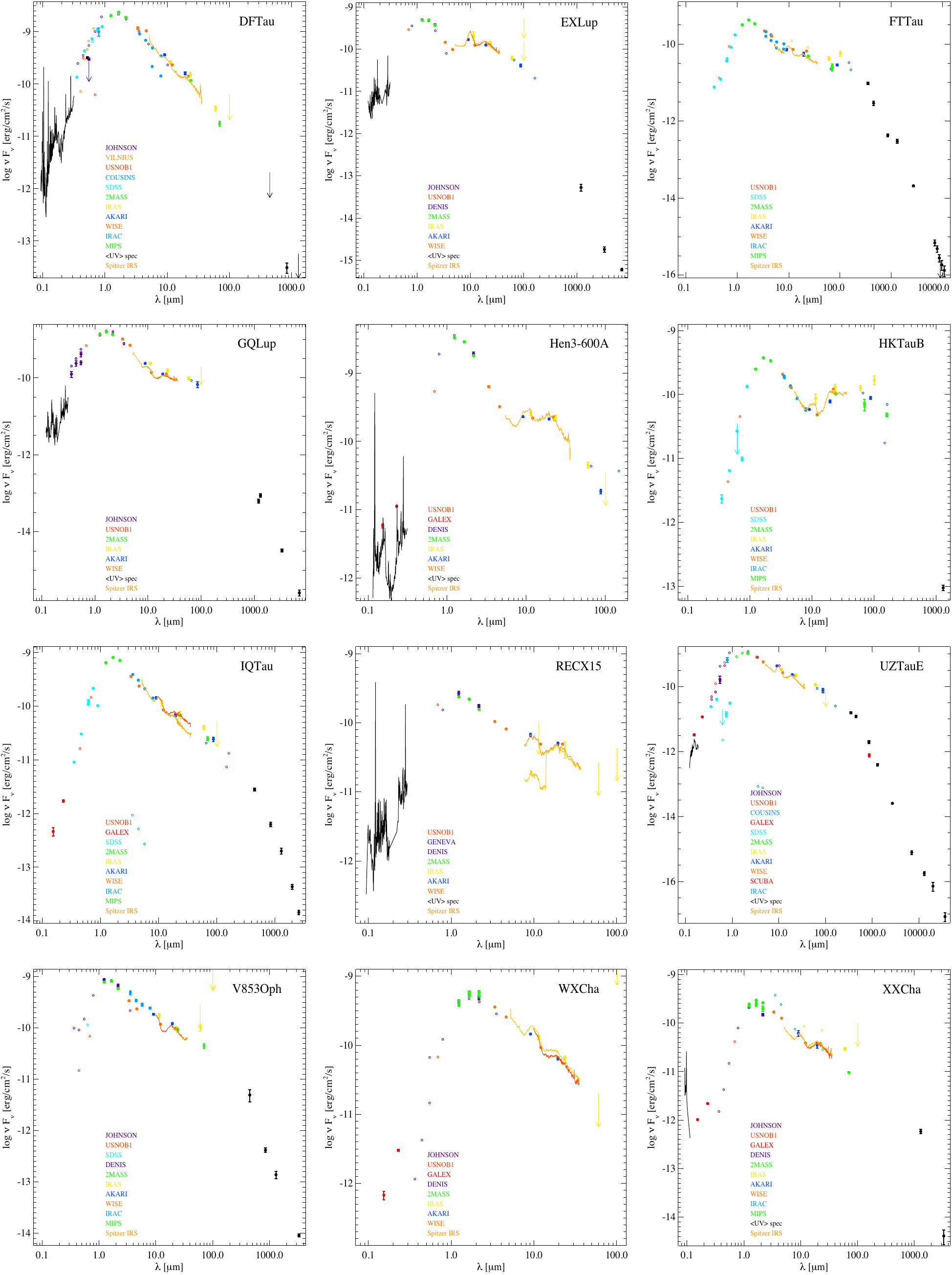}
\caption{(continued from page~\pageref{fig:2})} 
\end{figure*}

\setcounter{figure}{1}
\begin{figure*}
\centering
\includegraphics[width=0.98\textwidth]{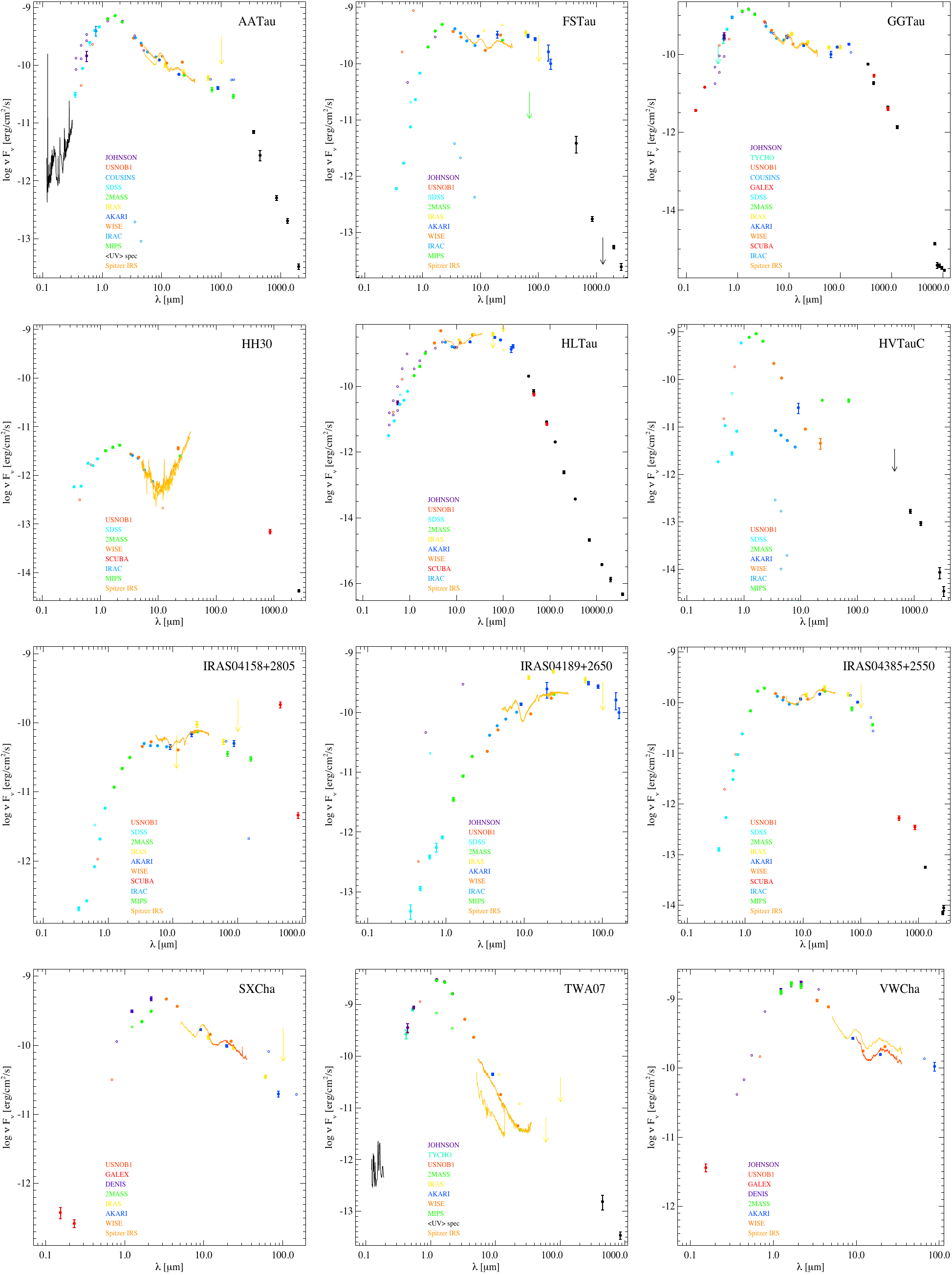}
\caption{(continued from page~\pageref{fig:2})}
\end{figure*}
}

In the following sections we provide a detailed account of the major facilities/resources used to assemble our data sample. An overview of the assembled photometric/spectroscopic datasets per wavelength regime along with information on the number of line fluxes and high resolution imaging information for each individual source is provided in Table~\ref{tab:1}.

\begin{longtab}
\begin{longtable}{l c c c c c c c c c c c c c c}
\caption{\label{tab:1}  Overview of  photometric and spectroscopic data collected per source and wavelength regime.}\\
\hline\hline
Source & Xray & \multicolumn{2}{c}{UV} & Visual & NIR & \multicolumn{2}{c}{MIR} & \multicolumn{2}{c}{FIR} &  \multicolumn{2}{c}{Sub-mm} & mm/cm & Gas & HiRes \\
name	   & spec.& phot.& spec. & phot.& phot.& phot.& spec.& phot. & spec. & phot. & spec. & phot.  & lines & img  \\ 
\hline
\endfirsthead
\caption{continued.}\\
\hline\hline
Source & Xray & \multicolumn{2}{c}{UV} & Visual & NIR & \multicolumn{2}{c}{MIR} & \multicolumn{2}{c}{FIR} &  \multicolumn{2}{c}{Sub-mm} & mm/cm & Gas & HiRes \\
name	   & spec.& phot.& spec. & phot.& phot.& phot.& spec.& phot. & spec. & phot. & spec. & phot.  & lines & img  \\
\hline
\endhead
\hline
\endfoot
\multicolumn{15}{c}{Herbig Ae/Be}\\ 
\hline
HD97048 & 1 & 5 & 12 & 27 & 7 & 9 & 1 & 6 & 1 & 0 & 1 & 1 & 43 & 0  \\
MWC480 & 1 & 9 & 16 & 25 & 3 & 9 & 0 & 8 & 0 & 4 & 0 & 0 & 12 & 1  \\
HD142666 & 1 & 2 & 9 & 19 & 2 & 7 & 0 & 6 & 1 & 4 & 0 & 0 & 7 & 1  \\
HD95881 & 0 & 1 & 8 & 15 & 4 & 7 & 1 & 2 & 0 & 0 & 0 & 0 & 10 & 0  \\
HD169142 & 1 & 8 & 5 & 26 & 2 & 8 & 0 & 8 & 1 & 4 & 0 & 0 & 30 & 1  \\
HD100546 & 1 & 8 & 34 & 28 & 4 & 12 & 0 & 12 & 1 & 0 & 1 & 1 & 50 & 1  \\
HD163296 & 1 & 13 & 46 & 35 & 5 & 19 & 2 & 4 & 1 & 4 & 1 & 0 & 53 & 2  \\
ABAur & 1 & 22 & 36 & 60 & 4 & 12 & 1 & 8 & 1 & 7 & 1 & 6 & 41 & 2 \\
HD141569 & 0 & 8 & 8 & 29 & 4 & 7 & 0 & 6 & 1 & 2 & 0 & 0 & 3 & 1  \\
HD104237 & 0 & 2 & 69 & 16 & 5 & 14 & 1 & 6 & 1 & 0 & 1 & 1 & 30 & 1  \\
HD144432 & 0 & 5 & 9 & 31 & 4 & 6 & 0 & 6 & 1 & 1 & 1 & 0 & 23 & 1  \\
V380Ori & 0 & 8 & 8 & 36 & 9 & 14 & 0 & 8 & 0 & 1 & 0 & 0 & 0 & 1  \\
HD150193 & 1 & 9 & 0 & 52 & 4 & 8 & 0 & 7 & 1 & 3 & 0 & 0 & 17 & 1  \\
\hline
\multicolumn{15}{c}{Transition Disks}\\ 
\hline

TCha & 1 & 1 & 10 & 7 & 5 & 18 & 3 & 6 & 0 & 0 & 0 & 8 & 21 & 1  \\
GMAur & 1 & 4 & 22 & 10 & 2 & 14 & 2 & 8 & 0 & 1 & 0 & 6 & 9 & 2 \\
DMTau & 1 & 2 & 8 & 9 & 5 & 9 & 2 & 6 & 0 & 2 & 0 & 4 & 9 & 1  \\
LkCa15 & 1 & 1 & 0 & 11 & 5 & 15 & 2 & 9 & 0 & 4 & 0 & 6 & 9 & 1  \\
49Cet & 0 & 10 & 7 & 39 & 3 & 7 & 1 & 6 & 0 & 0 & 0 & 0 & 9 & 0  \\
CoKuTau4 & 0 & 0 & 0 & 4 & 1 & 10 & 4 & 6 & 0 & 2 & 0 & 1 & 11 & 0  \\
UXTauA & 1 & 9 & 0 & 37 & 7 & 16 & 2 & 6 & 0 & 2 & 0 & 1 & 1 & 0  \\

\hline
\multicolumn{15}{c}{T-Tauri F-type}\\ 
\hline

HD142527 & 1 & 4 & 0 & 27 & 3 & 9 & 1 & 8 & 1 & 0 & 1 & 0 & 39 & 2  \\
HD135344B & 1 & 2 & 10 & 15 & 4 & 7 & 2 & 6 & 1 & 4 & 0 & 0 & 29 & 3  \\
RYTau & 1 & 5 & 113 & 21 & 5 & 16 & 2 & 5 & 0 & 3 & 1 & 4 & 26 & 1  \\
CQTau & 0 & 3 & 12 & 28 & 4 & 14 & 1 & 8 & 0 & 0 & 0 & 0 & 51 & 2  \\
HD181327 & 0 & 6 & 3 & 15 & 3 & 6 & 1 & 6 & 0 & 0 & 0 & 0 & 0 & 0  \\

\hline
\multicolumn{15}{c}{T-Tauri G-type}\\ 
\hline

DOTau & 1 & 3 & 0 & 10 & 4 & 19 & 2 & 8 & 0 & 4 & 0 & 8 & 6 & 4 \\
RULup & 1 & 5 & 69 & 33 & 4 & 7 & 3 & 6 & 1 & 0 & 0 & 3 & 21 & 1  \\
RYLup & 1 & 2 & 8 & 21 & 5 & 9 & 3 & 8 & 1 & 0 & 0 & 3 & 21 & 1 \\
V1149Sco & 0 & 1 & 3 & 17 & 4 & 9 & 1 & 8 & 0 & 0 & 0 & 0 & 0 & 1  \\
DLTau & 0 & 3 & 7 & 11 & 5 & 14 & 1 & 8 & 0 & 3 & 0 & 7 & 7 & 4  \\
RNO90 & 1 & 0 & 0 & 4 & 3 & 7 & 3 & 2 & 1 & 0 & 1 & 1 & 44 & 0  \\
RWAur & 1 & 6 & 44 & 29 & 6 & 15 & 2 & 8 & 0 & 3 & 0 & 1 & 2 & 0  \\
LkHa326 & 0 & 1 & 0 & 4 & 1 & 11 & 3 & 7 & 0 & 0 & 0 & 0 & 21 & 0  \\

\hline
\multicolumn{15}{c}{T-Tauri K-type}\\ 
\hline

VZCha & 1 & 2 & 2 & 1 & 5 & 7 & 3 & 2 & 0 & 0 & 0 & 0 & 0 & 0  \\
DNTau & 1 & 4 & 19 & 11 & 5 & 17 & 3 & 8 & 0 & 3 & 0 & 1 & 22 & 0  \\
TWCha & 1 & 3 & 1 & 4 & 4 & 7 & 3 & 2 & 0 & 0 & 0 & 0 & 21 & 0  \\
TWHya & 1 & 3 & 39 & 11 & 4 & 7 & 0 & 6 & 0 & 3 & 1 & 0 & 26 & 3  \\
BPTau & 1 & 5 & 83 & 18 & 6 & 14 & 2 & 3 & 0 & 3 & 0 & 0 & 4 & 4  \\
DRTau & 0 & 3 & 66 & 20 & 3 & 11 & 2 & 6 & 0 & 4 & 1 & 7 & 37 & 1  \\
Haro1-16 & 0 & 1 & 1 & 13 & 2 & 12 & 3 & 7 & 0 & 2 & 0 & 1 & 21 & 0  \\
CWTau & 0 & 4 & 9 & 9 & 3 & 14 & 2 & 7 & 0 & 3 & 0 & 6 & 2 & 0  \\
CITau & 0 & 5 & 0 & 11 & 5 & 12 & 1 & 7 & 0 & 3 & 0 & 7 & 4 & 4  \\
V4046Sgr & 0 & 1 & 25 & 15 & 4 & 9 & 1 & 8 & 0 & 0 & 0 & 0 & 0 & 2  \\
LkHa327 & 0 & 1 & 0 & 4 & 1 & 15 & 3 & 3 & 0 & 0 & 0 & 0 & 21 & 0  \\
PDS66 & 0 & 0 & 10 & 6 & 4 & 12 & 1 & 9 & 0 & 0 & 0 & 4 & 0 & 0  \\
UScoJ1604 & 0 & 0 & 0 & 4 & 3 & 7 & 1 & 8 & 0 & 0 & 0 & 1 & 2 & 0  \\
GOTau & 0 & 1 & 0 & 7 & 4 & 10 & 2 & 6 & 0 & 2 & 0 & 3 & 1 & 0  \\
V1121Oph & 0 & 0 & 4 & 9 & 3 & 7 & 3 & 6 & 0 & 0 & 0 & 1 & 21 & 1 \\
WWCha & 1 & 2 & 0 & 3 & 5 & 7 & 2 & 6 & 0 & 0 & 0 & 3 & 0 & 2  \\
FKSer & 1 & 0 & 0 & 9 & 4 & 11 & 1 & 4 & 0 & 0 & 0 & 0 & 0 & 1  \\
TTauN & 1 & 5 & 69 & 31 & 6 & 11 & 1 & 6 & 0 & 2 & 0 & 4 & 3 & 0  \\
AS205B & 1 & 0 & 2 & 4 & 3 & 8 & 3 & 4 & 1 & 3 & 0 & 1 & 21 & 3  \\
WaOph6 & 0 & 0 & 0 & 4 & 3 & 8 & 3 & 6 & 0 & 2 & 0 & 1 & 21 & 0  \\
HTLup & 1 & 3 & 6 & 15 & 5 & 10 & 3 & 6 & 1 & 0 & 0 & 4 & 21 & 1  \\
DoAr24E & 1 & 0 & 0 & 4 & 6 & 8 & 3 & 4 & 0 & 2 & 0 & 2 & 21 & 0  \\
UYAur & 1 & 5 & 3 & 17 & 5 & 13 & 2 & 8 & 0 & 3 & 0 & 2 & 3 & 0  \\
DGTau & 1 & 3 & 41 & 13 & 6 & 14 & 2 & 6 & 1 & 5 & 0 & 6 & 1 & 1  \\

\hline
\multicolumn{15}{c}{T-Tauri M-type}\\ 
\hline

IMLup & 1 & 1 & 3 & 6 & 3 & 8 & 3 & 6 & 0 & 0 & 0 & 2 & 21 & 1  \\
Haro6-13 & 0 & 1 & 0 & 6 & 3 & 17 & 2 & 8 & 0 & 5 & 0 & 2 & 5 & 4  \\
CYTau & 1 & 3 & 5 & 11 & 5 & 10 & 2 & 1 & 0 & 3 & 0 & 6 & 4 & 4  \\
DFTau & 0 & 3 & 54 & 25 & 7 & 15 & 1 & 3 & 0 & 2 & 0 & 1 & 2 & 0  \\
RECX15 & 1 & 1 & 12 & 10 & 3 & 7 & 1 & 2 & 0 & 0 & 0 & 0 & 0 & 0  \\
FTTau & 0 & 1 & 0 & 6 & 3 & 15 & 1 & 8 & 0 & 3 & 0 & 8 & 1 & 0  \\
EXLup & 0 & 0 & 3 & 2 & 5 & 9 & 3 & 7 & 0 & 0 & 0 & 3 & 21 & 0  \\
WXCha & 0 & 3 & 0 & 4 & 6 & 7 & 3 & 2 & 0 & 0 & 0 & 0 & 21 & 0  \\
VWCha & 0 & 2 & 0 & 3 & 6 & 5 & 3 & 2 & 0 & 0 & 1 & 0 & 44 & 0  \\
XXCha & 0 & 3 & 2 & 3 & 5 & 10 & 3 & 3 & 0 & 0 & 0 & 1 & 21 & 0  \\
GQLup & 0 & 3 & 2 & 8 & 3 & 7 & 3 & 4 & 0 & 0 & 0 & 4 & 21 & 1  \\
Hen3-600A & 0 & 2 & 7 & 1 & 3 & 7 & 1 & 5 & 0 & 0 & 0 & 0 & 0 & 0  \\
UZTauE & 1 & 5 & 0 & 13 & 5 & 9 & 1 & 5 & 0 & 4 & 0 & 6 & 2 & 0  \\
IQTau & 1 & 3 & 0 & 6 & 3 & 13 & 3 & 7 & 0 & 2 & 0 & 3 & 25 & 4  \\
HH30 & 0 & 1 & 0 & 6 & 3 & 7 & 2 & 0 & 0 & 1 & 0 & 1 & 0 & 0  \\
HKTauB & 0 & 1 & 0 & 6 & 3 & 17 & 1 & 9 & 0 & 0 & 0 & 1 & 1 & 0  \\
V853Oph & 0 & 1 & 0 & 6 & 5 & 13 & 3 & 5 & 0 & 2 & 0 & 2 & 21 & 0  \\
GGTau & 1 & 4 & 0 & 12 & 3 & 12 & 2 & 6 & 0 & 5 & 0 & 6 & 2 & 0  \\
SXCha & 0 & 2 & 0 & 1 & 3 & 7 & 3 & 5 & 0 & 0 & 0 & 0 & 21 & 0  \\
TWA07 & 1 & 0 & 0 & 6 & 3 & 7 & 2 & 2 & 0 & 2 & 0 & 0 & 0 & 1  \\
FSTau & 1 & 1 & 2 & 7 & 3 & 13 & 1 & 7 & 0 & 2 & 0 & 3 & 3 & 0  \\

\hline
\multicolumn{15}{c}{Edge-on Systems}\\ 
\hline

AATau & 1 & 3 & 12 & 11 & 5 & 14 & 2 & 8 & 0 & 3 & 0 & 2 & 14 & 5  \\
IRAS04158 & 0 & 1 & 0 & 5 & 3 & 11 & 1 & 7 & 0 & 2 & 0 & 0 & 0 & 0  \\
IRAS04385 & 0 & 1 & 0 & 6 & 3 & 11 & 1 & 8 & 0 & 2 & 0 & 3 & 0 & 0  \\

\hline
\multicolumn{15}{c}{Embedded Systems}\\ 
\hline

FlyingSaucer & 0 & 0 & 0 & 1 & 5 & 3 & 0 & 0 & 0 & 0 & 0 & 0 & 0 & 0  \\
HLTau & 0 & 3 & 0 & 13 & 5 & 13 & 1 & 8 & 0 & 5 & 0 & 7 & 3 & 0  \\
HVTauC & 0 & 1 & 0 & 6 & 3 & 12 & 0 & 1 & 0 & 2 & 0 & 3 & 0 & 0  \\
IRAS04189 & 0 & 1 & 0 & 6 & 3 & 11 & 2 & 6 & 0 & 0 & 0 & 0 & 0 & 0  \\

\hline
\end{longtable}
\end{longtab}


\begin{table}[!ht]
\vspace*{-4mm}
\caption{List of Xray observations.}
\label{tab:xray1}
\def\z{$\hspace*{-1mm}$}
\def\zz{$\hspace*{-2mm}$}
\def\zzz{$\hspace*{-3mm}$}
\vspace*{2mm}\hspace*{-2mm}
\resizebox{81mm}{!}{
\begin{tabular}{lccc}
\hline
&&&\\[-2.0ex]

Source	&	Instrument	& Obs-ID	&	Exposure \\
 & & & time (10$^4$ s)\\
&&&\\[-2.0ex]
\hline
&&&\\[-2.0ex]

DO Tau    &     XMM-Newton  &   0501500101  &   2.46 \\
DN Tau    &     XMM-Newton  &   0651120101  &   10.2 \\
VZ Cha    &     XMM-Newton  &   0300270201  &   10.9 \\
TW Cha    &     XMM-Newton  &   0152460301  &   2.61 \\
IM Lup    &     XMM-Newton  &   0303900301  &   2.49 \\
V806 Tau  &     XMM-Newton  &   0203540301  &   2.95 \\
RECX15    &     XMM-Newton  &   0605950101  &   4.00 \\
GM Aur    &     XMM-Newton  &   0652330201  &   3.07 \\
DM Tau    &     XMM-Newton  &   0554770101  &   3.37 \\
TW Hya    &     XMM-Newton  &   0112880201  &   2.38 \\
CY Tau    &     Chandra     &      3364     &   1.77 \\
UY Aur    &     XMM-Newton  &   0401870501  &   3.19 \\
UZ Tau E  &     XMM-Newton  &   0203541901  &   3.13 \\
IQ Tau    &     XMM-Newton  &   0203541401  &   2.84 \\
GG Tau    &     XMM-Newton  &   0652350201  &   1.43 \\
FS Tau    &     XMM-Newton  &   0203541101  &   3.43 \\
HL Tau    &     XMM-Newton  &   0109060301  &   4.86 \\
Haro 6-5B &     XMM-Newton  &   0203541101  &   3.43 \\
VW Cha    &     XMM-Newton  &   0002740501  &   2.78 \\
RW Aur    &     XMM-Newton  &   0401870301  &   3.02 \\
WW Cha    &     XMM-Newton  &   0203810101  &   2.30 \\
V709 CrA  &     XMM-Newton  &   0146390101  &   2.87 \\
FK Ser    &     XMM-Newton  &   0403410101  &   0.25 \\
T Tau N   &     XMM-Newton  &   0301500101  &   6.69 \\
DoAr 24E  &     Chandra     &      3761     &   9.11 \\
V853 Oph  &     Chandra     &      622      &   0.48 \\
HD97048   &     XMM-Newton  &   0002740501  &   2.80 \\
HD31648   &     Chandra     &      8939     &   0.98 \\
HD169142  &     Chandra     &      6430     &   0.99 \\
T Cha     &     XMM-Newton  &   0550120601  &   0.52 \\
HD142527  &     XMM-Newton  &   0673540501  &   1.08 \\
RU Lup    &     XMM-Newton  &   0303900301  &   2.49 \\
RY Lup    &     XMM-Newton  &   0652350501  &   0.49 \\
HD100546  &     Chandra     &      3427     &   0.26 \\
HD163296  &     Chandra     &      3733     &   1.92 \\
AB Aur    &     XMM-Newton  &   0101440801  &   12.3 \\
HD135344  &     Chandra     &      9927     &   3.17 \\
LkCa15    &     Chandra     &      10999    &   0.98 \\
HD150193  &     Chandra     &      982      &   0.29 \\
UX Tau A  &     Chandra     &      11001    &   0.50 \\
RNO90     &     XMM-Newton  &   0602731101  &   0.78 \\
AS205     &     XMM-Newton  &   0602730101  &   0.53 \\
Sz68      &     XMM-Newton  &   0652350401  &   0.69 \\
DG Tau    &     XMM-Newton  &   0203540201  &   2.49 \\
TWA7      &     Chandra     &      11004    &   0.14 \\
RY Tau    &     XMM-Newton  &   0101440701  &   4.09 \\
BP Tau    &     XMM-Newton  &   0200370101  &   11.5 \\
DR Tau    &     XMM-Newton  &   0406570701  &   0.96 \\
Haro1-16  &     XMM-Newton  &   0550120201  &   1.54 \\
GO Tau    &     XMM-Newton  &   0203542201  &   2.66 \\
CI Tau    &     XMM-Newton  &   0203541701  &   2.60 \\
EX Lup    &     XMM-Newton  &   0551640201  &   6.56 \\
WX Cha    &     XMM-Newton  &   0002740501  &   2.77 \\
XX Cha    &     XMM-Newton  &   0300270201  &   10.9 \\
AA Tau    &     XMM-Newton  &   0152680401  &   1.37 \\
HD142666  &     XMM-Newton  &   0673540801  &   0.85 \\

\hline
\end{tabular}}\\[0.5mm]
\end{table}

\subsection{X-rays}

While X-rays do not provide direct information about the disk, they can represent an important part of the total stellar radiation field which is directly affects the physical and chemical structure of the disc. We mined the XMM-Newton\footnote{http://xmm.esac.esa.int/xsa/} \citep{Jansen:01a} and Chandra\footnote{http://cxc.harvard.edu/cda/} \citep{Weisskopf:00a} mission-archives for X-ray observations of our target-list and obtained data for 56 sources (Table~\ref{tab:xray1}).  X-ray data was extracted by using the SAS software (version 12.0.1) for the
XMM-Newton data and the CIAO software (version 4.6.1) for the Chandra data.
The CALDB calibration data used for the spectral extraction of the Chandra
data were taken from version 4.6.2., while the XMM-Newton calibration data is
put on a rolling release and thus has no version number.
In order to get the source spectra, we selected a circular extraction region
around the center of the emission, while the background area
contained a large source-free area on the same CCD. The extraction tools (\textit{EVSELECT} for XMM and \textit{SPECEXTRACT} for
Chandra) delivered the source and background spectra as well as
the redistribution matrix and the ancillary response files. 

The spectra were modeled by using the package {\tt XSPEC}
\citep{Arnaud:96a}, assuming a plasma model ({\tt VAPEC} - an emission spectrum for collisionally ionized
diffuse plasma, based on the ATOMDB code [v.2.0.2]) combined with an absorption
column model ({\tt WABS}) based on the cross-sections from
\citet{Morrison:83a}. The element abundance values in the {\tt VAPEC} models were set to
typical values for pre-main sequence stars, as chosen by the XEST project
\citep[see also Table~\ref{tab:xray2},][]{Gudel:07a}, unless otherwise noted in
Table~\ref{tab:xray2}. Either a one component (1T), a two component (2T) or a three
component (3T) emission model is fitted to the data. Highly absorbed sources
or scarce data allow only for 1T fits. In some cases sources show such a high absorption that it is impossible to fix the higher
temperature due to low constraints on the slope of the harder (meaning more
energetic >1keV) part of the spectrum. In both these cases the higher temperature was fixed to 10 keV. 
The fit delivers the absorption column density towards the source $N_{H}$, the plasma
emission temperature $T_{X}$ for each
component. Finally, the unabsorbed spectrum is calculated after setting
the absorption column density parameter to zero, and the flux is
derived by integrating over the energy range from 0.3-10 keV.
Hardness is defined by $\frac{H-S}{H+S}$, with $H$ and $S$ denoting the hard part
(1-10 keV) and the soft part (0.3-1 keV) of the spectrum respectively. Thus
the hardness factor delivers a value between 1 and -1, showing a hard spectrum
in the case of $\sim$1 and a soft spectrum in the case of $\sim$ -1. Results from the fitting process are given in Table~4.

\begin{table}[!t]
\vspace*{-4mm}
\caption{Standard XEST abundances and deviations for particular sources used in the {\tt VAPEC} models.}
\label{tab:xray2}
\def\z{$\hspace*{-1mm}$}
\def\zz{$\hspace*{-2mm}$}
\def\zzz{$\hspace*{-3mm}$}
\vspace*{2mm}\hspace*{-2mm}
\begin{tabular}{cc|ccc}
\hline
&&&&\\[-2.0ex]

Element & XEST  & Source	&	Element	& Modified \\
    & abundance & &  & abundance \\
&&&&\\[-2.0ex]
\hline
&&&&\\[-2.0ex]

He         &    1       &  \multirow{2}{*}{TW Cha}     &    Mg       &    0.917  \\
C          &    0.450   &             &    Fe         &    0.222  \\                    
N          &    0.788   &  GM Aur     &    O          &    0.103  \\               
O          &    0.426   &  UZ Tau E   &    O          &    2.704  \\               
Ne         &    0.832   &  \multirow{4}{*}{HL Tau}     &    Mg       &    1.500  \\
Mg         &    0.263   &             &    S          &    1.500  \\               
Al         &    0.500   &             &    Ca         &    1.500  \\               
Si         &    0.309   &             &    Fe         &    0.740  \\               
S          &    0.417   &  \multirow{2}{*}{RW Aur}     &    FeI      &    0.058  \\
Ar         &    0.550   &             &    FeII       &    0.456  \\               
Ca         &    0.195   &  \multirow{3}{*}{V709 CrA}   &    O        &    0.308  \\
Fe         &    0.195   &             &    FeI        &    0.079  \\               
Ni         &    0.195   &             &    FeII       &    0.207  \\               
           &            &   FK Ser     &    O          &    0.098  \\               
           &            &   \multirow{3}{*}{T Tau N}    &    O        &    0.193  \\
           &            &              &    FeI        &    0.052  \\               
           &            &              &    FeII       &    0.074  \\               
           &            &   HD31648    &    Ne         &    0.056  \\               
           &            &   \multirow{2}{*}{RU Lup}     &    FeI      &    0.140  \\
           &            &              &    FeII       &    0.614  \\
           &            &   TWA7       &    Fe         &    0.121  \\
           &            &   BP Tau     &    Fe         &    0.047  \\
           &            &   WX Cha     &    Fe         &    0.098  \\
           &            &   AA Tau     &    Fe         &    0.491  \\

\hline
\end{tabular}\\[0.5mm]
\end{table}

\begin{table*}
\label{tab:xrayres}
\tiny
\centering
\caption{Results from the X-ray reduction; ``soft'' and ``hard'' subscripts correspond to the 0.3 - 1~KeV and 1 - 10~KeV spectral regions, respectively. Hardness is defined in the text.}
\resizebox{!}{10.0cm}{
\begin{tabular}{l c c c c c c c c c c c}
\hline\hline
Source         &  N$_H$      &  Flux    &  Flux$_{soft}$         &  Flux$_{hard}$        &  Hardness     &  F$_{abs}$         &  F$_{abs-soft}$         &  F$_{abs-hard}$         &  Hardness      &  T$_1$   &  T$_2$\\
	& (10$^{22}$ cm$^{-2}$) & \multicolumn{3}{c}{(10$^{-13}$erg cm$^{-2}$ s$^{-1}$)}  & & \multicolumn{3}{c}{(10$^{-13}$erg cm$^{-2}$ s$^{-1}$)}& & (10$^6$ K) & (10$^7$ K) \\
\hline
  HD169142   &     0.0     &     0.538    &     0.516    &     0.022    &    -0.920    &     0.538    &     0.516    &     0.022    &    -0.920    &     2.71    &     0.0     \\
    RY Lup   &     0.724   &    30.9      &    27.4      &     3.480    &    -0.774    &     2.230    &     0.577    &     1.660    &     0.483    &     2.46    &     1.22   \\
    TW Hya   &     0.063   &    62.3      &    53.6      &     8.650    &    -0.722    &    39.3      &    31.3      &     7.960    &    -0.595    &     2.27    &     0.795   \\
    AB Aur   &     0.153   &     2.530    &     2.130    &     0.4      &    -0.684    &     1.060    &     0.730    &     0.327    &    -0.380    &     2.02    &     0.767   \\
    FK Ser   &     0.287   &    33.9      &    27.2      &     6.710    &    -0.605    &     9.120    &     4.110    &     5.010    &     0.099    &     2.58    &     1.49   \\
   HD31648   &     0.397   &     2.210    &     1.750    &     0.457    &    -0.586    &     0.543    &     0.258    &     0.285    &     0.050    &     6.56    &     0.0     \\
    FS Tau   &     1.702   &    46.6      &    36.3      &    10.3      &    -0.557    &     5.410    &     0.028    &     5.380    &     0.990    &     2.7      &     3.45   \\
  HD142527   &     0.181   &     1.430    &     1.1      &     0.325    &    -0.545    &     0.663    &     0.374    &     0.289    &    -0.128    &     3.35    &    11.6     \\
    GM Aur   &     0.285   &     7.820    &     5.960    &     1.850    &    -0.526    &     2.430    &     0.942    &     1.490    &     0.225    &     2.75    &     2.34   \\
    EX Lup   &     0.364   &     1.830    &     1.340    &     0.489    &    -0.466    &     0.533    &     0.089    &     0.444    &     0.667    &     1.92    &    17.4     \\
  HD135344   &     0.0     &     1.360    &     0.987    &     0.371    &    -0.453    &     1.360    &     0.987    &     0.371    &    -0.453    &     7.05    &     0.0     \\
    VW Cha   &     0.453   &    22.2      &    14.7      &     7.5      &    -0.324    &     6.610    &     1.440    &     5.160    &     0.563    &     3.93    &     2.12   \\
    WW Cha   &     0.709   &    15.8      &    10.4      &     5.380    &    -0.319    &     3.610    &     0.405    &     3.2      &     0.775    &     4.46    &     2.84   \\
    AA Tau   &     2.174   &    12.8      &     8.360    &     4.450    &    -0.306    &     1.0      &     0.007    &     0.994    &     0.986    &    10.0      &     3.17   \\
    TW Cha   &     0.173   &     3.490    &     2.240    &     1.260    &    -0.280    &     1.880    &     0.782    &     1.1      &     0.167    &     3.95    &     2.51   \\
      Sz68   &     0.362   &    11.6      &     7.410    &     4.210    &    -0.276    &     4.040    &     1.120    &     2.920    &     0.444    &     4.32    &     1.55   \\
  HD163296   &     0.001   &     2.720    &     1.730    &     0.996    &    -0.268    &     2.710    &     1.710    &     0.995    &    -0.265    &     6.3      &    12.6     \\
    LkCa15   &     0.233   &     7.060    &     4.340    &     2.720    &    -0.228    &     3.520    &     1.170    &     2.350    &     0.337    &     4.24    &     6.77   \\
    DN Tau   &     0.072   &     6.170    &     3.710    &     2.450    &    -0.204    &     4.510    &     2.220    &     2.280    &     0.013    &     5.35    &     2.28   \\
    CY Tau   &     0.0     &     0.274    &     0.161    &     0.113    &    -0.174    &     0.274    &     0.161    &     0.113    &    -0.174    &    11.0      &     0.0     \\
   UX TauA   &     0.104   &     7.360    &     4.250    &     3.110    &    -0.156    &     5.060    &     2.270    &     2.790    &     0.102    &     8.09    &     1.67   \\
    DM Tau   &     0.196   &    10.5      &     6.080    &     4.440    &    -0.156    &     5.560    &     1.780    &     3.780    &     0.359    &     3.56    &     1.91   \\
    WX Cha   &     0.411   &     9.260    &     5.140    &     4.120    &    -0.111    &     3.840    &     0.552    &     3.290    &     0.712    &     3.69    &     3.44   \\
   UZ TauE   &     0.264   &     1.790    &     0.980    &     0.812    &    -0.094    &     0.897    &     0.244    &     0.652    &     0.455    &    10.3      &     2.08   \\
  Haro1-16   &     0.276   &     5.720    &     3.070    &     2.650    &    -0.074    &     2.890    &     0.737    &     2.150    &     0.489    &     7.34    &     2.91   \\
    GG Tau   &     0.084   &     1.730    &     0.914    &     0.820    &    -0.054    &     1.3      &     0.532    &     0.773    &     0.184    &     5.26    &     3.84   \\
     T Cha   &     0.987   &    21.6      &    11.3      &    10.3      &    -0.044    &     5.560    &     0.231    &     5.330    &     0.917    &     9.71    &     2.32   \\
    DO Tau   &     1.127   &     1.370    &     0.703    &     0.663    &    -0.029    &     0.274    &     0.010    &     0.264    &     0.926    &    12.5      &     0.0     \\
    UY Aur   &     0.071   &     1.680    &     0.854    &     0.823    &    -0.019    &     1.320    &     0.544    &     0.779    &     0.178    &     9.2      &     3.08   \\
  HD100546   &     0.107   &     1.2      &     0.605    &     0.599    &    -0.005    &     0.845    &     0.310    &     0.535    &     0.266    &    13.0      &     0.0     \\
    BP Tau   &     0.086   &     7.940    &     3.930    &     4.010    &     0.010    &     5.9      &     2.140    &     3.760    &     0.275    &     5.6      &     2.92   \\
    GO Tau   &     0.344   &     0.781    &     0.379    &     0.403    &     0.030    &     0.389    &     0.065    &     0.324    &     0.666    &     6.29    &     3.49   \\
    IM Lup   &     0.093   &    10.3      &     4.950    &     5.380    &     0.042    &     7.750    &     2.750    &     5.0      &     0.291    &     9.54    &     2.49   \\
    XX Cha   &     0.272   &     2.250    &     1.080    &     1.170    &     0.043    &     1.230    &     0.260    &     0.968    &     0.576    &     9.05    &     2.78   \\
      TWA7   &     0.0     &    36.5      &    17.1      &    19.5      &     0.066    &    36.5      &    17.1      &    19.5      &     0.066    &     7.7      &     5.4     \\
  V806 Tau   &     1.227   &     1.510    &     0.703    &     0.808    &     0.069    &     0.385    &     0.007    &     0.378    &     0.962    &    12.5      &    11.8     \\
    T TauN   &     0.265   &    34.0      &    15.1      &    18.9      &     0.111    &    19.2      &     3.210    &    16.0      &     0.665    &     5.6      &     3.09   \\
     RNO90   &     0.631   &    22.7      &     9.790    &    12.9      &     0.137    &     9.6      &     0.574    &     9.030    &     0.880    &     8.89    &     2.82   \\
    RU Lup   &     0.144   &     8.360    &     3.6      &     4.760    &     0.139    &     5.890    &     1.520    &     4.370    &     0.485    &     7.33    &     5.01 \\
  V709 CrA   &     0.227   &    91.3      &    38.2      &    53.1      &     0.164    &    56.5      &     9.960    &    46.5      &     0.647    &     5.86    &     3.7     \\
    DR Tau   &     0.214   &     1.650    &     0.674    &     0.972    &     0.181    &     1.060    &     0.205    &     0.857    &     0.614    &     9.15    &     3.51   \\
    RECX15   &     0.073   &     0.492    &     0.197    &     0.295    &     0.2      &     0.408    &     0.123    &     0.284    &     0.394    &     7.39    &     6.58   \\
  V853 Oph   &     0.043   &     9.850    &     3.830    &     6.020    &     0.222    &     8.620    &     2.760    &     5.850    &     0.359    &    33.0      &     1.14   \\
    IQ Tau   &     0.533   &     4.530    &     1.720    &     2.810    &     0.242    &     2.240    &     0.139    &     2.1      &     0.876    &    11.5      &     3.14   \\
  HD150193   &     0.0     &     8.2      &     2.950    &     5.250    &     0.281    &     8.2      &     2.950    &     5.250    &     0.281    &     6.48    &     4.97   \\
    RY Tau   &     0.567   &    19.6      &     6.910    &    12.7      &     0.295    &    10.4      &     0.456    &     9.940    &     0.912    &     5.89    &     4.13   \\
    HL Tau   &     2.547   &    14.0      &     4.840    &     9.140    &     0.308    &     4.110    &     0.001    &     4.110    &     1.0      &    23.8      &     0.398   \\
    CI Tau   &     0.339   &     0.774    &     0.237    &     0.537    &     0.386    &     0.488    &     0.035    &     0.453    &     0.856    &    79.6      &     1.97   \\
     AS205   &     1.739   &     2.430    &     0.722    &     1.7      &     0.405    &     0.979    &     0.001    &     0.978    &     0.998    &    33.1      &     0.0     \\
    RW Aur   &     0.209   &    57.9      &    17.0      &    40.9      &     0.413    &    42.4      &     4.810    &    37.6      &     0.773    &     7.18    &     7.26   \\
   HD97048   &     0.162   &     0.201    &     0.054    &     0.148    &     0.467    &     0.156    &     0.018    &     0.138    &     0.764    &    40.7      &     0.0     \\
    VZ Cha   &     0.211   &     5.710    &     1.520    &     4.190    &     0.468    &     4.3      &     0.444    &     3.860    &     0.794    &    10.4      &     6.01   \\
  DoAr 24E   &     1.071   &     3.730    &     0.864    &     2.870    &     0.538    &     2.130    &     0.010    &     2.120    &     0.991    &    53.6      &     0.0     \\
 Haro 6-5B   &     1.814   &     1.080    &     0.236    &     0.848    &     0.564    &     0.565    &     0.0      &     0.565    &     0.999    &    60.3      &     0.0     \\
    DG Tau   &     0.043   &     1.520    &     0.205    &     1.310    &     0.730    &     1.460    &     0.153    &     1.310    &     0.791    &     4.46    &     2.4     \\ 
       \hline
\end{tabular}
}
\end{table*}

\subsection{Ultraviolet}


Ultraviolet data were collected from different resources. Spectra were obtained from the archives of the International Ultraviolet Explorer (IUE)\footnote{https://archive.stsci.edu/iue/}, the Far Ultraviolet Spectroscopic Explorer (FUSE)\footnote{https://archive.stsci.edu/fuse/} and the Hubble Space Telescope (HST)\footnote{https://archive.stsci.edu/hst/}. Hubble data originate from three instruments, namely the Space Telescope Imaging Spectrograph (STIS), the Cosmic Origins Spectrograph (COS) and the Advanced Camera for Surveys (ACS).  

All multi-instrument data is integrated over a number of wavelength bins and then combined with weights as 1/$\sigma^2$, where sigma is the given instrument error after integration.  An iterative procedure is carried out where the number of retrieved spectral points is lowered step-by-step, until statistically relevant data is obtained ($F_k  >  3*\sigma_k$), as described in detail in Appendix~\ref{App:A}. The idea for this procedure is from  \citet{Valenti:00a, Valenti:03a}, but we have modified it to include multi-instrument data, and we have added the idea to lower number of bins until statistically significant data is obtained. Below we summarize the main characteristics of each data type used and its applicability in our data collection.

\begin{itemize}
\setlength{\itemsep}{1.3pt}
\setlength{\parskip}{0pt}
\setlength{\topsep}{0pt}
\setlength{\parsep}{0pt}
\setlength{\partopsep}{0pt}
\item{ IUE's short and long wavelength spectroscopic cameras provided low resolution ($R\!\approx\!400$) spectra covering the $1150-1980$\,\AA\ and $1850-3350$\,\AA\  windows, respectively. Often, a large number of data files exists per source in the archive\footnote{http://sdc.cab.inta-csic.es/cgi-ines/IUEdbsMY}, however we have typically used the first $\sim$ 20 with the longest integration times for each source. IUE averaged spectra as treated in \citet{Valenti:00a, Valenti:03a} were collected for comparisons but not used, as we combine IUE data along with spectra from other instruments.}

\item{ The Far Ultraviolet Spectroscopic Explorer (FUSE) covers
  the important $900-1190$\,\AA\ band in high resolution ($R\!\approx\!20000$). The FUSE
  data may be affected by a number of emission lines due to the residual Earth atmosphere, also known as ``airglow''. At first, FUSE data on faint disk sources may appear quite noisy, however combining and processing as described in App.~\ref{App:A} can lead to high quality data in the very important region around $1000$\,\AA\ .}

\item The {HST Cosmic Origins Spectrograph (COS)} and {Space
  Telescope Imaging Spectrograph (STIS)} cover wavelengths
  $1150-3600$\,\AA\ and for our purposes the range $1150-3200$\,\AA\ in very
  different resolutions up to $R\ga10\,000$. High resolution data come in chunks that rarely cover large wavelength ranges, so they need to be combined. Combined HST datasets including lower resolution ACS data, are provided in \citet{Yang:12a}\footnote{http://archive.stsci.edu/prepds/ttauriatlas/table.html}.

\end{itemize}

\begin{figure}[!ht]
\centering
\includegraphics[width=0.45\textwidth]{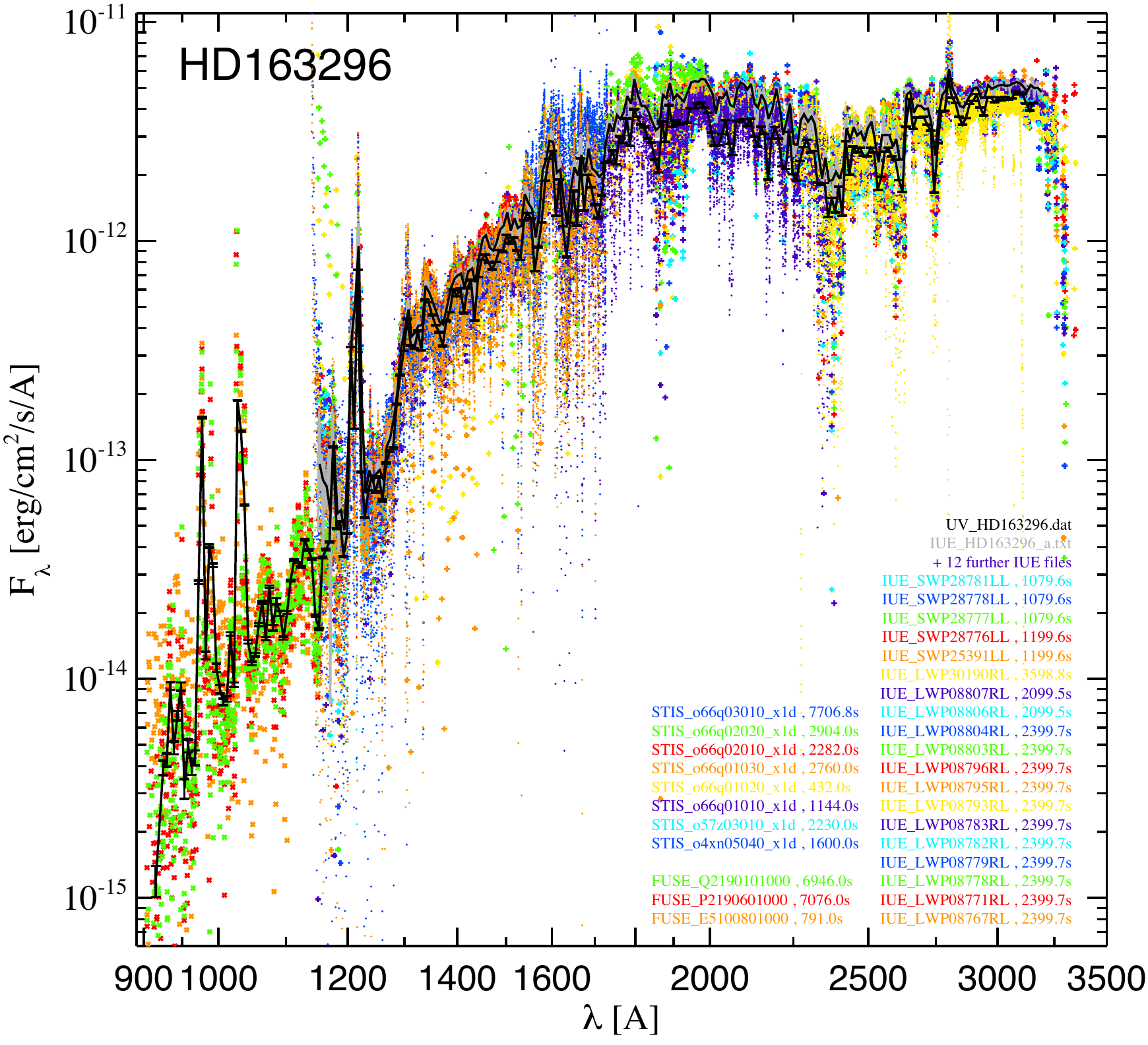}
\caption{Ultraviolet spectrum of HD163296, consisting of a series of individual observations from HST, FUSE and IUE. Black line represents the co-added spectrum of all observations as described in Appendix~\ref{App:A}.} 
\label{fig:3}
\end{figure}

\begin{table}
\label{ResTable}
\caption{Number of archival UV data files collected.} 
\bigskip
\resizebox{!}{12.0cm}{
\begin{tabular}{l|cccccp{8cm}}
		& IUE	& FUSE	& STIS	& COS	& aux\\
\hline
{\bf HAeBe}\\
\hline
HD 97048	& 12	& 0	& 0	& 0	& 0   \\
  
MWC 480	        & 8	& 6	& 0	& 0     & 0  \\
HD 142666	& 7	& 2	& 0	& 0     & 0   \\
HD 95881	& 4	& 4	& 0	& 0     & 0  \\ 
HD 169142	& 5	& 0	& 0	& 0	& 0  \\
HD 100546       & 9	& 6	& 8	& 0     & 0   \\ 
HD 163296       & 31	& 6	& 8	& 0     & 0  \\
AB Aur	        & 29	& 4	& 1	& 0     & 0  \\
HD 141569	& 5	& 2	& 0	& 0     & 0  \\
HD 104237       & 54 	& 8  	& 7 	& 0     & 0  \\
HD 144432       & 9 	& 0  	& 0 	& 0     & 0  \\
V380 Ori        & 8     & 0     & 0     & 0     & 0  \\

\hline
{\bf trans.\,discs}\\
\hline
T Cha	        & 0	& 4	& 6	& 0     & 0  \\ 
GM Aur	        & 10    & 4     & 3     & 5     & 2  \\
DM Tau	        & 0	& 0	& 3	& 5     & 1  \\
LkCa 15	        & 0	& 0	& 0	& 0     & 1   \\
49 Cet	        & 2	& 4	& 1	& 0     & 0  \\
\hline
{\bf F-type}\\
\hline
HD 135344B	& 0     & 2     & 0     & 8     & 0 \\
RY Tau	        & 107	& 0	& 6	& 0     & 1  \\
CQ Tau          & 12    & 0     & 0     & 0     & 1  \\
HD 181327	& 0 	& 3 	& 0 	& 0     & 0   \\ 
\hline
{\bf G-type T Tauri}\\
\hline
DO Tau	        & 0	& 0	& 0	& 0     & 1  \\
RU Lup          & 51 	& 2 	& 6	& 0     & 1 \\
RY Lup	        & 4	& 0	& 4	& 0	& 1   \\
V1149 Sco	& 3	& 0	& 0	& 0     & 0  \\
DL Tau          & 7     & 0     & 0     & 0     & 1  \\
RNO 90          & -     & -     & -     & -     & -  \\
RW Aur          & 44    & 0     & 0     & 0     & 1  \\
LkHa 326        & -     & -     & -     & -     & -  \\
\hline
{\bf K-type T Tauri}\\
\hline
VZ Cha          & 2 	& 0 	& 0 	& 0     & 0  \\
DN Tau          & 19 	& 0 	& 0 	& 0     & 1  \\
TW Cha          & 1 	& 0 	& 0 	& 0     & 0  \\
TW Hya	        & 16	& 6	& 0	& 0 	& 2 \\
BP Tau 	        & 81	& 0	& 2	& 0	& 1  \\ 
DR Tau    	& 54	& 0	& 7	& 2     & 1 \\	
Haro 1-16	& 1 	& 0 	& 0  	& 0     & 0  \\
CW Tau          & 5	& 0	& 4	& 0     & 0 \\
CI Tau          & 0 	& 0 	& 0 	& 0     & 1 \\
V4046 Sgr       & 15 	& 2 	& 0	& 8     & 1 \\
PDS 66          & 2 	& 8  	& 0 	& 0     & 0 \\
V1121 Oph       & 4     & 0     & 0     & 0     & 0 \\
T TauN          & 67    & 2     & 0     & 0     & 2 \\
AS 205B         & 2     & 0     & 0     & 0     & 0 \\
HT Lup          & 6     & 0     & 0     & 0     & 0 \\
UY Aur          & 3     & 0     & 0     & 0     & 0 \\
DG Tau          & 24    & 0 	& 15 	& 0     & 0 \\
\hline
{\bf M-type T Tauri}\\
\hline
IM Lup          & 3 	& 0 	& 0 	& 0     & 0  \\
CY Tau          & 0 	& 0 	& 5  	& 0     & 2  \\
DF Tau          & 33 	& 2 	& 11 	& 8     & 1 \\
RECX 15         & 0 	& 4 	& 1 	& 5     & 3  \\
EX Lup          & 3 	& 0 	& 0 	& 0     & 0  \\
XX Cha          & 0	& 2 	& 0 	& 0     & 0  \\
GQ Lup          & 2 	& 0 	& 0 	& 0     & 0 \\
Hen 3-600A      & 1 	& 0 	& 6 	& 0     & 1 \\ 
UZ Tau E        & 0 	& 0 	& 0 	& 0     & 1  \\
TWA 7           & 0 	& 0 	& 0 	& 0     & 1  \\
FS Tau          & 0 	& 0 	& 2 	& 0     & 0 \\
\hline
{\bf edge-on discs}\\
\hline
AA Tau          & 7 	& 0 	& 1 	& 4     & 1  \\
\end{tabular}}
\end{table}

\noindent The original datasets are therefore inhomogeneous, as they originate from different instruments with different resolutions, integration times and sensitivities. Moreover, some sources were targeted multiple times with a number of different instruments (see also Table~\ref{ResTable}). Intrinsic variation in the UV spectra as a result of changing accretion rates is expected, it is however beyond the scope of this study. As a first step, exceedingly noisy spectra were discarded after visual inspection. We note that IUE data shortward of the  Ly\,$\alpha$ ($\lambda\!\approx\!1215\,$\AA) show abnormally high fluxes when compared to HST/COS spectra and were consequently not used. IUE data longward of  about 3100\,\AA\ can become exceedingly noisy and were also disregarded.  We also note that while UV data of high quality exist for sources with spectral type ranging from $A$ to $M$, data for the $K$ and $M$-type stars either are sparse or do not exist. 

 An example of a co-added UV spectrum is presented in Fig.~\ref{fig:3} for HD163296, while more plots for all other sources are provided as online material in Figure~\ref{fig:uvonl1}.   

\onlfig{
\setcounter{figure}{3}
\begin{figure*}
\centering
\includegraphics[width=0.98\textwidth]{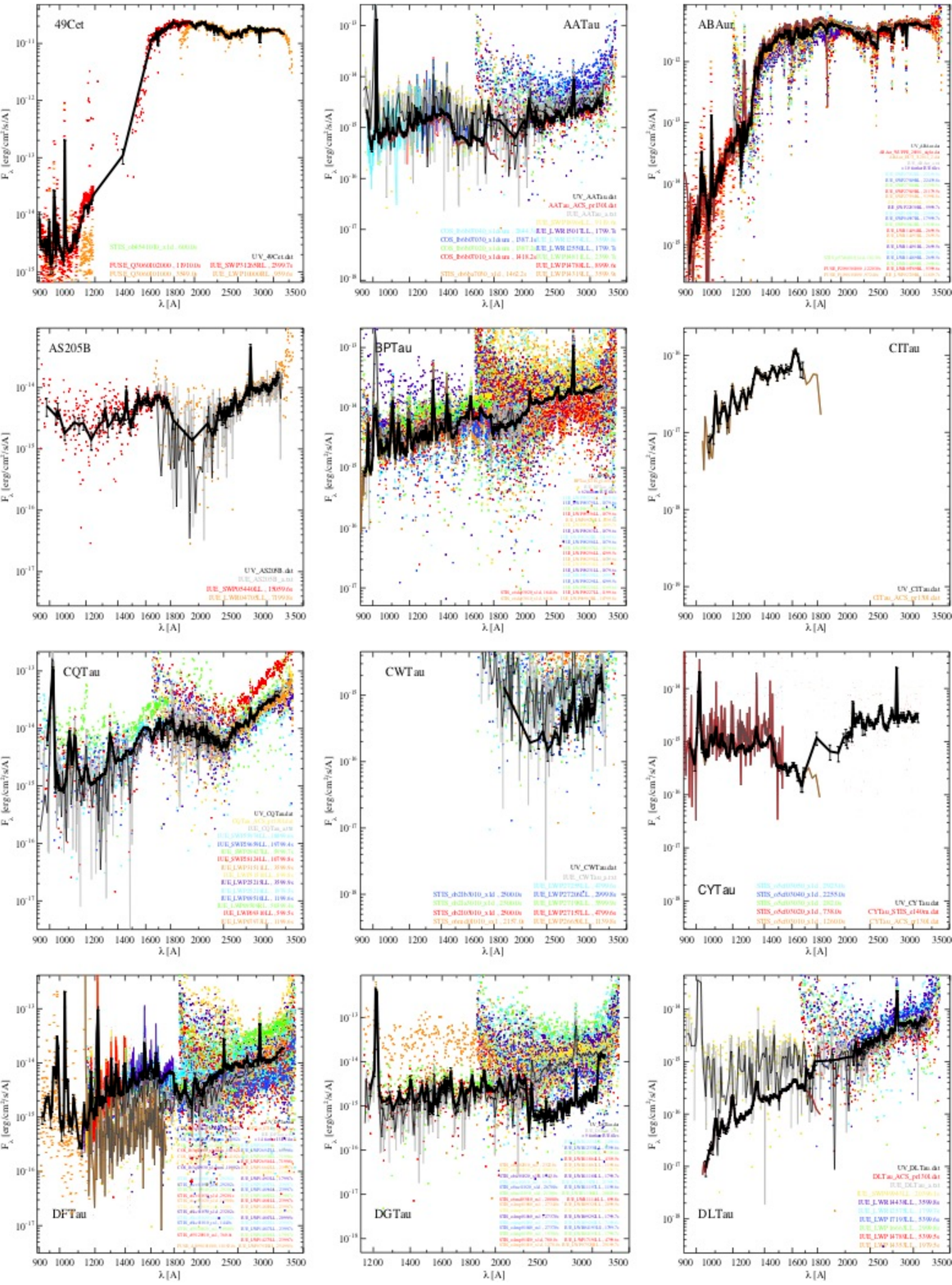}
\caption{Plots of UV averaged spectra for all sources with available UV observations (see also Fig.~\ref{fig:3} and App.~\ref{App:A})} 
\label{fig:uvonl1}
\end{figure*}

\setcounter{figure}{3}
\begin{figure*}
\centering
\includegraphics[width=0.98\textwidth]{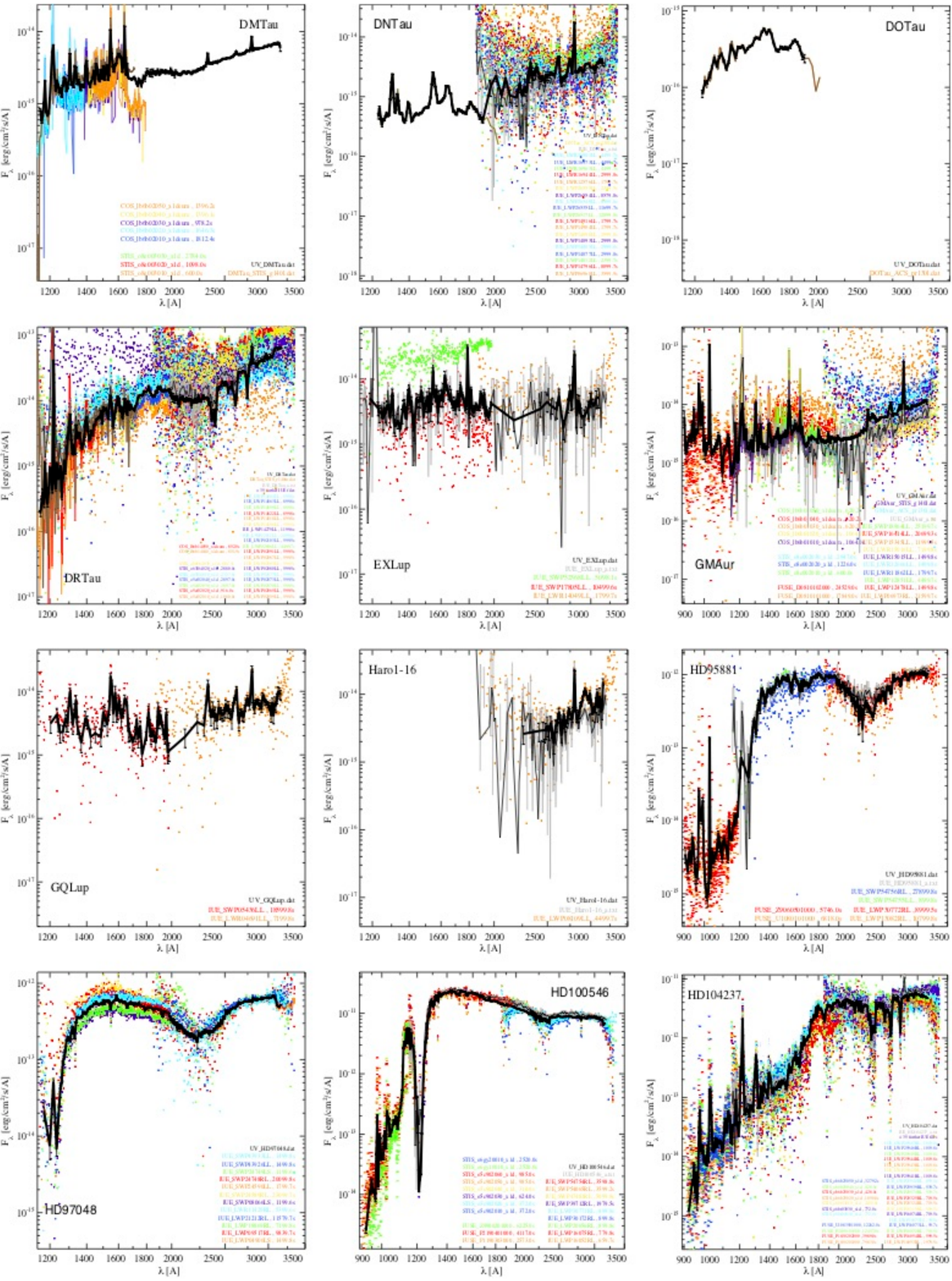}
\caption{(continued from page~\pageref{fig:uvonl1})} 
\end{figure*}

\setcounter{figure}{3}
\begin{figure*}
\centering
\includegraphics[width=0.98\textwidth]{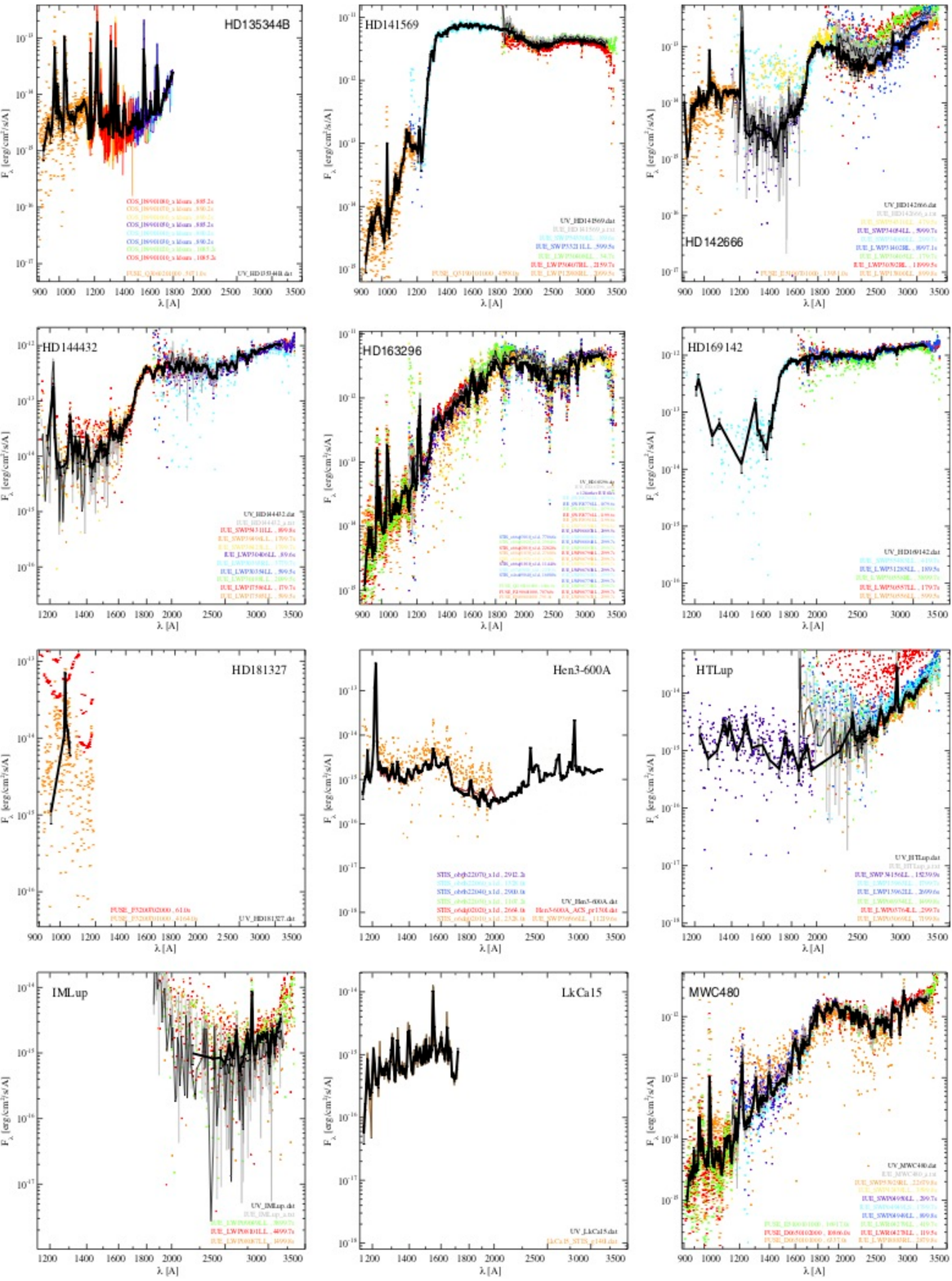}
\caption{(continued from page~\pageref{fig:uvonl1})} 
\end{figure*}

\setcounter{figure}{3}
\begin{figure*}
\centering
\includegraphics[width=0.98\textwidth]{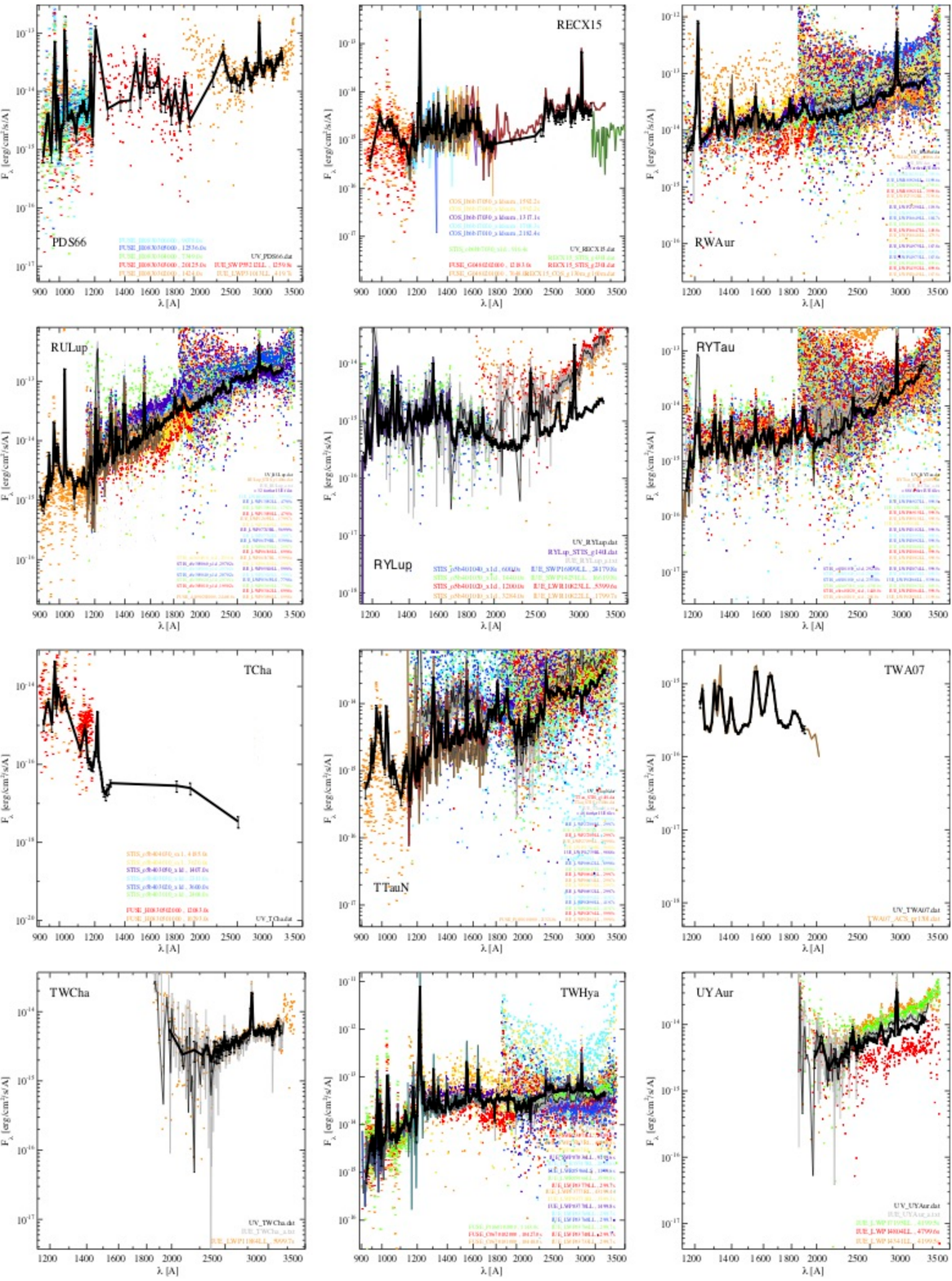}
\caption{(continued from page~\pageref{fig:uvonl1})} 
\end{figure*}

\setcounter{figure}{3}
\begin{figure*}
\centering
\includegraphics[width=0.98\textwidth]{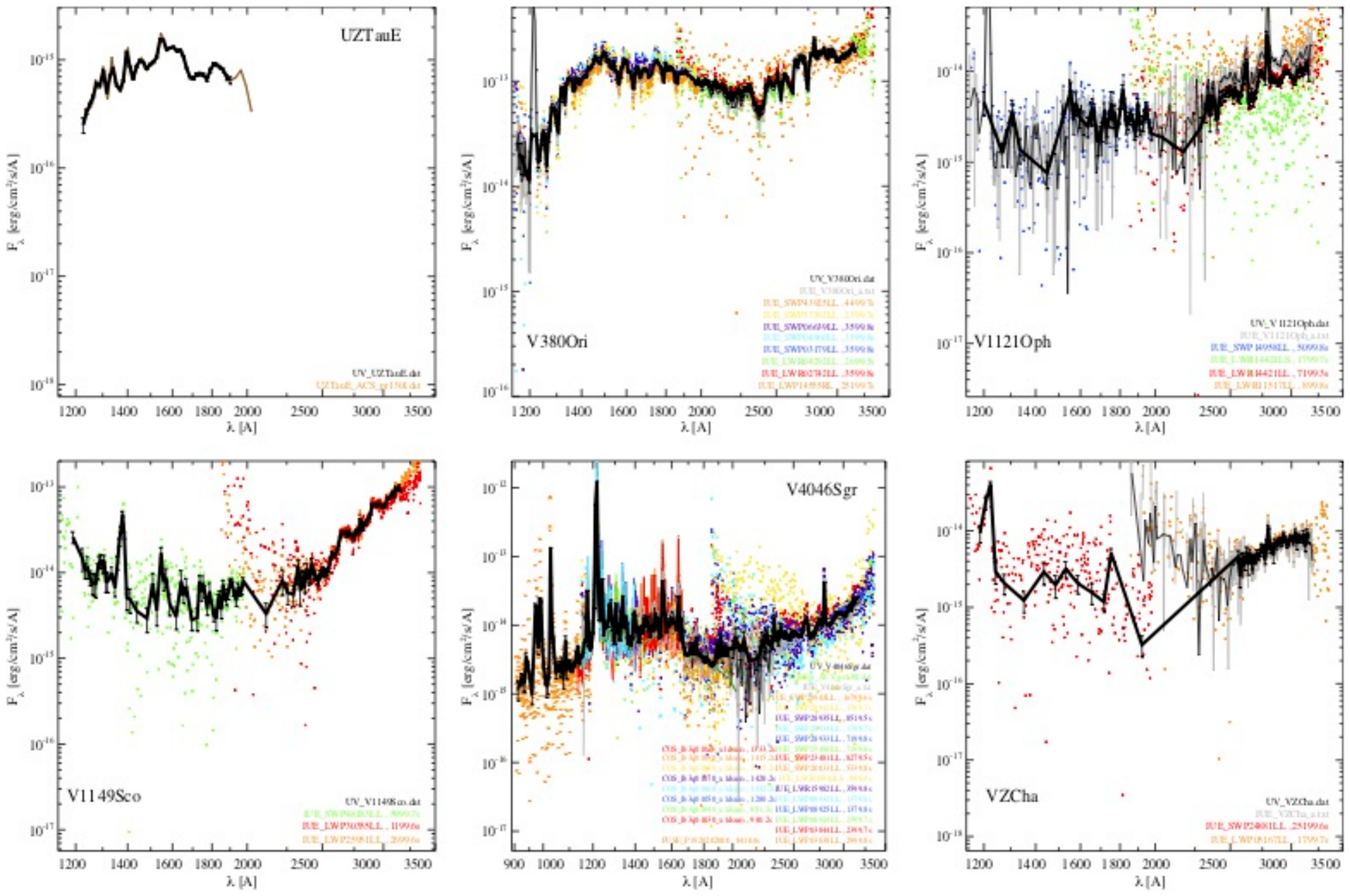}
\caption{(continued from page~\pageref{fig:uvonl1})} 
\end{figure*}
}

For the cases that UV spectra were not available, we have collected photometric data points from a number of space facilities, namely: 

\begin{itemize}
\setlength\itemsep{1em}
\item{The Ultraviolet Sky Survey Telescope (UVSST) onboard the TD1 satellite \citep{Humphries:76a}, provides photometry down to 10th mag in four UV 4 bands at 1565~$\AA$, 1965~$\AA$, 2365~$\AA$ and 2740~$\AA$. }

\item{The ultraviolet photometer of the Astronomical Netherlands Satellite (ANS) having 5 bands at 1500 $\AA$, 1800 $\AA$, 2500 $\AA$ and 3300 $\AA$  \citep{Wesselius:82a}.}

\item{ The Galaxy Evolution Explorer (GALEX) mission provided wide band photometry in two windows; the FUV channel between 1350 and 1750 $\AA$ and the NUV channel between 1750 and 2800 $\AA$ \citep{Morrissey:07a}.}

\end{itemize}


\subsection{Visual}

Visual data are considered the photometric data in all major photometric systems that can traditionally be observed from ground based facilities. Visual data have been collected using customized query scripts that scan and automatically retrieve data from online data archives. Such resources include:


\begin{itemize}
\setlength{\itemsep}{1.3pt}
\setlength{\parskip}{0pt}
\setlength{\topsep}{0pt}
\setlength{\parsep}{0pt}
\setlength{\partopsep}{0pt}
\item{The Amateur Sky Survey (TASS) of the Northern Sky, measured Mark IV magnitudes which are then converted to Johnson-Cousins V- and I-magnitudes \citep{Richmond:07a}.}

\item{General Catalogue of Photometric Data II (GCPD), was queried for standard photometric systems \citep{Mermilliod:97a}.} 

\item{Sloan Digital Sky Survey Photometric Catalog, release 8 \citep{Adelman-McCarthy:11a} and release 6 \citep{Adelman-McCarthy:08a}.}

\item{DENIS J-K photometry \citep{Kimeswenger:04a}.}

\item{USNO-B1 All Sky Catalogue \citep{Monet:03a}.}

\item{VizieR Online Data Catalog: Homogeneous Means in the UBV System \citep{Mermilliod:06a}.} 

\item{The Geneva-Copenhagen survey of the solar neighbourhood. III. Improved distances, ages, and kinematics \citep{Holmberg:09a}.}

\item{Catalogue of stars measured in the Geneva Observatory photometric system \citep{Rufener:88a}.}

\item{VizieR Online Data Catalog: Catalogue of Stellar Photometry in Johnson's 11-color system \citep{Ducati:02a}.}

\item{All-sky compiled catalogue of 2.5 million stars, comprising data from HIPPARCHOS, Tycho, PPM and CMC11 catalogues \citep{Kharchenko:01a}.}

\item{UBVRIJKLMNH photoelectric photometric catalogue \citep{Morel:78a}.}

\item{Uvby $\beta$ photoelectric photometric catalogue \citep{Hauck:98a}.}

\item{Uvby $\beta$ photometry of 1017 stars earlier than G0 in the Centaurus-Crux-Musca-Chamaeleon direction \citep{Corradi:95a}.}

\item{Tycho-2 bright source catalogue \citep{Hog:00a}.}

\item{The HIPPARCOS and TYCHO catalogues. \citep{Hipparchos:97a}.}

\item{SDSS g,r,i,z filters calculated from HIPPARCHOS and TYCHO data \citep{Ofek:08a}. }

\item{Catalogue of photoelectric photometry in the Vilnius system \citep{Straizys:89a}.}



\item{Hipparchos catalogue photometric filters \citep{Perryman:97a}}

\end{itemize}

The offset positions for different sets of observations along with proper motion vectors were visually inspected and subsequently selected/deselected by hand. In order to maintain homogeneity in our datasets, fluxes and corresponding errors were converted from original units to Jy. Data from different catalogues were cross-correlated and checked against and flags were applied according to their quality. If no flux errors were given in the original catalogues, a nominal 10\% error was assumed, which sometimes was increased to 30\% for particularly unreliable passbands.

There are some noticeable trends among the collected visual datasets. The SDSS data, for example, are of high quality but the survey was designed to be deep, so that background sources are sometimes confused with our intended targets. Such cases are easily identifiable and corrected. Photometric data from DENIS/VLTI are often saturated for rather bright sources, and in such cases data are flagged as unreliable. Data from the USNO-B1 survey suffer from rather high uncertainties, estimated between 30 and 50\%, and the photometric filters of the survey are not well defined \citep{Monet:03a}.


\subsection{Near infrared}

For the purposes of the present data collection, near-infrared lies between 0.8 (i.e. the Johnson $I$ band) and $\sim$2.2 $\micron$  (K$_S$ band). In addition to the references from the Visual wavelengths that also apply here in some cases (the DENIS/ VLTI datasets, for example),  near infrared data were additionally collected from the following resources: 

\begin{itemize}
\setlength{\itemsep}{1.3pt}
\setlength{\parskip}{0pt}
\setlength{\topsep}{0pt}
\setlength{\parsep}{0pt}
\setlength{\partopsep}{0pt}
\item{Two Micron All Sky Survey (2MASS) \citep{Cutri:12a, Cutri:03a}.}

\item{The Cosmic Background Explorer (COBE) Diffuse Infrared Background Experiment  (DIRBE) Point Source Catalog  \citep{Smith:04a}.}

\item{ J, H, and Ks for sources in Chameleon were retrieved from \citet{Carpenter:02a}.}

\end{itemize}


\subsection{Mid and far-IR}

Mid- and far-infrared refers here to photometric and spectroscopic data data between 5 and 200 $\micron$, observed mainly with space facilities. Data collection in this wavelength range consists of already reduced and previously published data, and quite often different reductions of the same dataset exist. The wavelength range is of particular importance for the proper modeling of the dust content in disks. Therefore special care has been taken in order to evaluate the different datasets and reductions, in order to provide high quality data of silicate features, especially the most intense one centered at $\sim$10\,$\micron$. 

The mid- and far-infrared data were collected from the following resources:



\begin{itemize}
\setlength{\itemsep}{1.3pt}
\setlength{\parskip}{0pt}
\setlength{\topsep}{0pt}
\setlength{\parsep}{0pt}
\setlength{\partopsep}{0pt}
\item{The Faint Source Catalogue \citep{Moshir:90a} of the Infrared Astronomical Satellite \citep[IRAS][]{Iras:88a}}

\item{Spitzer spectra from "\textit{Dust Evolution in Protoplanetary Disks Around Herbig Ae/Be Stars}" \citep{Juhasz:10a}}

\item{Spitzer data from the \textit{Cores 2 Disks (c2d)} Survey \citep{Evans:03a, Wahhaj:10a}. }

\item{\bf Smoothed ISO spectra for a sample of Herbig Ae/Be systems \citep{Meeus:01a}.}

\item{Spectra from the "\textit{Spitzer Infrared Spectrograph Survey of T-Tauri Stars in Taurus}" \citep{Furlan:11a}. }

\item{Spitzer IRAC data from "\textit{Galactic Legacy Infrared Midplane Survey Extraordinaire (GLIMPSE)}"  \citep{Spitzer-Science:09a}.}

\item{Spitzer IRAC and MIPS, data from  "\textit{The Disk Population of the Taurus Star-Forming Region}"  \citep{Luhman:10a}.}

\item{Spitzer IRAC data from "\textit{Taurus Spitzer Survey: New Candidate Taurus Members Selected Using Sensitive Mid-Infrared Photometry}" \citep{Rebull:10a}.}

\item{Spitzer spectrophotometric data from "\textit{The Formation and Evolution of Planetary Systems: Placing Our Solar System in Context with Spitzer}", \citep{Meyer:06a}.}

\item{Data from the "\textit{The Cornell Atlas of Spitzer/IRS Sources (CASSIS\footnote{http://cassis.astro.cornell.edu/atlas/index.shtml})}"  \citep{Lebouteiller:11a}. }

\item{Data from the Spitzer Map of the Taurus Molecular Clouds  \citep{Padgett:06a}.} 

\item{AKARI/IRC mid-infrared all-sky survey \citep{Murakami:07a, Ishihara:10a}.}

\item{Spitzer/IRS data from the "\textit{The Different Evolution of Gas and Dust in Disks around Sun-Like and Cool Stars}" project \citep{Pascucci:09a}.}

\item{Midcourse Space Experiment (MSX) Infrared Point Source Catalog\footnote{http://irsa.ipac.caltech.edu/applications/Gator/GatorAid/MSX/ readme.html} \citep{Egan:03a}.}

\item{Wide-field Infrared Survey Explorer (WISE\footnote{http://wise2.ipac.caltech.edu/docs/release/allwise/}) catalogue \citep{Cutri:12b}.}

\item{Herschel/PACS spectra for sources in the  Upper Scorpius star-forming region \citep{Mathews:13a}. }

\item{Herschel/PACS spectra from the "\textit{Gas in Protoplanetary Systems Survey}" (GASPS) \citep{Meeus:12a, Dent:13a}.}

\item{Herschel/PACS spectra from the "\textit{Dust, Ice and Gas in Time Survey}"  \citep[DIGIT,][]{Green:16a, Fedele:13a, Meeus:12a, Meeus:13a, Cieza:13a}. }

\item{\bf Herschel/SPIRE spectra sample of Herbig Ae/Be systems from \citet{vanderWiel:14a}. }

\end{itemize}

\subsection{Submillimeter and millimeter wavelength data (continuum)}

Continuum data in the (sub)-millimeter come from a large number of facilities, including both single-dish telescopes and interferometers, and were mainly compiled from published articles. In the following we give a complete description of these resources per wavelength band. 

\begin{itemize}
\setlength{\itemsep}{1.3pt}
\setlength{\parskip}{0pt}
\setlength{\topsep}{0pt}
\setlength{\parsep}{0pt}
\setlength{\partopsep}{0pt}
\item{350~$\micron$: \citet{Andrews:07a, Mannings:94a, Carpenter:05a, Mannings:94b, Dent:98a}}

\item{450-850~$\micron$: the SCUBA Legacy Catalogues \citet{Di-Francesco:08a}  and from individual papers \citet{Sandell:11a, Andrews:07a, Mannings:94a, Mannings:94b, Dent:98a, Beckwith:91a, Nilsson:09a}.}


\item{1.0 - 2.0 mm: \citet{Beckwith:91a, Mannings:94b, Dent:98a, Henning:93a, Henning:94a, Nuernberger:97a, Guilloteau:11a, Schaefer:09a, Mannings:94a, Carpenter:05a, Andre:94a, Osterloh:95a, Mannings:94b, Motte:98a, Lommen:07a}.}

\item{2.0 - 5.0 mm: \citet{Mannings:94a, Kitamura:02a, Schaefer:09a, Dutrey:96a, Guilloteau:11a, Carpenter:05a, Ubach:12a, Ricci:10a}.}


\item{7 mm: \citet{Ubach:12a, Lommen:07a, Rodmann:06a}.}


\end{itemize}

Data were also hand-picked from papers focusing on the study of individual sources. Examples of such resources include:

\begin{itemize}
\setlength{\itemsep}{1.3pt}
\setlength{\parskip}{0pt}
\setlength{\topsep}{0pt}
\setlength{\parsep}{0pt}
\setlength{\partopsep}{0pt}
\item{mm and cm observations of PDS 66 from \citet{Cortes:09a}}

\item{7mm observations of DO Tau from \citet{Koerner:95a}}

\item{CARMA observations of RY Tau and DG Tau at wavelengths of 1.3 mm and 2.8 mm from \citet{Isella:10a}}

\item{mm and cm ATCA observations of WW Chamaeleontis, RU Lupi, and CS Chamaeleontis  from \citet{Lommen:09a}}

\item{ 850 and 450 micron observations of the TWA 7 debris disk from \citet{Matthews:07a}}

\item{Millimeter Continuum Image of the  disk around the Haro 6-5B from \citet{Yokogawa:01a}}

\item{Multi-wavelength observations of the HV Tau C disk from \citet{Duchene:10a}}

\end{itemize}

\subsection{Gas lines}

Fluxes for gas lines along with spectral line profiles have been retrieved from a limited number of gas-line surveys of protoplanetary disks. More lines were handpicked for individual sources and from articles focusing on the modeling of gas lines with thermochemical codes \citep[e.g.][]{Carmona:14a, Woitke:18a}.

\begin{itemize}
\setlength{\itemsep}{1.3pt}
\setlength{\parskip}{0pt}
\setlength{\topsep}{0pt}
\setlength{\parsep}{0pt}
\setlength{\partopsep}{0pt}


\item{CO $J$=1-0, 2-1 transitions from \citet{Schaefer:09a} }

\item{The Herschel/DIGIT and GASPS line surveys ([OI], [CII], H$_2$O, OH, CH$^+$ and CO transitions), \citep{Fedele:13a, Meeus:12a, Meeus:13a, Mathews:10a, Mathews:13a, Dent:13a}}

\item{Herschel SPIRE lines \citep{vanderWiel:14a}}


\item{Spitzer lines \citep{Pontoppidan:10a, Salyk:11a, Boogert:08a, Oberg:08a, Pontoppidan:08a, Bottinelli:10a}.}

\end{itemize}

Space-born data was complemented by data and/or line measurements from  ground-based high-spectral resolution near- and mid-IR surveys:

\begin{itemize}
\setlength{\itemsep}{1.3pt}
\setlength{\parskip}{0pt}
\setlength{\topsep}{0pt}
\setlength{\parsep}{0pt}
\setlength{\partopsep}{0pt}

\item{CO ro-vibrational data from the ESO-VLT/CRIRES large program {\it "The planet-forming zones of disks around solar-mass stars"}  (PI. van Dishoeck) \footnote{ESO-program 179.C-0151 http://www.stsci.edu/~pontoppi/} \citep{Pontoppidan:11a, Brown:13a, Banzatti:17a}}.
\item{ CO ro-vibrational line-measurements from \citep{Najita:03a, Blake:04a, Carmona:14a}}
\item{Near- and mid-IR H$_2$ emission in Herbig Ae/Be stars \citep{Carmona:11a, Bitner:08a, Carmona:08a, Martin-Zaidi:10a}}


\item{Millimeter and submillimeter line surveys \citep{Dutrey:96a, Oberg:10a, Oberg:11a, Guilloteau:12a, Fuente:10a, Bergin:13a, Cleeves:15a}.}




\end{itemize}

\section{Auxiliary data and model results.}


As a starting point for modeling efforts, we have collected descriptive parameters of the central protostar from $\sim$60 refereed articles. A detailed account of these records is given in Table~\ref{tab:carla}, along with corresponding references. Stellar parameters along with the interstellar extinction are used as starting points for dust radiative transfer and thermochemical models.


\begin{longtab}
\begin{longtable}{l c c c c c c c c c c c c c c c c}
\caption{\label{tab:carla}  Stellar parameters}\\
\hline\hline
Source & Spectral & Av & log(L)& Mass &  Teff & log(age)  & L$_X$ & log[M$_{acc}$]  & L$_{FUV}$ & D  \\
name & Type &  (mag) &  (L$_{\odot}$) & (M$_{\odot}$) & (K)  &  (yr) &   (L$_{\odot}$) & (M$_{\odot}$/yr) & (L$_{\odot}$) &(pc) \\ 
\hline
\endfirsthead
\caption{continued.}\\
\hline\hline
Source & Spectral & Av & log(L)& Mass &  Teff & log(age)  & L$_X$E- & log[M$_{acc}$]  & Lfuv & D  \\
name & Type &  (mag) &  (L$_{\odot}$) & (M$_{\odot}$) & (K)  &  (yr) &   (L$_{\odot}$) & (M$_{\odot}$/yr) & (L$_{\odot}$) &(pc) \\ 
\hline
\endhead
\hline
\endfoot
\multicolumn{17}{c}{Herbig Ae/Be}\\
\hline
\multirow{3}{*}{HD97048} & B9.5$^{(1)}$ & 0.87$^{(2)}$  &1.84$^{(1)}$ &2.5$^{(3)}$  & 10000$^{(1)}$ & 6.8$^{(4)}$& 8.26E-5$^{(5)}$ &  &  &175$^{(2)}$ \\
 & A0V $^{(2)}$ & 1.15$^{(4)}$ &1.46$^{(2)}$ &  &  &  &  &  &  &158$^{(4)}$ \\
 & A0$^{(3)}$ & 1.24$^{(3)}$ &1.53$^{(3)}$ &  &  &  &  &  &  &  \\
HD142666 & A8V $^{(3)}$ & 0.8$^{(3)}$ &0.81$^{(3)}$ &1.6$^{(3)}$ & 7590$^{(6)}$ & 5.1$^{(6)}$&  &-6.73$^{(6)}$ &  &145$^{(6)}$ \\
\multirow{3}{*}{HD100546} & B9V $^{(2)}$ & 0.09$^{(7)}$ &1.47$^{(2)}$ &2.5$^{(3)}$ & 10470$^{(4)}$ &  & 2.08E-5$^{(5)}$ &  &  &103$^{(2)}$ \\
 &  & 0.15$^{(2)}$ &1.36$^{(4)}$ &  & 11412$^{(3)}$ &  &  &  &  &97$^{(4)}$ \\
 &  & 0.36$^{(3)}$ &1.63$^{(3)}$ &  &  &  &  &  &  &  \\
\multirow{2}{*}{HD163296} & A4$^{(3)}$ & 0.5$^{(3)}$ &1.58$^{(3)}$ &2.47$^{(3)}$ & 8907$^{(3)}$ &  &  &  &  &122$^{(3)}$ \\
 & A1(Ve) $^{(3)}$ &  &  &  &  &  &  &  &  &  \\
\multirow{3}{*}{ABAur} & A1$^{(8)}$ & 0.55$^{(8)}$ &1.39$^{(8)}$ &2.4$^{(9)}$ & 9840$^{(9)}$ & 6.6$^{(8)}$& 8.26E-5$^{(5)}$ &-6.9$^{(9)}$ &  &140$^{(9)}$ \\
 & A0$^{(9)}$ & 0.5$^{(9)}$ &1.68$^{(9)}$ &2.31$^{(8)}$ &  &  &  &  &  &  \\
 &  & 0.25$^{(4)}$ &  &  &  &  &  &  &  &  \\
\multirow{2}{*}{HD141569} & B9.5$^{(9)}$ & 0.37$^{(4)}$ &1.36$^{(9)}$ &2.2$^{(9)}$ & 9550$^{(6)}$ & 6.7$^{(6)}$& <3.29E-6$^{(5)}$ &-6.89$^{(6)}$ &  &99$^{(6)}$ \\
 &  &  &1.47$^{(4)}$ &  &  &  &  &-8.13$^{(9)}$ &  &116$^{(4)}$ \\
\multirow{3}{*}{HD104237} & A0V $^{(2)}$ & 0.08$^{(7)}$ &1.72$^{(2)}$ &  & 8550$^{(4)}$ &6.74$^{(4)}$& 4.14E-4$^{(5)}$ &  &  &114$^{(4)}$ \\
 & A4-5V $^{(4)}$ & 0.56$^{(2)}$ &1.45$^{(4)}$ &  &  &  &  &  &  &  \\
 &  & 0.16$^{(4)}$ &  &  &  &  &  &  &  &  \\
\multirow{2}{*}{HD144432} & A5V $^{(2)}$ & 0.62$^{(2)}$ &1.68$^{(2)}$ &2$^{(9)}$ & 7410$^{(6)}$ & 5.3$^{(6)}$&  &<-7.22$^{(6)}$ &  &145$^{(6)}$ \\
 &  &  &1.17$^{(9)}$ &  &  &  &  &-7.69$^{(9)}$ &  &253$^{(2)}$ \\
\multirow{2}{*}{V380Ori} & B8/A2$^{(10)}$ &  &1.99$^{(11)}$ &  &  &  &  &  &  &510$^{(11)}$ \\
 & A1e $^{(11)}$ &  &  &  &  &  &  &  &  &  \\
\multirow{3}{*}{HD150193} & A1V $^{(2)}$ & 1.15$^{(2)}$ &1.26$^{(2)}$ &2.2$^{(9)}$ & 8970$^{(6)}$ & 5$^{(6)}$ & 1.00E-4$^{(5)}$ &-6.12$^{(6)}$ &  &203$^{(6)}$ \\
 & B9.5Ve $^{(10)}$ & 1.55$^{(4)}$ &1.69$^{(4)}$ &  &  &6.58$^{(4)}$&  &  &  &150$^{(2)}$ \\
 &  &  &  &  &  &  &  &  &  &216$^{(4)}$ \\
 \hline
\multicolumn{17}{c}{Transition Disks}\\
\hline
TCha &  &  &  &  &  &  & 2.87E-4$^{(12)}$ &  &  &  \\
GMAur &  &  &  &  &  &  & 4.18E-4$^{(12)}$ &  &  &  \\
\multirow{3}{*}{DMTau} & M3$^{(8)}$ & 0.1$^{(8)}$ &-0.89$^{(8)}$ &0.62$^{(9)}$ & 3700$^{(9)}$ & 6.6$^{(8)}$& 5.23E-4$^{(12)}$ &-8.20$^{(13)}$ &8.30E-3$^{(14)}$ &140$^{(15)}$ \\
 & M1$^{(9)}$ & 0.6$^{(15)}$ &-0.49$^{(9)}$ &0.35$^{(8)}$ &  &  &  &-8.54$^{(16)}$ &4.02E-3$^{(15)}$ &  \\
 &  & 0$^{(17)}$ &-0.6$^{(15)}$ &  &  &  &  &-7.95$^{(12)}$ &6.75E-4$^{(18)}$ &  \\
\multirow{2}{*}{49Cet} & A1V $^{(19)}$ & 0.22$^{(4)}$ &1.34$^{(20)}$ &2.17$^{(21)}$ & 9970$^{(21)}$ &0.94$^{(19)}$&  &  &  &59$^{(20)}$ \\
 & A4 V $^{(4)}$ &  &  &  &  &6.9$^{(4)}$ &  &  &  &  \\
CoKuTau4 & M1.1$^{(8)}$ & 1.75$^{(8)}$ &-0.5$^{(8)}$ &0.53$^{(8)}$ & 3720$^{(22)}$ &6.5$^{(8)}$ & <5.49E-5$^{(23)}$ &<-10.0$^{(12)}$ &  &  \\
\multirow{3}{*}{UXTauA} & K2$^{(24)}$ & 0.51$^{(7)}$ &0$^{(24)}$ &1.5$^{(25)}$ & 5856$^{(17)}$ &6.43$^{(27)}$& 0.00052$^{(7)}$ &-8.018$^{(25)}$ &6.75E-4$^{(18)}$ &140 \\
 &  & 0.2$^{(24)}$ &0.338$^{(25)}$ &  &  & 6.1$^{(17)}$&  &  &  &  \\
 &  & 1.3$^{(25)}$ &  &  &  &  &  &  &  &  \\
  \hline
\multicolumn{17}{c}{T-Tauri, F-type}\\
\hline
\multirow{2}{*}{HD142527} & F6IIIe $^{(28)}$ & 0.6$^{(27)}$ &1.18$^{(27)}$ &2.2$^{(28)}$ & 6250$^{(27)}$ & 6.7$^{(27)}$&  &-6.85$^{(30)}$ &  &198$^{(2)}$ \\
 &  & 0.37$^{(2)}$ &1.32$^{(2)}$ &  &  & 6.3$^{(4)}$&  &-7.02$^{(9)}$ &  &233$^{(4)}$ \\
\multirow{3}{*}{HD135344B} & F4V $^{(30)}$ & 0.4$^{(30)}$ &0.91$^{(3)}$ &1.65$^{(30)}$ & 6620$^{(31)}$ &  &  &  &  &140$^{(66;67}$ \\
 & F8V $^{(3)}$ &  &  &1.7$^{(3)}$ & 6950$^{(3)}$ &  &  &  &  &  \\
 & F5V $^{(3)}$ &  &  &  &  &  &  &  &  &  \\
\multirow{4}{*}{RYTau} & G0$^{(8)}$ & 1.84$^{(26)}$ &1.03$^{(8)}$ &2.24$^{(26)}$ & 5770$^{(6)}$ & 6.7$^{(8)}$& 1.44E-3$^{(31)}$ &-7.19$^{(33)}$ &0.16$^{(14)}$ &134$^{(6)}$ \\
 & F8$^{(15)}$ & 1.8$^{(15)}$ &0.89$^{(15)}$ &  & 5496$^{(3)}$ & 5.6$^{(17)}$&  &  &0.072$^{(15)}$ &131$^{(8)}$ \\
 & K1$^{(17)}$ & 2.2$^{(3)}$ &1.04$^{(3)}$ &  &  &  &  &  &  &140$^{(3)}$ \\
 & F8V $^{(3)}$ &  &  &  &  &  &  &  &  &  \\
\multirow{3}{*}{CQTau} & A8$^{(33)}$ & 2.85$^{(34)}$ &0.82$^{(15)}$ &1.5$^{(35)}$ & 6750$^{(35)}$ & 7$^{(35)}$ &  &<-8.30$^{(6)}$ &0.094$^{(15)}$ &100$^{(33)}$ \\
 & F2$^{(15)}$ & 1.9$^{(15)}$ &1.05$^{(36)}$ &  & 7200$^{(36)}$ & 6.6$^{(4)}$&  &  &  &140$^{(36)}$ \\
 &  & 1.4$^{(4)}$ &0.53$^{(4)}$ &  &  &  &  &  &  &113$^{(4)}$ \\
HD181327 & F5/6$^{(37)}$ &  &0.522$^{(38)}$ &1.36$^{(38)}$ &  &7.08$^{(37)}$& <6.56E-5$^{(5)}$ &  &  &51.9$^{(38)}$ \\
 \hline
\multicolumn{17}{c}{T-Tauri, G-type}\\
\hline
\multirow{3}{*}{DOTau} & M0.3$^{(8)}$ & 0.78$^{(8)}$ &-0.64$^{(8)}$ &0.56$^{(26)}$ & 3850$^{(9)}$ & 6.9$^{(8)}$&  &-6.84$^{(39)}$ &0.066$^{(14)}$ &140$^{(9)}$ \\
 & M0$^{(17)}$ & 1.35$^{(24)}$ &0$^{(15)}$ &0.37$^{(9)}$ & 3777$^{(3)}$ &5.73$^{(17)}$&  &-7.28$^{(9)}$ &0.0325$^{(15)}$ &  \\
 & M6$^{(3)}$ & 2.3$^{(15)}$ &0.11$^{(3)}$ &0.66$^{(17)}$ &  &  &  &  &  &  \\
\multirow{3}{*}{RYLup} & G8$^{(40)}$ & 0.65$^{(41)}$ &0.1$^{(41)}$ &1.71$^{(40)}$ & 4590$^{(42)}$ & 7.08$^{(40)}$  &   & &2.2E-3$^{(15)}$ &108$^{(41)}$ \\
 & K4$^{(41)}$ & 0.44$^{(2)}$ &-0.4$^{(2)}$ &1.38$^{(3)}$ & 5200$^{(3)}$ &  &  &  &  &150$^{(11)}$ \\
 & G0V $^{(11)}$ & 2.48$^{(3)}$ &0.42$^{(15)}$ &  &  &  &  &  &  &120$^{(3)}$ \\
\multirow{2}{*}{V1149Sco} & K0III $^{(42)}$ & 1.6$^{(15)}$ &0.4$^{(9)}$ &  & 5088$^{(44)}$ &  &  &  &0.0817$^{(15)}$ &186$^{(43)}$ \\
 & G6$^{(15)}$ &  &  &  &  &  &  &  &  &145$^{(15)}$ \\
\multirow{2}{*}{DLTau} & K5.5$^{(8)}$ & 1.8$^{(8)}$ &-0.3$^{(8)}$ &0.92$^{(8)}$ & 4000$^{(9)}$ & 6.7$^{(8)}$&  &-7.41$^{(9)}$ &0.0041$^{(14)}$ &140$^{(9)}$ \\
 & K7$^{(9)}$ & 1.3$^{(15)}$ &0.06$^{(9)}$ &0.76$^{(9)}$ &  &  &  &  &9.6E-4$^{(15)}$ &  \\
RNO90 & G5$^{(44)}$ & 4.2$^{(44)}$ &0.82$^{(45)}$ &1.6$^{(9)}$ & 5660$^{(9)}$ &  &  &-7.4$^{(9)}$ &  &120$^{(9)}$ \\
RWAur &  &  &  &  &  &  &  &  &  &140$^{(8)}$ \\
\multirow{3}{*}{RWAurA} & K0$^{(8)}$ & -0.25$^{(8)}$ &-0.14$^{(8)}$ &1.13$^{(8)}$ & 4900$^{(36)}$ & 7.2$^{(8)}$& 4.1E-4$^{(7)}$ &-7.7$^{(45)}$ &  &140$^{(8)}$ \\
 & K4$^{(15)}$ & 0.5$^{(16)}$ &  &1.4$^{(36)}$ &  &5.85$^{(36)}$&  &  &  &  \\
 &  & 1.2$^{(15)}$ &  &  &  &  &  &  &  &  \\
\multirow{2}{*}{RWAurB} & K6.5$^{(8)}$ & 0.1$^{(8)}$ &-0.35$^{(8)}$ &0.85$^{(8)}$ & 4350$^{(17)}$ & 6.7$^{(8)}$&  &  &  &140$^{(8)}$ \\
 & K4$^{(15)}$ & 1.2$^{(15)}$ &  &0.86$^{(17)}$ &  & 6.4$^{(17)}$&  &  &  &  \\
LkHa326 & M0$^{(47)}$ &  &  &  &  &5.48$^{(47)}$&  &  &  &250$^{(9)}$ \\
 \hline
\multicolumn{17}{c}{T-Tauri, K-type}\\
\hline
\multirow{2}{*}{VZCha} & K7e $^{(49)}$ & 0.44$^{(2)}$ &-0.54$^{(49)}$ &0.9$^{(49)}$ & 3990$^{(49)}$ &  & 1.38E-4$^{(12)}$ &-8.28$^{(12)}$ &  &168$^{(2)}$ \\
 & M0V $^{(49)}$ &  &  &  &  &  &  &-7.39; $^{(49)}$ &  &150$^{(9)}$ \\
DNTau & K6V $^{(3)}$ & 0.5$^{(3)}$ &-0.1$^{(3)}$ &0.65$^{(3)}$ & 3904$^{(3)}$ &  &  &-8.00$^{(45)}$ &  &140$^{(3)}$ \\
\multirow{2}{*}{DFTau} & K8$^{(1)}$ & 2.11$^{(2)}$ &-0.24$^{(1)}$ &1.15$^{(49)}$ & 3990$^{(1)}$ & 3.59$^{(39)}$ & 3.66E-4$^{(12)}$ &-9.55; $^{(49)}$ &  &168$^{(2)}$ \\
 & K0$^{(2)}$ &  &  &  &  &  &  &  &  &  \\
\multirow{2}{*}{BPTau} &  &  &  &  &  &  & 3.60E-4$^{(12)}$ &  &  &140 \\
\multirow{3}{*}{DRTau} & K6.0$^{(8)}$ & 0.45$^{(8)}$ &-0.51$^{(8)}$ &0.87$^{(8)}$ & 4060$^{(36)}$ & 7$^{(8)}$ & 0.0001$^{(7)}$ &-7.28$^{(45)}$ &0.0083$^{(14)}$ &140$^{(9)}$ \\
 & K7$^{(15)}$ & 0.95$^{(24)}$ &0.23$^{(24)}$ &0.4$^{(9)}$ &  &5.29$^{(17)}$&  &-6.86$^{(9)}$ &0.0011$^{(15)}$ &  \\
 &  & 1.2$^{(15)}$ &  &  &  &  &  &  &  &  \\
Haro1-16 & K2$^{(13)}$ & 1.7$^{(13)}$ &-0.194$^{(13)}$ &0.97$^{(13)}$ & 4900$^{(13)}$ &  & 3.40E-4$^{(12)}$ &-8.2$^{(13)}$ &0.0125$^{(15)}$ &125$^{(13)}$ \\
CWTau & K3$^{(8)}$ & 1.8$^{(8)}$ &-0.35$^{(8)}$ &1.01$^{(8)}$ & 4730$^{(9)}$ & 7.2$^{(8)}$& 7.43E-4$^{(31)}$ &-7.99$^{(31)}$ &  &140$^{(9)}$ \\
\multirow{3}{*}{CITau} & K5.5$^{(8)}$ & 1.9$^{(8)}$ &-0.2$^{(8)}$ &1.53$^{(8)}$ & 4000$^{(9)}$ & 6.6$^{(8)}$& 5.10E-5$^{(31)}$ &-7.59$^{(31)}$ &0.0016$^{(14)}$ &140$^{(9)}$ \\
 & K7$^{(24)}$ & 1.2$^{(24)}$ &-0.08$^{(15)}$ &0.74$^{(17)}$ &  &5.96$^{(17)}$& 8.26E-5$^{(5)}$ &-7.19$^{(31)}$ &9.2E-4$^{(15)}$ &  \\
 &  & 1.8$^{(15)}$ &  &  &  &  &  &  &  &  \\
\multirow{2}{*}{V4046Sgr} & K5$^{(15)}$ & 0.04$^{(7)}$ &-0.41x2$^{(11)}$ &0.9$^{(50)}$ & 4250$^{(50)}$ &1.08$^{(50)}$& 3.14E-4$^{(52)}$ &-9.3$^{(50)}$ &7.23E-4$^{(15)}$ &73$^{(51)}$ \\
 &  & 0$^{(15)}$ &  &  &  &  &  &  &1.51E-3$^{(18)}$ &83$^{(18)}$ \\
LkHa327 & K2$^{(46)}$ &  &0.96$^{(53)}$ &  &  & 5.$^{(50)}$ &  &  &  &250$^{(9)}$ \\
\multirow{2}{*}{PDS66} & K1$^{(16)}$ & 0.2$^{(16)}$ &-0.046$^{(16)}$ &1.1$^{(16)}$ &  &  & 3.82E-4$^{(51)}$ &-9.89$^{(16)}$ &2.46E-2$^{(15)}$ &103$^{(51)}$ \\
 &  & 1.2$^{(15)}$ &0.1$^{(15)}$ &  &  &  &  &-9.1$^{(51)}$ &  &  \\
UScoJ1604 & K2$^{(53)}$ & 1$^{(54)}$ &-0.118$^{(54)}$ &1$^{(53)}$ & 4549$^{(56)}$ & 6.7$^{(53)}$ & 5.21E-4$^{(5)}$ &  &  &145$^{(54)}$ \\
\multirow{2}{*}{GOTau} & M2.3$^{(8)}$ & 1.5$^{(8)}$ &-0.7$^{(8)}$ &0.42$^{(8)}$ & 3850$^{(36)}$ &6.6$^{(8)}$ & 6.51E-5$^{(31)}$ &-8.42$^{(31)}$ &  &140$^{(8)}$ \\
 & M0$^{(17)}$ & 1.2$^{(17)}$ &  &0.72$^{(17)}$ &  &6.81$^{(17)}$&  &  &  &  \\
\multirow{2}{*}{V1121Oph} & K5$^{(15)}$ & 1.2$^{(15)}$ &-0.06$^{(2)}$ &1.4$^{(9)}$ & 4400$^{(9)}$ &  &  &-7.52$^{(9)}$ &0.00876$^{(15)}$ &95$^{(2)}$ \\
 &  &  &  &  &  &  &  &  &  &160$^{(9)}$ \\
WWCha & K5$^{(1)}$ & 2.31$^{(2)}$ &0.74$^{(1)}$ &  & 4350$^{(1)}$ &  &  &  &  &168$^{(2)}$ \\
FKSer & K6IVe $^{(10)}$ &  &0.2$^{(11)}$ &  &  &  &  &  &  &32$^{(11)}$ \\
\multirow{3}{*}{TTauN} & K0$^{(15)}$ & 1.5$^{(15)}$ &0.86$^{(9)}$ &2.11$^{(9)}$ & 5250$^{(9)}$ & 5.9$^{(8)}$& 2.1E-3$^{(31)}$ &-7.5$^{(31)}$ &0.1$^{(14)}$ &147$^{(8)}$ \\
 & K1$^{(24)}$ & 1.44$^{(24)}$ &1.03$^{(24)}$ &1.99$^{(8)}$ &  &  &  &-7.24$^{(31)}$ &0.061$^{(15)}$ &  \\
 &  & 1.25$^{(8)}$ &0.85$^{(8)}$ &  &  &  &  &  &  &  \\
\multirow{3}{*}{AS205B} & M0.1$^{(8)}$ & 2.4$^{(8)}$ &0.05$^{(8)}$ &0.55$^{(8)}$ & 3450$^{(9)}$ & 6.1$^{(8)}$&  &  &  &145$^{(2)}$ \\
 & K5$^{(2)}$ & 1.09$^{(2)}$ &0.34$^{(2)}$ &0.3$^{(9)}$ &  &  &  &-6.68$^{(9)}$ &  &160$^{(9)}$ \\
 & M3$^{(9)}$ &  &  &  &  &  &  &  &  &121$^{(8)}$ \\
WaOph6 & K $^{(9)}$ &  &-0.17$^{(9)}$ &  &  &  &  &  &  &120$^{(9)}$ \\
\multirow{3}{*}{HTLup} & K2$^{(41)}$ & 1.45$^{(41)}$ &0.78$^{(41)}$ &2.5$^{(9)}$ & 4890$^{(41)}$ &  &  &-7.78$^{(9)}$ &  &159$^{(2)}$ \\
 &  & 0.28$^{(2)}$ &0.44$^{(2)}$ &  &  &  &  &  &  &  \\
 &  &  &1.16$^{(9)}$ &  &  &  &  &  &  &  \\
\multirow{2}{*}{DoAr24E} & K5$^{(2)}$ & 2.16$^{(2)}$ &-0.45$^{(2)}$ &0.47$^{(23)}$ &  &  & 1.46E-4$^{(12)}$ &-8.46$^{(23)}$ &  &120$^{(9)}$ \\
 &  &  &0.1$^{(23)}$ &  &  &  & 1.31E-4$^{(23)}$ &  &  &  \\
UYAur & K7$^{(8)}$ & 1$^{(8)}$ &-0.07$^{(8)}$ &0.7$^{(8)}$ &  & 6.3$^{(8)}$& 1.04E-4$^{(12)}$ &-7.18$^{(39)}$ &  &140$^{(9)}$ \\
UYAurA & M0$^{(17)}$ & 0.6$^{(17)}$ &0.49$^{(22)}$ &0.66$^{(17)}$ & 3850$^{(17)}$ &5.56$^{(17)}$&  &  &  &  \\
UYAurB & M2.5$^{(17)}$ & 2.7$^{(17)}$ &  &0.62$^{(17)}$ & 3485$^{(17)}$ &5.84$^{(17)}$&  &  &  &  \\
\multirow{2}{*}{DGTau} & K7$^{(8)}$ & 1.6$^{(8)}$ &-0.31$^{(8)}$ &0.77$^{(8)}$ & 4200$^{(9)}$ & 6.6$^{(8)}$& *2.1E-3$^{(5)}$ &-7.49$^{(9)}$ &  &140$^{(9)}$ \\
 &  & 1$^{(24)}$ &0.18$^{(24)}$ &  &  &  &  &  &  &  \\
 \hline
\multicolumn{17}{c}{T-Tauri, M-type}\\
\hline
IMLup & K6$^{(8)}$ & 0.4$^{(8)}$ &-0.03$^{(8)}$ &0.78$^{(8)}$ &  & 6.3$^{(8)}$& 8.36E-4$^{(12)}$ &  &  &  \\
Haro6-13 &  & 5.43$^{(56)}$ &-0.159$^{(56)}$ &  & 3850$^{(56)}$ &  & 4.14E-5$^{(5)}$ &  &  &140$^{(5)}$ \\
Haro6-13E & M1.6$^{(8)}$ & 2.2$^{(8)}$ &-0.57$^{(8)}$ &0.48$^{(8)}$ & 3800$^{(9)}$ & 6.5$^{(8)}$&  &-7.02$^{(9)}$ &  &140$^{(9)}$ \\
Haro6-13W & K5.5$^{(8)}$ & 2.25$^{(8)}$ &-0.04$^{(8)}$ & $^{(8)}$ &  & 6.3$^{(8)}$&  &  &  &  \\
\multirow{2}{*}{CYTau} & M1$^{(26)}$ & 0.1$^{(26)}$ &-0.4$^{(26)}$ &0.48$^{(26)}$ & 3628$^{(3)}$ &6.37$^{(26)}$& 3.47E-5$^{(31)}$ &-8.86$^{(31)}$ &  &140$^{(3)}$ \\
 & M2V $^{(3)}$ &  &  &  &  &  & 5.07E-5$^{(31)}$ &-8.12$^{(31)}$ &  &  \\
\multirow{2}{*}{DFTau} & M2.7$^{(8)}$ & 0.1$^{(8)}$ &-0.35$^{(8)}$ &0.32$^{(8)}$ & 3470$^{(9)}$ & 5$^{(8)}$ & 3.30E-5$^{(5)}$ &-6.75$^{(39)}$ &6.9E-4$^{(15)}$ &140$^{(9)}$ \\
 & M0.5$^{(9)}$ & 0.6$^{(15)}$ &0.29$^{(9)}$ &0.27$^{(9)}$ &  &  &  &  &  &  \\
DFTauA & M2$^{(17)}$ & 0.6$^{(17)}$ &  &0.61$^{(17)}$ & 3560$^{(17)}$ & 5.14$^{(17)}$  &  &  &  \\
DFTauB & M2.5$^{(17)}$ & 0.8$^{(17)}$ &  &0.65$^{(17)}$ & 3485$^{(17)}$ & 5.74$^{(17)}$  &  &  &  \\
RECX15 & M3$^{(16)}$ & 0.02$^{(7)}$ &-0.3$^{(16)}$ &0.3$^{(16)}$ &  &  &  &-9.1$^{(16)}$ &2.95E-4$^{(18)}$ &97$^{(18)}$ \\
EXLup & M0.5$^{(57)}$ & 0$^{(57)}$ &-0.4$^{(57)}$ &0.5$^{(57)}$ & 3800$^{(41)}$ &  & 3.92E-4$^{(12)}$ &  &  &147$^{(2)}$ \\
\multirow{2}{*}{WXCha} & M1.25$^{(1)}$ & 1.99$^{(2)}$ &-0.076$^{(1)}$ &1.05$^{(48)}$ & 3700$^{(1)}$ &  & 1.20E-3$^{(12)}$ &-8.47$^{(12)}$ &  &168$^{(2)}$ \\
 &  &  &-0.37$^{(2)}$ &  &  &  &  &  &  &  \\
XXCha & M2$^{(1)}$ &  &-0.43$^{(1)}$ &0.57$^{(49)}$ & 3560$^{(1)}$ &  & 2.87E-4$^{(12)}$ &-9.07$^{(12)}$ &  &  \\
\multirow{2}{*}{GQLup} & K5$^{(8)}$ & 1.6$^{(8)}$ &-0.04$^{(8)}$ &0.89$^{(8)}$ & 4000$^{(9)}$ &6.4$^{(8)}$ & 1.93E-4$^{(12)}$ &-7.5$^{(12)}$ &  &100$^{(9)}$ \\
 & K7$^{(9)}$ &  &  &  &  &  &  &-8.15$^{(9)}$ &  &150$^{(11)}$ \\
\multirow{2}{*}{Hen3-600A} & M4Ve $^{(10)}$ & 0$^{(61)}$ &-1.1$^{(58}$ &0.37$^{(58}$ & 3350$^{(58)}$ & 7$^{(58)}$ & 4.14E-5$^{(5)}$ &-9.6$^{(58}$ &  &42$^{(5)}$ \\
 & M3$^{(58}$ &  &  &  &  &  &  &  &  &34$^{(59}$ \\
UZTauE & M2-3$^{(15)}$ & 0.3$^{(15)}$ &0.2$^{(22)}$ &  & 3720$^{(37)}$ & 5.3$^{(22)}$& 2.33E-4$^{(31)}$ &-8.70; $^{(31)}$ &5.6E-4$^{(15)}$ &140$^{(5)}$ \\
\multirow{2}{*}{IQTau} & M1.1$^{(8)}$ & 0.85$^{(8)}$ &-0.72$^{(8)}$ &0.56$^{(8)}$ & 3775$^{(17)}$ & 6.9$^{(8)}$& 1.09E-4$^{(31)}$ &-8.32$^{(31)}$ &  &140 \\
 & M0.5$^{(17)}$ & 1.3$^{(17)}$ &  &0.64$^{(17)}$ &  &5.93$^{(17)}$& 8.26E-5$^{(5)}$ &  &  &  \\

HKTauB & M1$^{(17)}$ & 2.3$^{(17)}$ &-0.33$^{(22)}$ &0.41$^{(47}$ & 3705$^{(17)}$ & 7.7$^{(17)}$& 2.07E-5$^{(31)}$ &-7.65$^{(31)}$ &  &  \\
V853Oph & M2.5$^{(2)}$ & 0.14$^{(2)}$ &-0.42$^{(2)}$ &0.42$^{(24)}$ &  &  & 8.10E-4$^{(12)}$ &-8.31$^{(12)}$ &  &128$^{(2)}$ \\
GGTauA & K7.5$^{(8)}$ & 1.05$^{(8)}$ &0.14$^{(8)}$ &0.62$^{(8)}$ & 4000$^{(22)}$ & 6$^{(8)}$ &  &  &  &  \\
GGTauAB & K7.5$^{(8)}$ & 1.05$^{(8)}$ &0.12$^{(8)}$ &0.63$^{(8)}$ &  & 6$^{(8)}$ & 1.04E-5$^{(5)}$ &-7.76$^{(39)}$ &  &140$^{(5)}$ \\
GGTauB & M5.8$^{(8)}$ & 0$^{(8)}$ &-1.13$^{(8)}$ &0.07$^{(8)}$ & 3760$^{(22)}$ &5.2$^{(8)}$ &  &  &  &  \\
\multirow{2}{*}{SXCha} & M0$^{(1)}$ & 0.79$^{(2)}$ &-0.38$^{(1)}$ &  & 3850$^{(1)}$ &  &  &-8.37$^{(12)}$ &  &168$^{(2)}$ \\
 & M3.5$^{(1)}$ &  &-1.4$^{(1)}$ &  & 3340$^{(1)}$ &  &  &  &  &  \\
\multirow{3}{*}{TWA07} & M3.2$^{(8)}$ & -0.1$^{(8)}$ &-0.94$^{(8)}$ &0.32$^{(8)}$ &  & 6.7$^{(8)}$& 1.04E-4$^{(5)}$ &  &5.98E-6$^{(15)}$ &34$^{(11)}$ \\
 & M1$^{(15)}$ & 0$^{(15)}$ &-0.3$^{(16)}$ &0.5$^{(16)}$ &  &  &  &  &  &  \\
 &  &  &-1.09$^{(15)}$ &  &  &  &  &  &  &  \\
FSTau & M2.4$^{(8)}$ & 2.95$^{(8)}$ &-0.84$^{(8)}$ &0.41$^{(8)}$ &  &6.7$^{(8)}$ & 8.43E-4$^{(31)}$ &-9.50; $^{(31)}$ &  &  \\
FSTauA & M0$^{(17)}$ & 5$^{(17)}$ &  &0.66$^{(17)}$ & 3850$^{(17)}$ &5.67$^{(17)}$&  &  &  &  \\
FSTauB & M3.5$^{(17)}$ & 5.2$^{(17)}$ &  &0.32$^{(17)}$ & 3340$^{(17)}$ & 6.4$^{(17)}$&  &  &  &  \\
 \hline
\multicolumn{17}{c}{Edge-on systems}\\
\hline
\multirow{3}{*}{AATau} & M0.6$^{(8)}$ & 0.4$^{(8)}$ &-0.36$^{(8)}$ &0.58$^{(8)}$ & 4000$^{(9)}$ & 6.4$^{(8)}$& 3.24E-4$^{(31)}$ &-8.48$^{(39)}$ &0.016$^{(14)}$ &140$^{(9)}$ \\
 & K7$^{(9)}$ & 0.74$^{(15)}$ &-0.15$^{(24)}$ &0.74$^{(17)}$ & 4030$^{(3)}$ &6.16$^{(17)}$& 2.61E-4$^{(5)}$ &-7.82$^{(16)}$ &0.002$^{(16)}$ &  \\
 & M0V $^{(3)}$ & 0.93$^{(24)}$ &-0.97$^{(3)}$ &0.85$^{(3)}$ &  &5.98$^{(40)}$&  &  &0.0022$^{(18)}$ &  \\
 &  & 1.64$^{(3)}$ &  &  &  &  &  &  &  &  \\
\multirow{2}{*}{IRAS04158+2805} & M6$^{(17)}$ & 8.6$^{(17)}$ &-0.39$^{(22)}$ &0.1$^{(17)}$ & 2990$^{(17)}$ &5.91$^{(17)}$& 2.30E-4$^{(31)}$ &< -9.5$^{(31)}$ &  &140$^{(5)}$ \\
 & M3$^{(5)}$ &  &  &  & 3470$^{(23)}$ &  &  &  &  &  \\
\multirow{2}{*}{IRAS04385+2550} & M0.5$^{(17)}$ & 10.2$^{(17)}$ &-0.425$^{(56)}$ &0.64$^{(17)}$ & 3775$^{(17)}$ &5.91$^{(17)}$& 1.05E-4$^{(31)}$ &-8.11$^{(31)}$ &  &140$^{(5)}$ \\
 &  & 5.24$^{(56)}$ &  &  &  &  &  &  &  &  \\
 \hline
\multicolumn{17}{c}{Embedded sources}\\
\hline
\multirow{2}{*}{HVTauC} & M0$^{(60)}$ & 1.72$^{(56)}$ &-1.56$^{(56)}$ &0.5-1$^{(60)}$ & 4205$^{(57)}$ &  & *0.0000971$^{(5)}$ &  &  &140$^{(5)}$ \\
 &  &  &  &0.72$^{(52}$ &  &  &  &  &  &  \\
FlyingSaucer &  & 2.1$^{(61)}$ &-0.85$^{(61)}$ &  & 3500$^{(60)}$ &  &  &  &  &140$^{(61)}$ \\
HLTau & K3$^{(8)}$ & 2.5$^{(8)}$ &-0.84$^{(8)}$ &0.55$^{(9)}$ & 4350$^{(9)}$ &5.93$^{(17)}$& 1.00E-3$^{(31)}$ &-8.83; $^{(32)}$ &  &140$^{(9)}$ \\
IRAS04189+2650 & K5$^{(10)}$ & 9.96$^{(47}$ &-1.33$^{(47}$ &1.2$^{(47}$ & 4395$^{(47}$ &  &  &-6.76$^{(47}$ &  &  \\
\multirow{3}{*}{VWCha} & K5$^{(57)}$ & 2.39$^{(57)}$ &0.62$^{(57)}$ &1.4$^{(57)}$ & 3955$^{(1)}$ &  & 2.17E-3$^{(12)}$ &-6.95$^{(12)}$ & 0.041$^{(14)}$ &150$^{(9)}$ \\
 & K8$^{(1)}$ & 1.91$^{(2)}$ &0.48$^{(1)}$ &0.6$^{(9)}$ &  &  &  &-7.5$^{(9)}$ &  &168$^{(2)}$ \\
 &  &  &0.22$^{(2)}$ &  &  &  &  &  &  &  \\
\end{longtable}
\tablebib{(1)~\citet{Luhman:07a};
(2)~\citet{Sartori:03a}; (3)~\citet{Antonellini:15a}; (4)~\citet{Meeus:12a};
(5)~\citet{Dent:13a}; (6)~\citet{Mendigutia:12a}; (7)~\citet{McJunkin:14a};
(8)~\citet{Herczeg:14a}; (9)~\citet{Salyk:13a}; (10)~\citet{Skiff:14a};
(11)~\citet{Kraus:11a}; (12)~\citet{Gudel:10a}; (13)~\citet{Manara:14a}; (14)~\citet{Ingleby:11a}; (15)~\citet{Yang:12a}; (16)~\citet{Ingleby:13a}; (17)~\citet{Kraus:09a}; (18)~\citet{Schindhelm:12a}; (19)~\citet{Roberge:13a}; (20)~\citet{Montesinos:09a}; (21)~\citet{Carmona:07a}; (22)~\citet{White:04a}; (23)~\citet{Salyk:11a}; (24)~\citet{Cabrit:90a}; (25)~\citet{Espaillat:07a}; (26)~\citet{Bertout:07a}; (27)~\citet{Verhoeff:11a}; (28)~\citet{Biller:12a}; (29)~\citet{Mendigutia:14a}; (30)~\citet{Carmona:14a}; (31)~\citet{Gudel:07a}; (32)~\citet{Calvet:04a}; (33)~\citet{Trotta:13a}; (34)~\citet{Alecian:13a}; (35)~\citet{Testi:03a}; (36)~\citet{Chapillon:08a}; (37)~\citet{Stark:14a}; (38)~\citet{Lebreton:12a}; (39)~\citet{Gullbring:98a}; (40)~\citet{Manset:09a}; (41)~\citet{Hughes:94a}; (42)~\citet{Pickles:10a}; (43)~\citet{Ammons:06a}; (44)~\citet{Chen:95a}; (45)~\citet{Ingleby:13a}; (46)~\citet{Pontoppidan:10a};  (47)~\citet{Lahuis:07a};  (48)~\citet{Torres:06a};  (49)~\citet{Frasca:15a};  (50)~\citet{Donati:11a};  (51)~\citet{Sacco:12a};  (52)~\citet{Cohen:89a};  (53)~\citet{Zhang:14a};  (54)~\citet{Mathews:12a};  (55)~\citet{Palla:02a};  (56)~\citet{Howard:13a};  (57)~\citet{Johns-Krull:00a};  (58)~\citet{Herczeg:09a};  (59)~\citet{Duchene:10a};  (60)~\citet{Grosso:03a};  (61)~\citet{Andrews:10a}.
}
\end{longtab}


Along with the observational data collection, we employ the same database infrastructure to also provide results from models that were run on a subset of sources. These results include accurate SED fits to 27 sources along with consistent 18 dust and gas models using the DiscAnalysis standards as described in \citet{Woitke:18a}.

Modeling is divided into three major phases. The first phase involves  fitting of stellar and extinction properties, using the UV to near-IR data. Xray-derived extinction data was partly used, but only to see which range of extinction data it supports, in the case of multiple, degenerate extinction estimations. The second phase involves modeling of the SED alone using MCFOST, while the third phase involves the DIANA-standard fitting, using either a combination of MCFOST with ProDiMo,  MCMax with ProDiMo, or just ProDiMo alone. We mention that all codes employed have been benchmarked for consistency \citep{Woitke:16a}. 

Duiring the first modeling phase, additional photometric data are searched for, and initial values for e.g. $T_{\rm eff}$, extinction, distance and luminosity values are looked up in previous spectral analysis papers.  The fitting is then made by varying $T_{\rm eff}$, $L_{\rm star}$ and $A_{\rm V}$ by a genetic algorithm (evolutionary strategy) until a good fit with all selected photometric and (for Herbig Ae) soft UV data is obtained. The fit uses standard PHOENIX photospheric model spectra (which have no additional hot components).  $M_{\rm star}$ and $log(g)$ are found by using stellar evolutionary tracks from \citet{Siess:00a}. In some cases it is necessary to connect the photospheric model  with the UV observations by a power-law. In other cases (mostly for Herbig Ae stars) there is a good overlap. Other groups proceed in a different way here, using early-type template spectra from selected sources, which  have veiled photospheric emission already built-in. During the first phase  $M_{\rm star}$, $L_{\rm star}$, $T_{\rm eff}$, log(g), $A_{\rm V}$, spectral type, distance and  age are estimated as a result of the modeling process. However, initial values for a subset of these parameters can be collected from the rich literature which then are used as  a starting point for the modeling (e.g. see Table~\ref{tab:carla}). 

\begin{figure*}
  \hspace*{-7mm}
  \begin{tabular}{rr}
  \hspace*{-5mm}\includegraphics[width=90mm,height=60mm]{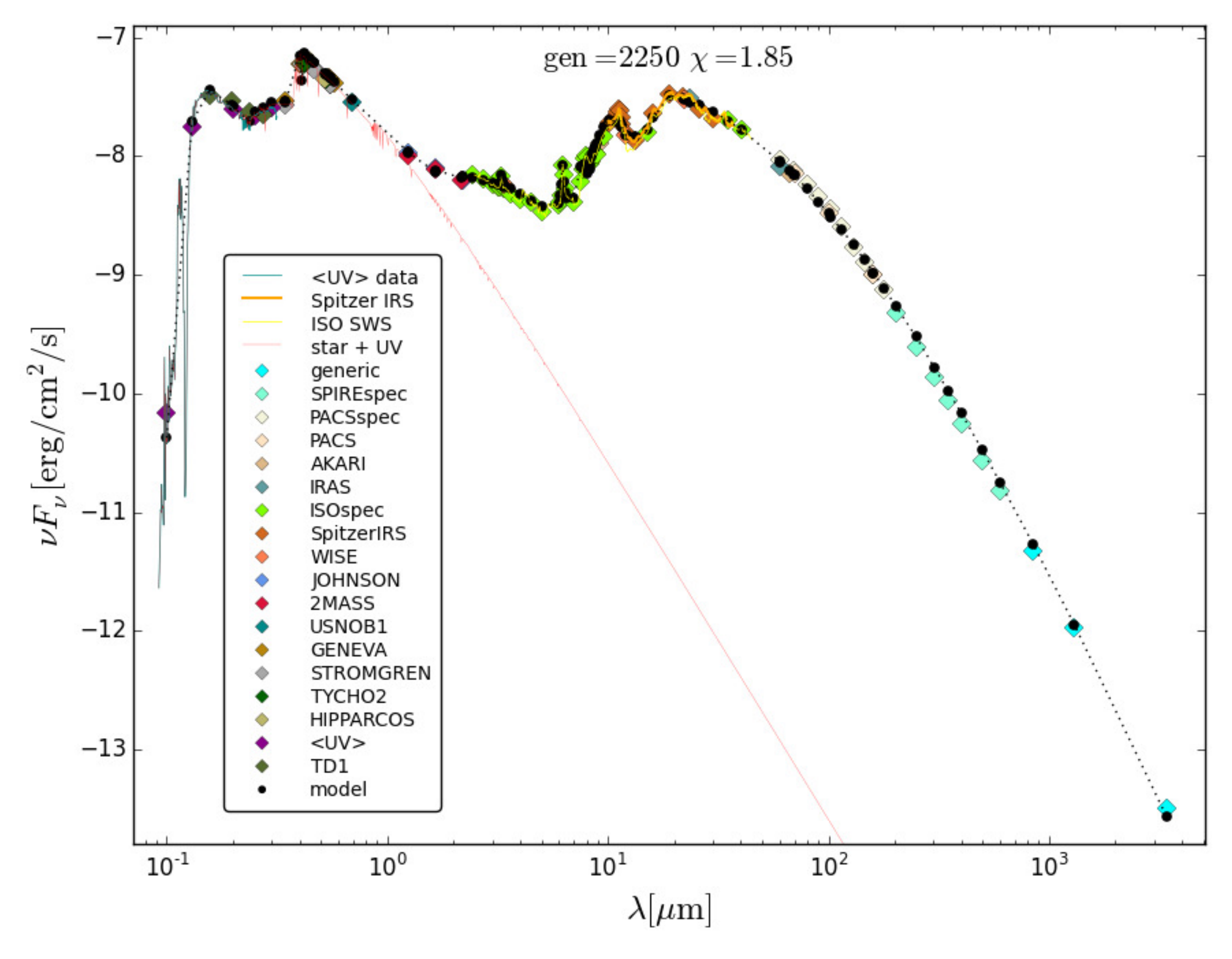} &
 \hspace*{-5mm}\includegraphics[width=90mm,height=60mm]{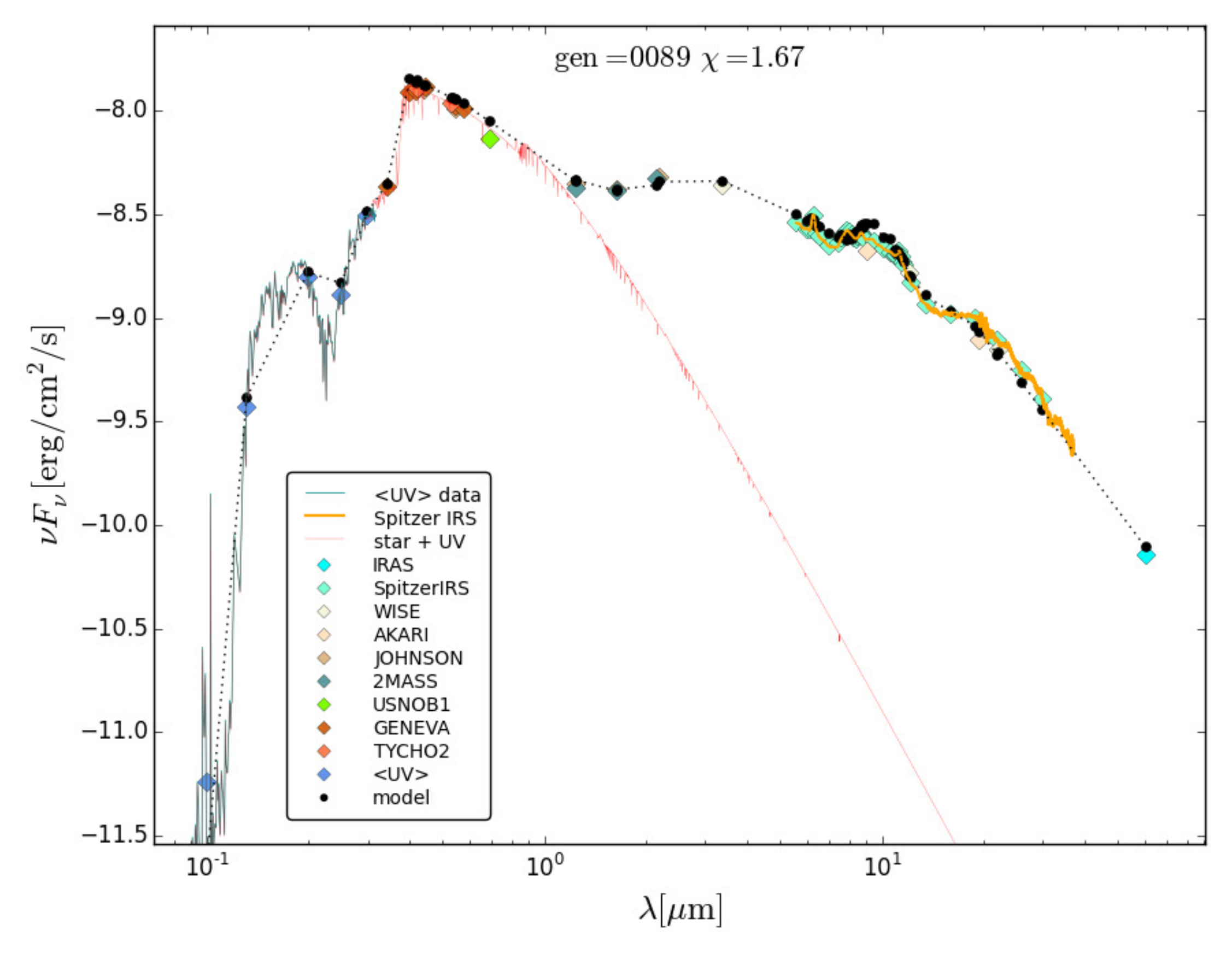}\\[-55mm]
    \sf HD\,100546\hspace*{3mm} & \sf HD\,95881\hspace*{3mm}\\[49mm]
 \end{tabular}
  \hspace*{3mm}\resizebox{18.0cm}{!}{\parbox{17.5cm}{\caption{Examples
        of results from the SED-fits included in the database. The red line is the assumed
        photospheric + UV spectrum of the star. The black dots are the
        fluxes computed by MCFOST, at all wavelength points where we
        could find observations. The other coloured dots and lines and
        observational data as indicated in the embedded
        legends.}\label{fig:SEDs}}}
\end{figure*}

The second phase involves a collection of additional photometric points, extending from the near IR to millimeter wavelengths, including far-IR lines from Herschel and ISO. If PAH features are apparent in the Spitzer/IRS spectra, we include the PAH fitting (amount and average charge of PAHs) consistently in the SED fitting. As a result from the fitting process, the dust mass, disk size and shape, the dust settling and the dust-grain size parameters are constrained and if applicable, the amount and charge of PAHs. The result for 27 SED-fitted sources are included in the database and examples are presented in Fig.~\ref{fig:SEDs}.

In the third phase, line fluxes, line profiles, with resolved images from ALMA  and NICMOS or visibility data from PIONIER and MIDI are included. The modeller decides which data to trust and which not (for example because the data is contaminated by backgroud/foreground cloud emission), assigns fitting weight to each observation, then follows the most appropriate fitting strategy, e.g. genetic fitting algorithm or by-hand-fitting. Fitting in phases two and three starts assuming a single-zone disk without gaps. If this fails, then a two-zone disk model is employed with a possible gap between the two zones. During the third phase, all phase 2 data refitted, where in particular the radial extension, tapering parameters and shadow-casting from the inner to the outer zones can now be fitted using line observations, while the gas/dust ratio is constrained.

 The methods are detailed in \citet{Woitke:16a, Woitke:18a}, and the results are listed in
Table~\ref{tab:stellar} for {\bf 27} sources.  The second step of the modelling is to determine the disc shape, dust and
PAH properties by means of highly automated SED-fits. The result for 27 SED-fitted sources are included in the database and examples are presented in Fig.~\ref{fig:SEDs}.



\begin{table*}
\caption{\bf Stellar parameters, and UV and X-ray irradiation properties, for 27 protoplanetary disks.}
\label{tab:stellar}
\def\z{$\hspace*{-1mm}$}
\def\zz{$\hspace*{-2mm}$}
\def\zzz{$\hspace*{-3mm}$}
\vspace*{2mm}\hspace*{-2mm}
\resizebox{180mm}{!}{
\begin{tabular}{c|ccc|ccc|c|llll}
\hline
&&&&&&&&&\\[-2.0ex]
object & \z SpTyp$^{(1)}$\zzz
       & $d$\,[pc]\z 
       & $A_V^{(15)}$
       & $T_{\rm eff}\rm[K]$\z 
       & \z$L_\star\rm[L_\odot]^{(15)}$\zz
       & \z$M_\star\rm[M_\odot]^{(1)}$\zz
       & \z age\,[Myr]$^{(1)}$\zz
       & $L_{\rm UV,1}^{(2)}$\zz & $L_{\rm UV,2}^{(3)}$\zz  
       & $L_{\rm X,1}^{(4)}$\zz & $L_{\rm X,2}^{(5)}$\zz\\ 
&&&&&&&&&\\[-2.0ex]
\hline
&&&&&&&&&\\[-2.0ex]
HD\,100546 & B9$^{(7)}$& 103 & 0.22 & 10470 & 30.5   & 2.5 & $>\!4.8^{(7)}$& 8.0     & 1.6(-2) & 4.9(-5) & 2.0(-5)\\
HD\,97048  & B9$^{(7)}$& 171 & 1.28 & 10000 & 39.4   & 2.5 & $>\!4.8^{(7)}$& 7.2     & 1.9(-2) & 2.1(-5) & 1.4(-5)\\
HD\,95881  & B9$^{(7)}$& 171 & 0.89 &  9900 & 34.3   & 2.5 & $>\!4.8^{(7)}$& 4.9     & 8.0(-2) 
                                                                          & 2.0(-5)$^{(11)}$\zz & 1.3(-5)$^{(11)}$\zz\zz\zz\\
AB\,Aur    & B9$^{(6)}$& 144 & 0.42 &  9550 & 42.1  & 2.5  & $>\!4.5^{(6)}$& 4.0     & 9.6(-3) & 2.3(-4) & 2.6(-5)\\
HD\,163296 & A1       & 119 & 0.48 &  9000 & 34.7   & 2.47 & 4.6          & 2.1     & 1.8(-2) & 1.5(-4) & 4.4(-5)\\
49\,Cet    & A2       & 59.4& 0.00 &  8770 & 16.8   & 2.0  & 9.8          & 1.0     & 1.7(-4) & 2.6(-4) & 5.3(-5)\\
MWC\,480   & A5       & 137 & 0.16 &  8250 & 13.7   & 1.97 & 11           & 5.6(-1) & 3.8(-3) & 1.5(-4) & 2.5(-5)\\
HD\,169142 & A7       & 145 & 0.06 &  7800 &  9.8   & 1.8  & 13           & 2.2(-1) & 1.6(-5) & 4.8(-5) & 1.4(-6)\\
HD\,142666 & F1$^{(12)}$\zz& 116 & 0.81 &  7050 &  6.3   & 1.6  & $>\!13^{(12)}$ & 3.7(-2)$^{(10)}$\zz & 5.6(-9)$^{(10)}$\zz
                                                                                                  & 1.6(-4) & 1.1(-5)\\
\zz HD\,135344B\z& F3 & 140 & 0.40 &  6620 &  7.6   & 1.65 & 12           & 3.2(-2) & 6.3(-3) & 2.4(-4) & 5.3(-5)\\
V\,1149\,Sco & F9     & 145 & 0.71 &  6080 &  2.82  & 1.28 & 19           & 5.1(-2) & 1.4(-2) & 3.7(-4) & 2.8(-5)\\
Lk\,Ca\,15 & K5$^{(16)}$ &140 &1.7  &  4730 &  1.2   & 1.0  & $\approx\!2^{(16)}$  
                                                                          & 5.1(-2) & 6.3(-3) & 5.5(-4) & 1.7(-4)\\
\zzz USco\,J1604-2130\z& K4& 145& 1.0& 4550& 0.76   & 1.2  & 10           
                           & 4.0(-3)$^{(17)}$\zz & 3.1(-4)$^{(17)}$\zz & 2.6(-4)$^{(18)}$\zz & 5.3(-5)$^{(18)}$\zz\zz\zz \\
RY\,Lup    & K4       & 185 & 0.29 &  4420 &  2.84  & 1.38 & 3.0          & 2.4(-3) & 1.5(-4) & 4.3(-3) & 3.6(-4)\\
CI\,Tau    & K6       & 140 & 1.77 &  4200 &  0.92  & 0.90 & 2.8          & 2.0(-3) & 8.7(-5) & 5.0(-5) & 1.0(-5)\\
TW\,Cha    & K6       & 160 & 1.61 &  4110 &  0.594 & 1.0  & 4.3          & 7.2(-2) & 4.4(-3) & 3.4(-4) & 1.0(-4)\\
RU\,Lup    & K7       & 150 & 0.00 &  4060 &  1.35  & 1.15 & 1.2          & 1.4(-2) & 9.0(-4) & 7.1(-4) & 3.4(-4)\\
AA\,Tau    & K7       & 140 & 0.99 &  4010 &  0.78  & 0.85 & 2.3          & 2.3(-2) & 5.8(-3) & 1.1(-3) & 3.2(-4)\\
TW\,Hya    & K7       & 51  & 0.20 &  4000 &  0.242 & 0.75 & 13           & 1.1(-2) & 4.2(-4) & 7.7(-4) & 7.0(-5)\\
GM\,Aur    & K7       & 140 & 0.30 &  4000 &  0.6   & 0.7  & 2.6          & 6.6(-3) & 2.8(-3) & 7.0(-4) & 1.2(-4)\\
BP\,Tau    & K7       & 140 & 0.57 &  3950 &  0.89  & 0.65 & 1.6          & 1.3(-2) & 1.1(-3) & 5.9(-4) & 2.5(-4)\\ 
DF\,Tau$^{(14)}$ & K7  & 140 & 1.27 &  3900 &  2.46  & 1.17 & $\approx\!2.2^{(14)}$ 
                                                                          & 3.6(-1) & 2.9(-1) & $-^{(13)}$ & $-^{(13)}$ \\
DO\,Tau    & M0       & 140 & 2.6  &  3800 &  0.92  & 0.52 & 1.1          & 1.3(-1) & 2.7(-2) & 1.1(-4) & 4.1(-5)\\
DM\,Tau    & M0       & 140 & 0.55 &  3780 &  0.232 & 0.53 & 6.0          & 7.0(-3) & 6.3(-4) & 8.4(-4) & 2.9(-4)\\
CY\,Tau    & M1       & 140 & 0.10 &  3640 &  0.359 & 0.43 & 2.2          & 7.3(-4) & 7.1(-5) & 2.1(-5) & 6.9(-6)\\
FT\,Tau    & M3       & 140 & 1.09 &  3400 &  0.295 & 0.3  & 1.9          & 5.2(-3)$^{(8)}$\zz & 8.4(-4)$^{(8)}$\zz 
                                                                              & 2.3(-5)$^{(9)}$ & 7.0(-6)$^{(9)}$\zz\zz\zz\\
RECX\,15   & M3       & 94.3& 0.65 &  3400 &  0.091 & 0.28 & 6.5          & 6.3(-3) & 4.0(-4) & 1.7(-5) & 8.2(-6)\\[1mm]
\hline
\end{tabular}}\\[2mm]
\hspace*{4mm}\resizebox{160mm}{!}{\parbox{175mm}{The table shows
    spectral type, distance $d$, interstellar extinction $A_V$, 
    effective temperature $T_{\rm eff}$, stellar luminosity $L_\star$,
    stellar mass $M_\star$, age, and UV and X-ray luminosities
    without extinction, i.\,e.\ as seen by the disk. Numbers written
    $A(-B)$ mean $A\times 10^{-B}$.  The UV and X-ray luminosities 
    are listed in units of $[L_\odot]$.\\[0.5mm]
$^{(1)}$: spectral types, ages and stellar masses are consistent with evolutionary
    tracks for solar-metallicity pre-main sequence stars \hspace*{4mm} by
    \citet{Siess:00a}, using $T_{\rm eff}$ \& $L_\star$ as input,\\
$^{(2)}$: FUV luminosity from 91.2 to 205\,nm, as seen by the disk,\\
$^{(3)}$: hard FUV luminosity from 91.2 to 111\,nm, as seen by the disk,\\
$^{(4)}$: X-ray luminosity for photon energies $>\!0.1\,$keV, as seen by the disk,\\
$^{(5)}$: hard X-ray luminosity from 1\,keV to 10\,keV, as seen by the disk,\\
$^{(6)}$: no matching track, values from closest point at $T_{\rm eff}\!=\!9650\,$K 
         and $L_\star\!=\!42\rm\,L_{\odot}$,\\ 
$^{(7)}$: no matching track, values from closest point at $T_{\rm eff}\!=\!10000\,$K 
         and $L_\star\!=\!42\rm\,L_{\odot}$,\\ 
$^{(8)}$: no UV data, model uses an UV-powerlaw with $f_{\rm UV}\!=\!0.025$ 
         and $p_{\rm UV}\!=\!0.2$
         \citep[see][]{Woitke:16a}\\
$^{(9)}$: no detailed X-ray data available, model uses a
         bremsstrahlungs-spectrum with $L_X\!=8.8\times 10^{28}$\,erg/s and
         $T_X\!=\!20\,$MK,  \hspace*{4mm} based on archival XMM survey data
         (M.~G{\"u}del, priv.\,comm.),\\
$^{(10)}$: ``low-UV state'' model, where a purely photospheric spectrum is assumed,\\
$^{(11)}$: no X-ray data available, X-ray data taken from HD\,97048,\\
$^{(12)}$: no matching track, values from closest point at $T_{\rm eff}\!=\!7050\,$K
         and $L_\star\!=\!7\rm\,L_{\odot}$,\\
$^{(13)}$: no X-ray data available,\\
$^{(14)}$: resolved binary, 2$\times$ spectral type M1, luminosities $0.69\,L_\odot$ and $0.56\,L_\odot$, 
          separation $0.094''\approx 13\,$AU \\  \hspace*{5mm} \citep{Hillenbrand:04a},\\
$^{(15)}$: derived from fitting our UV, photometric optical and X-ray data\\
$^{(16)}$: no matching track, values taken from \citep{Drabek:16a,Kraus:12a},\\
$^{(17)}$: no UV data, model uses $f_{\rm UV}\!=\!0.01$ and $p_{\rm UV}\!=\!2$ 
          \citep[see][App.~A for explanations]{Woitke:16a},\\
$^{(18)}$: no X-ray data, model uses $L_X\!=\!10^{30}$erg/s and $T_X\!=\!20\,$MK 
          \citep[see][App.~A for explanations]{Woitke:16a}.}}
\end{table*}

\section{Summary}\label{sec:4}

In this paper we presented a large sample of Class II and III, T Tauri and Herbig Ae systems with spectral types ranging from B9 to M3 which cover ages between 1 and 10 Myr. The sample of 85 sources in expected to include another 30-40 sources in the near future, rendering this one of the largest and most complete collections of its kind. The collection was assembled combining data from more than 50 observational facilities and 100 published articles in a transparent manner, so that each dataset can be back-traced to each original resources. In addition, 27 of the sources in the collection have their SEDs consistently modeled with dust radiative transfer models (MCFOST, MCMAX and ProDiMo)\footnote{All SED input files and output models available at:  http://www-star.st-and.ac.uk/~pw31/DIANA/SEDfit/ }, and a subset of 18 that have both dust and gas consistently modeled with ProDiMo\footnote{Gas line input files and output models available at:  http://www-star.st-and.ac.uk/~pw31/DIANA/DIANAstandard/ }. The user interface and the supporting DIOD database provide the user with the flexibility to compare different characteristics among the sample sources and models, but also directly download data for further use. We believe that this collection with its future extensions will provide a reference point, facilitating observational and modeling studies of protoplanetary disks.

\begin{acknowledgements}

The research leading to these results has received funding from the European Union Seventh Framework Programme FP7-2011 under grant agreement no 284405. OD acknowledges support from the Austrian Research Promotion Agency (FFG) for the Austrian Space Applications Program (ASAP) project JetPro* (FFG-854025).

 \end{acknowledgements}

\begin{tiny}

\bibliographystyle{aa}
\bibliography{diana_bdp}
\end{tiny}


\Online

\begin{appendix}

\section{The DiscAnalysis Object Database (DIOD).}\label{app:db}

The multi-wavelength datasets available through the \textit{DIANA Object Database} repository are processed end-products, as opposed to raw data, that can be directly compared to models. Deriving meaningful physical quantities from raw astrophysical data is a complex process that requires good knowledge of the instrumentation involved, together with adequate experience and advanced programming skills. Therefore the dissemination of processed end-products has a greater impact to the community as they can be directly used for analysis and interpretation. As an example we mention the SDSS survey that has been up to now the most prolific project in terms of scientific outcome when compared  to cost. We therefore anticipate that along with the advancements and increasing interest in the field, our database will become a point of reference for the study of protoplanetary disks.

\subsection{The end-user interface}

The database functionalities for the end-user are limited to searching, inspecting and downloading datasets and models for either single or multiple sources.

\begin{figure}
\centering
 \includegraphics[width=80mm]{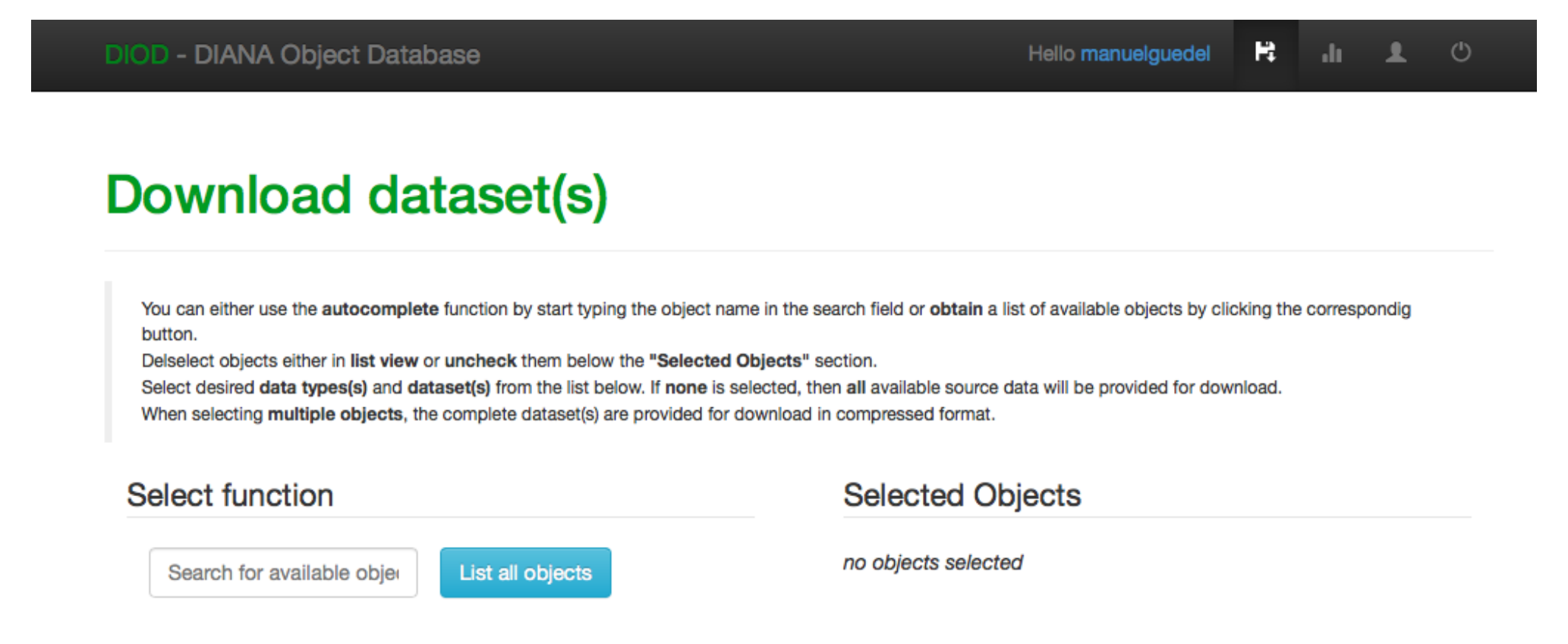} 
\caption{\small The data search/download window, without any sources selected. The user can either search for a source by its name, or get a list of all available sources when selecting the \textit{List all sources} button.}
\label{fig:4a}
\end{figure}

\begin{figure}
\centering
 \includegraphics[width=80mm]{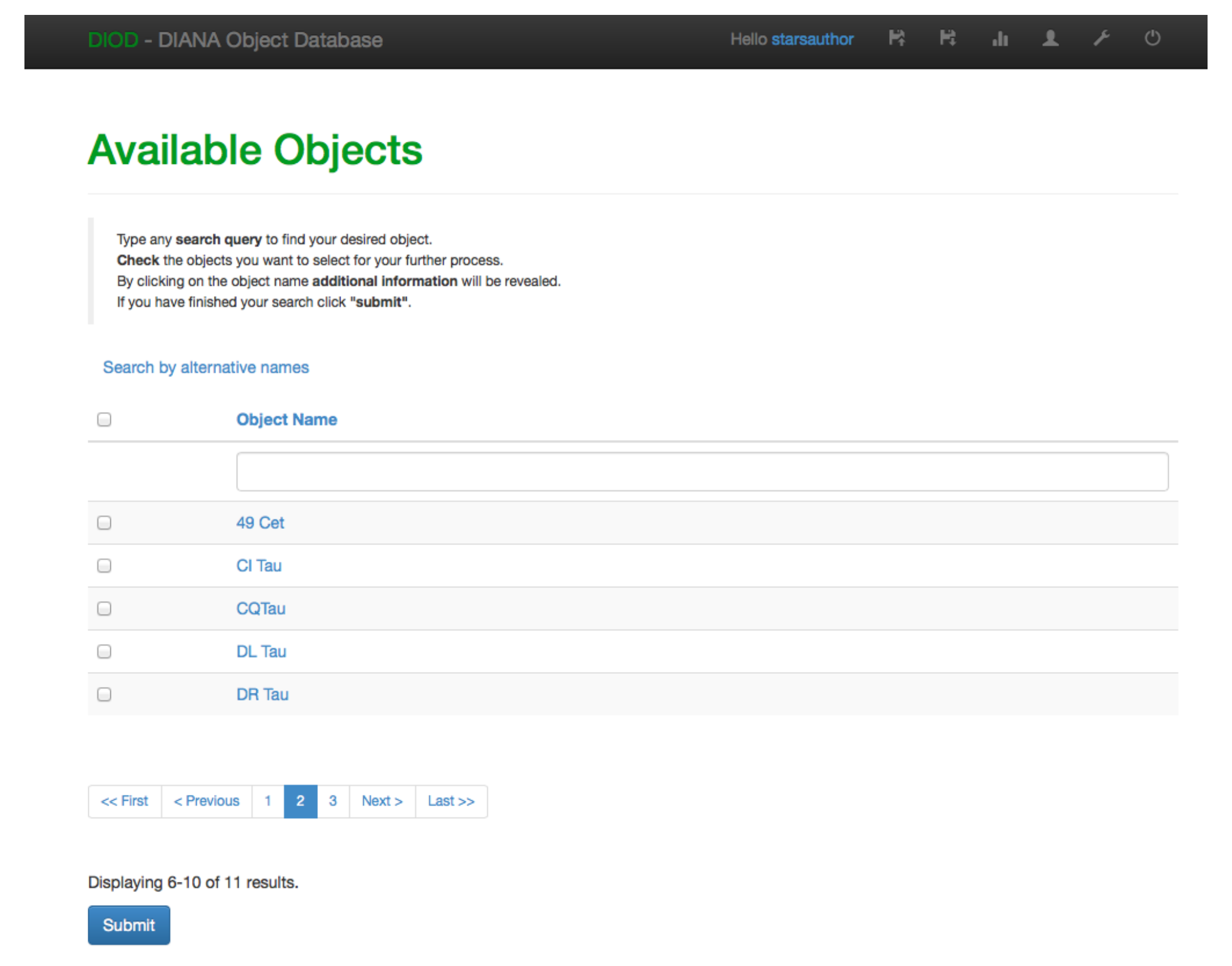} 
\caption{\small List of available sources that can be selected for further investigation/download by marking the checkbox on the right of the source name. On the top the user can search the source list with alternative names, as provided by source name resolvers. Direct selection of a source name presents information retrieved by three online name resolver services. (see Fig.~\ref{fig:6a})}
\label{fig:5a}
\end{figure}

\begin{figure}
\centering
 \includegraphics[width=80mm]{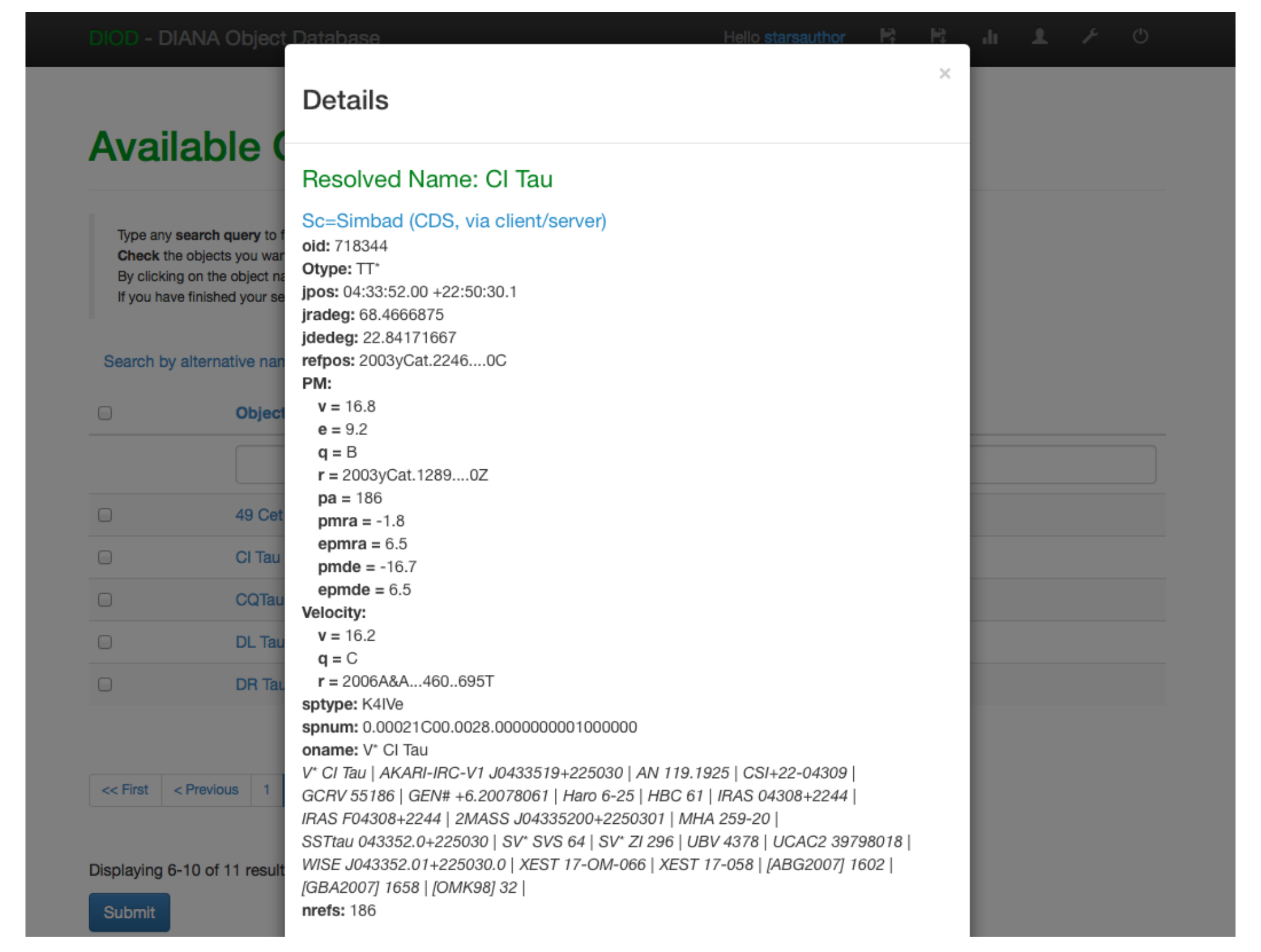} 
\caption{\small When selecting a source, a pop-up window presents its details as retrieved from SIMBAD, NED and VizieR/CDS name resolvers.}
\label{fig:6a}
\end{figure}

When the \textit{Download window} is selected in the welcome page, the user is redirected to the data search and retrieval interface. As in most pages of the interface, a short description of the available functions are displayed on the top part. At this stage the user is provided with two options: either search for a source by name, or select sources from a list that is returned when the \textit{List all objects} button is selected (Fig.~\ref{fig:4a}). In the list window, presented in Fig.~\ref{fig:5a} the user can search the database for alternative source names as provided by online name resolvers. In addition, selection of a source name brings into the foreground a pop-up window, which provides detailed information on the source as retrieved by the SIMBAD, NED and VizieR/CDS name servers (Fig.~\ref{fig:6a}).

\begin{figure}
\centering
 \includegraphics[width=80mm]{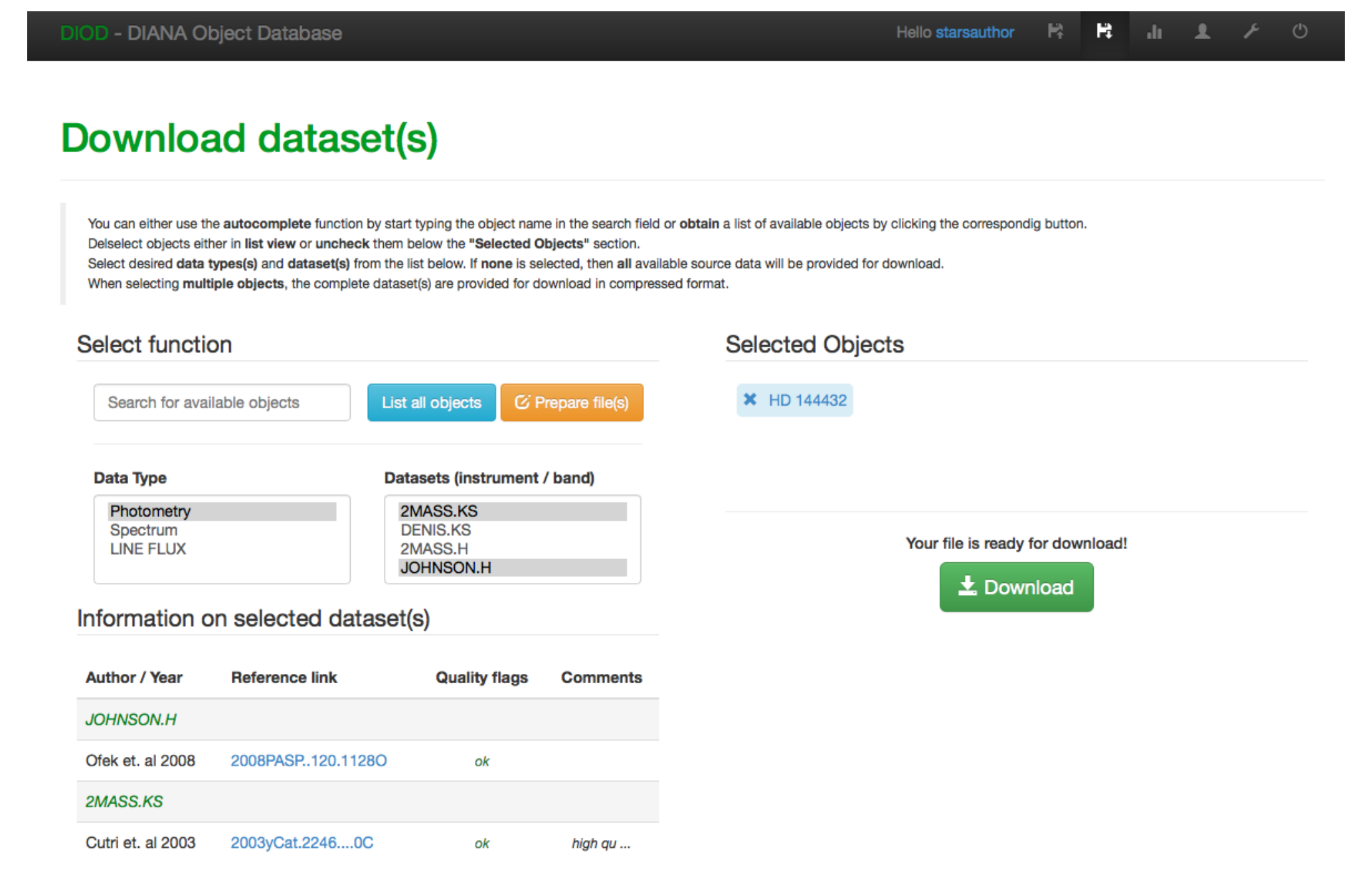} 
\caption{\small  When a single source is selected, the user can examine all available data by selecting first the \textit{Data Type} and then the desired \textit{Dataset}. At the bottom of the screen a detailed description of the selected dataset(s) appears, providing active links to the original work where the data is retrieved, along with quality flags and additional comments.}
\label{fig:7a}
\end{figure}

Once one or more sources are selected, they appear on the right panel of the search and retrieval window (Fig.~\ref{fig:7a}), under the \textit{Selected Objects} section. At the same section, sources can be removed by selecting the \textit{``X''} symbol next to each source name. Below the \textit{Select function} pane, two lists appear providing the possibility to select first the \textit{Data Type}  (e.g. photometric points, spectra, line fluxes, line shapes, etc). For each data type, the \textit{Dataset} box is populated with available options (instruments, bands, etc). Information on all selected datasets is listed below under the pane bearing the same title. Here, the user can find information on the original source where the data is retrieved from and an active link to the original publication (if available). In addition, the user is informed on the data quality by relevant flags and comments on the data. Comments are truncated to save space, but  the full comments are displayed once the user ``hoovers over'' the mouse pointer.

\begin{figure}
\centering
 \includegraphics[width=80mm]{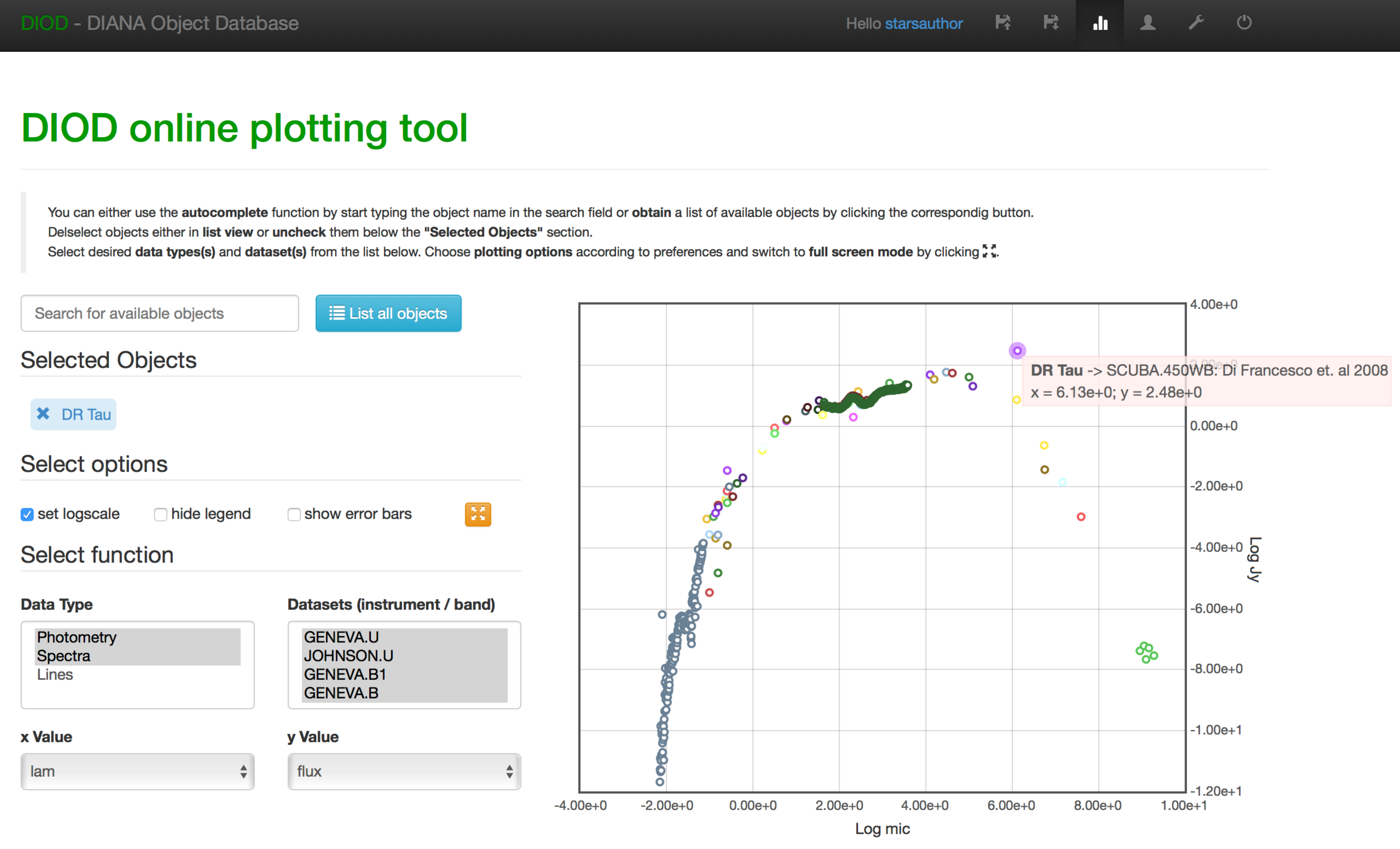} 
\caption{\small  The basic plotting/overview functionality of {\tt DIOD}; when desired datasets are selected, they can be plotting through a number of pre-selected display modes. Color encoding of the data points in the plot corresponds to different datasets/objects. When the mouse pointer is over a specific data point, a popup window displays relevant information on its provenance.}
\label{fig:8a}
\end{figure}

\begin{figure}
\centering
 \includegraphics[width=80mm]{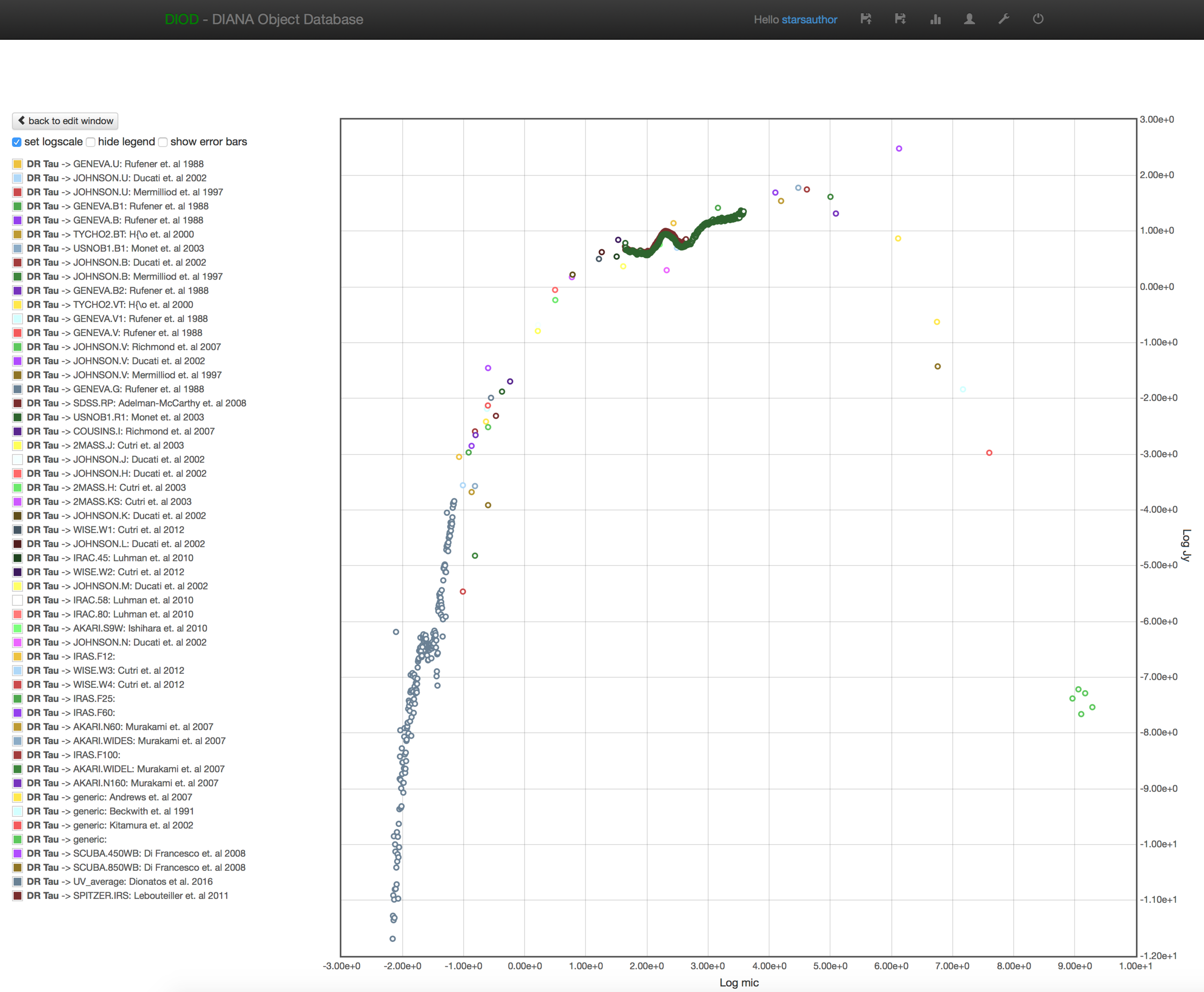} 
\caption{\small  Full screen plotting window, displaying on the left-hand-side a complete list on the provenance of all data points.}
\label{fig:8a}
\end{figure}

\newpage

\subsection{Data preview \& retrieval}

Selected datasets can be previewed and compared on the plotting window. Once desired datasets/models are selected for one or more sources they are plotted using a color-encoded scheme that allows to distinguish the data origin into individual sources, datasets and instruments. For a ``mouse-over'' action a pop-up window displays the provenance of the specific data point (see Fig.~\ref{fig:8a} ). When the full screen plotting mode is selected, then a complete list on the provenance of all data points is provided on the left-hand side of the plot.  Selected data can be prepared for direct download when the user selects the \textit{Prepare files} button. If no selections under the \textit{Data Type} and \textit{Dataset(s)} boxes have been made, then the full dataset for the selected source is prepared. After the \textit{Prepare files} button has been pressed, a \textit{Download} button appears under the \textit{Selected Objects} pane, and when pressed it provides the user with the selected data. 

The option to inspect and select parts of the available data is only provided when a single object is selected. For multiple objects, the user can only retrieve the full datasets, which are provided as a compressed file.

\section{UV data co-adding and post-processing}\label{App:A}

Here we provide a detailed account of the process followed to co-add the UV data from different instruments and observing runs. The process is divided in three basic steps, as follows:

{\bf STEP 1:\ \ } We first define wavelength-bins as
\begin{equation}
  \lambda_{j+1} = \lambda_j \left(1+\frac{1}{R}\right)
\end{equation}
starting with $j\!=\!0$ (where $\lambda_0\!=\!950$\,\AA\ if FUSE data
is available, otherwise $\lambda_0\!=\!1150$\,\AA). Resolution
$R\approx200$ is an adjustable parameter. Index $j$ is then increased until
$j\!=\!J$ where $\lambda_J>3350$\,\AA\ is achieved, forming $J$
wavelength bins. 

\bigskip\noindent {\bf STEP 2:\ \ } Each observational data set $d$ 
is used to (partly) fill these wavelength bins by integration as
\begin{eqnarray}
  \lambda^c_j &=& \frac{1}{2}\big(\lambda_{j-1}+\lambda_{j}\big) \\
  F^d_j       &=& \int_{\lambda_{j-1}}^{\lambda_{j}} F^d_\lambda d\lambda
                  \Bigg{/}
		  \int_{\lambda_{j-1}}^{\lambda_{j}} d\lambda \ .
  \label{Fint}
\end{eqnarray}
, where $c$ is the center wavelength of the bin $j$. However, even one data set $d$ may come in several, partly overlapping chunks 
(echelle spectra) and some points may be flagged and hence 
need to be discarded. To deal with all these special cases, every
valid original data point $\{\lambda^d_i,F^d_i\}$ is assigned a spectral width as 
\begin{eqnarray}
  \Delta\lambda^d_i &=& \lambda^{d,c}_{i}-\lambda^{d,c}_{i+1}\\
  \lambda^{d,c}_{i}   &=& \frac{1}{2}\big(\lambda^d_{i-1}+\lambda^d_{i}\big)\\
  \lambda^{d,c}_{i+1} &=& \frac{1}{2}\big(\lambda^d_{i+1}+\lambda^d_{i}\big)\\
\end{eqnarray}
This width $\Delta\lambda^d_i$, or precisely speaking the overlap of
$\Delta\lambda^d_{i,j}$ with the interval $[\lambda_{j-1},\lambda_j]$,
is used to numerically calculate the integrals in
Eq.\,(\ref{Fint}), replacing them by simple sums as
\begin{eqnarray}
  F^d_j      &=& \sum_i F^d_i \Delta\lambda^d_{i,j} 
        \Bigg{/} \sum_i \Delta\lambda^d_{i,j} \\
  \sigma^d_j &=& \sqrt{\sum_i \big(\sigma^d_i \Delta\lambda^d_{i,j}\big)^2} 
              \Bigg{/} \sum_i \Delta\lambda^d_{i,j} 
\end{eqnarray}
Step 2 results in a number of spectra
$\{\lambda^c_j,F^d_j,\sigma^d_j\}$ on the same low-resolution
wavelength grid for every object.  However, usually only a (small)
subset of the spectral bins $p$ are populated by a single
observational data set.

\bigskip\noindent {\bf STEP 3:\ \ } {\it All} available spectra
sets $d$ are co-added, using the inverse square of the bin uncertainty 
as summation weight
\begin{eqnarray}
  w^d_j &=& 1\Big{/}(\sigma^d_j)^2        \\
  \bar{F}_j     &=& \sum_d w^d_j F^d_j \Bigg{/} \sum_d w^d_j \label{flux} \\
  \bar\sigma_j  &=& \sqrt{\sum_d \left(w^d_j\,\sigma^d_j\right)^2}
                                      \Bigg{/} \sum_d
                                      w^d_j \label{sigma} \ ,
\end{eqnarray}
where the uncertainties $\bar\sigma_j$ follow from Gauss' error propagation law
applied to Eq.\,(\ref{flux}), i.e. $\bar\sigma_j$ is the error of the
mean value
\begin{eqnarray}
  (\bar\sigma_j)^2 = \Delta^2 \bar{F}_j 
  &=& \sum_d \left(\frac{\partial}{\partial F^d_j} 
             \left(\frac{\sum_{d'} w^{d'}_j F^{d'}_j}
                        {\sum_{d'} w^{d'}_j}\right) \sigma^d_j\right)^2 \\
  &=& \frac{1}{\big(\sum_{d'} w^{d'}_j\big)^2} 
      \sum_d \left(\sum_{d'} w^{d'}_j\,\delta_{dd'}\,\sigma^d_j\right)^2 \\
  &=& \frac{1}{\big(\sum_{d} w^{d}_j\big)^2} 
      \sum_d \left(w^d_j\,\sigma^d_j\right)^2 
\end{eqnarray} 
This procedure results in one set of co-added and resolution-decreased spectral data points
$\{\lambda^c_j,\bar{F}_j,\bar\sigma_j\}\ (j=1,...,J)$ for each object.

\bigskip\noindent {\bf STEP 4:\ \ } After steps 1 and 2, the data can
still be very noisy, negative or otherwise statistically irrelevant,
and it would be an error to use such data for disc irradiation. The
idea in the following is to join neighboring $\lambda$-bins, until
statistically relevant data results:

\begin{eqnarray}
  w_j       &=& 1\Big{/}(\bar\sigma_j)^2 \\    
  \lambda_k &=& \frac{1}{\sum_{j=j_0}^{j_0-1+N(k)} w_j} 
                \sum_{j=j_0}^{j_0-1+N(k)} w_j\,\lambda^c_j  \\
  F_k       &=& \frac{1}{\sum_{j=j_0}^{j_0-1+N(k)} w_j} 
                \sum_{j=j_0}^{j_0-1+N(k)} w_j\,\bar{F}_j \\
  \sigma_k  &=& 1\Big{/}{\sqrt{\sum_{j=j_0}^{j_0-1+N(k)} w_j}} 
\end{eqnarray}
We start at $j_0\!=1\!$. The number of joined bins $N(k)$ is increased
from 1 to any number, until
\begin{equation}
  F_k > \alpha\cdot\sigma_k
\end{equation}
is achieved, where $\alpha\approx 3$ is an adjustable parameter. Once
$N(k)$ is found, the new data point $\{\lambda_k, F_k, \sigma_k\}$ is added,
$j_0$ is incremented by $N(k)$, and $k$ is incremented by one. This procedure 
gives co-added, further resolution-decreased, but statistically relevant
$\langle\rm UV\rangle$-spectrum $\{\lambda_k, F_k, \sigma_k\}\ (k=1,...,K)$.

\end{appendix}

\end{document}